\documentclass[aps,twocolumn,pra,superscriptaddress,amsmath,amssymb]{revtex4-1}

\ifx\pdfoutput\undefined
\usepackage[dvipdfmx]{graphicx}
\usepackage[dvipdfmx]{hyperref}
\usepackage[dvipdfmx]{color}
\else
\usepackage{graphicx}
\usepackage{hyperref}
\usepackage{color}
\fi

\usepackage{epsfig,epstopdf}
\usepackage{dcolumn}
\usepackage{bm}
\usepackage{color}
\usepackage{comment}
\usepackage{soul}
\usepackage{lineno}

\renewcommand{\figurename}{{\bf Fig.}}

\begin{document}

\title{Aperiodic approximants bridging quasicrystals and modulated structures}
\author{Toranosuke Matsubara}
\author{Akihisa Koga}
\affiliation{Department of Physics, Tokyo Institute of Technology, Meguro, Tokyo 152-8551, Japan}
\author{Atsushi Takano}
\affiliation{Department of Molecular and Macromolecular Chemistry, Nagoya University, Nagoya, Aichi 464-8603, Japan}
\author{Yushu Matsushita}
\affiliation{Toyota Physical and Chemical Research Institute, Nagakute, Aichi 480-1192, Japan}
\author{Tomonari Dotera}
\affiliation{Department of Physics, Kindai University, Higashi-Osaka, Osaka 577-8502, Japan}

\begin{abstract}
Aperiodic crystals constitute a fascinating class of materials that includes incommensurate (IC) modulated structures~\cite{WJJ1981,Bak1982} and quasicrystals (QCs)~\cite{shechtman_1984,Penrose,Levine1984,Janssen2007,Janssen2014,deBissieu2019}. Although these two categories share a common foundation in the concept of superspace, the relationship between them has remained enigmatic and largely unexplored. 
Here, we show ``any metallic-mean" QCs~\cite{Beenker,Dotera2017,Sam_big}, surpassing the confines of Penrose-like structures, and explore their connection with IC modulated structures. In contrast to periodic approximants of QCs~\cite{Deguchi,RevGoldman}, our work introduces the pivotal role of ``aperiodic approximants"~\cite{Nakakura2019}, articulated through a series of $k$-th metallic-mean tilings serving as aperiodic approximants for the honeycomb crystal, while simultaneously redefining this tiling as a metallic-mean IC modulated structure, highlighting the intricate interplay between these crystallographic phenomena. We extend our findings to real-world applications, discovering these unique tiles in a terpolymer/homopolymer blend~\cite{Izumi2015} and applying our QC theory to a colloidal simulation displaying planar IC structures~\cite{Engel2011,Schoberth2016}. In these structures, domain walls are viewed as essential components of a quasicrystal, introducing additional dimensions in superspace. Our research provides a fresh perspective on the intricate world of aperiodic crystals, shedding light on their broader implications for domain wall structures across various fields~\cite{Tendeloo1976,Landuyt1985}.
\end{abstract}

\maketitle

\noindent
Prior to the discovery of quasicrystals (QCs) as the advent of aperiodicity in materials science, incommensurate (IC) modulated structures and IC composite structures were investigated, wherein IC spatial modulations were added to the background crystalline structures~\cite{WJJ1981,Bak1982}. Then, the concept of superspace and additional degrees of freedom known as phasons were introduced. After Shechtman's discovery~\cite{shechtman_1984}, aperiodic crystals, including IC modulated structures and QCs, emerged as an important class of materials~\cite{Penrose,Levine1984,Janssen2007,Janssen2014,deBissieu2019}. Aperiodicity is characterized by irrational numbers, thereby making a distinction between the two. In QCs, the irrational numbers are locked by two-length scales~\cite{Dotera2014,Beenker,Dotera2017,Sam_big} in geometry, whereas in IC modulated structures, these numbers remain unlocked. QCs typically consist of concentric shell clusters arranged quasiperiodically, locking in the golden mean in icosahedral QCs. In certain alloys, such as Au-Al-Yb~\cite{Deguchi}, periodic approximants are synthesized, where the clusters are arranged periodically. Consequently, ``periodic approximants" have been extensively studied to gain a better understanding of QCs. A crucial aspect of periodic approximants is that they exhibit local quasiperiodicity (resembling QCs), but globally display periodicity. Moreover, as the degree of approximation increases, these periodic approximants converge towards QCs~\cite{RevGoldman}.

A complementary treatment has been explored where a quasiperiodic structure approaches the periodic one by varying the characteristic irrational. Such treatments are known as ``aperiodic approximants"~\cite{Nakakura2019}. An elementary example of aperiodic approximants is the generalized Fibonacci sequence, which comprises two letters, $A$ and $B$. The sequence is generated by the substitution rules: $ A \rightarrow AA\cdots AB(=A^kB)$ and $B \rightarrow A$, where $k$ is a natural number. The numbers of the letters $A$ and $B$ at iteration $n$ ($N_A^{(n)}$ and $N_B^{(n)}$) satisfy 
\begin{equation}
\left(\begin{array}{l}
N_A^{(n+1)} \\
N_B^{(n+1)}
\end{array}\right)=\left(\begin{array}{ll}
k & 1 \\
1 & 0
\end{array}\right)\left(\begin{array}{l}
N_A^{(n)} \\
N_B^{(n)}
\end{array}\right),
\end{equation}
where the maximum eigenvalue of the matrix is given by the metallic-mean: $\tau_k = (k + \sqrt{k^2 + 4})/2$. When $k=1$, the sequence is the conventional Fibonacci one with the golden mean. The eigenvector of the matrix is given by $(\tau_k, 1)^T$, indicating $N_A^{(n)}/N_B^{(n)}\rightarrow\tau_k$ as $n\rightarrow\infty$, where the sequence is filled with the letter $A$ for large values of $k$. In the limit $k\rightarrow\infty$, the sequence converges to a crystal consisting of consecutive ``$A$"s. Hence, the generalized Fibonacci sequence with the metallic-mean can be considered as the aperiodic approximants of the one-dimensional crystal $AAA\cdots$. 

\begin{figure}[htbp]
{\small
\centering
\includegraphics[width=\linewidth]{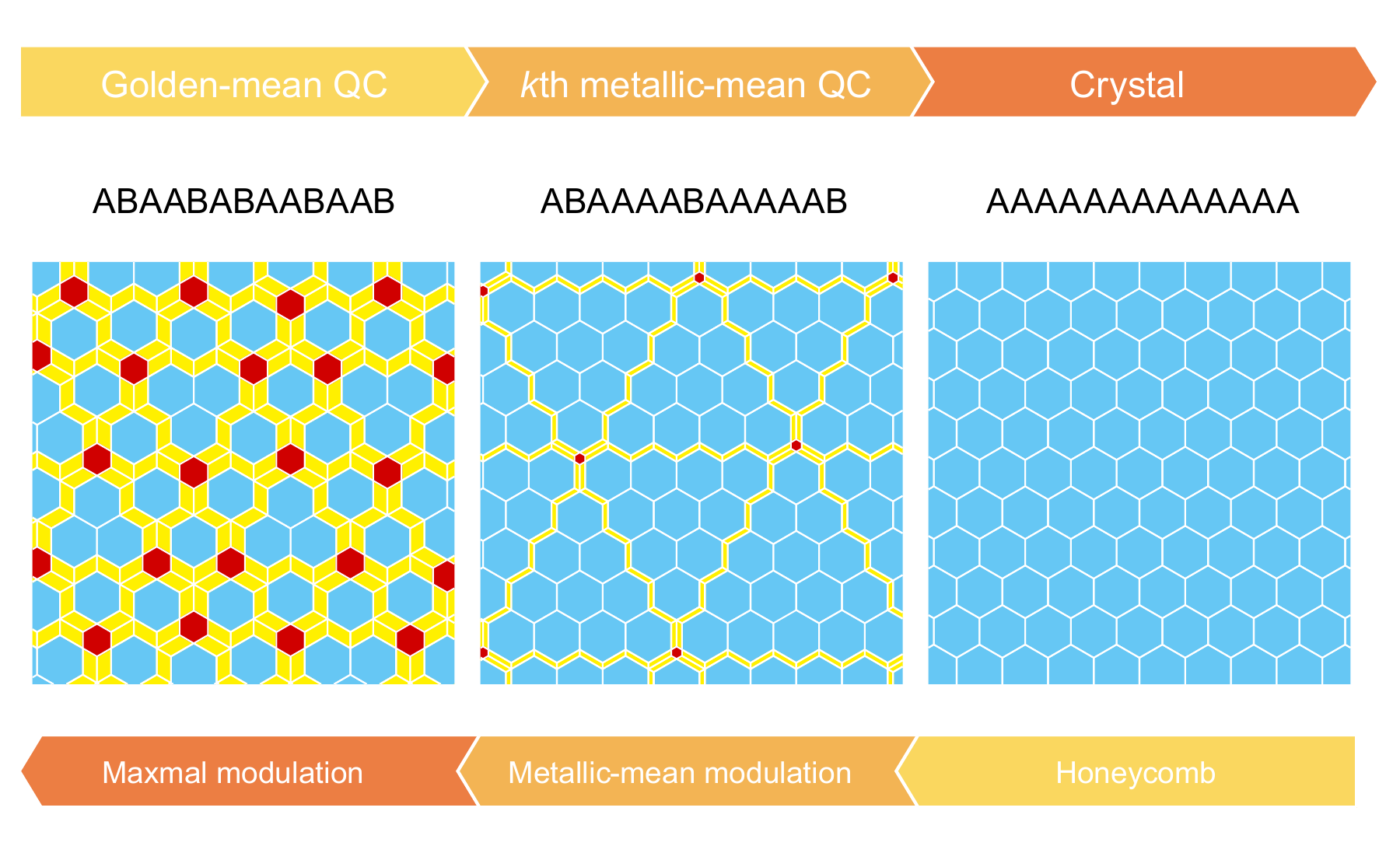}
\caption{{\bf Aperiodic approximants.} Schematic showing the role of aperiodic approximants as a link between quasicrystals and periodic crystals. The link is a series of $k$-th metallic-mean tilings as an aperiodic approximant of the honeycomb crystal (top arrow), which is simultaneously regarded as a metallic-mean (incommensurate) modulated honeycomb crystal (bottom arrow).
}
\label{fig1}
}
\end{figure}

Similarly, aperiodic approximants of triangular lattices were proposed. These metallic-mean quasiperiodic tilings start from the bronze-mean tilings~\cite{Dotera2017}. Majority tiles increase with increasing $k$, and eventually, the systems converge to the triangular lattices in the limit $k\rightarrow\infty$. A crucial aspect of these is that they are locally periodic, but globally quasiperiodic, in other words, they are considered as planar IC modulated structures.

Here we present hexagonal metallic-mean approximants of the honeycomb lattice, which bridge the gap between QCs and IC modulated structures. Schematic of our view is presented in Fig.~\ref{fig1}. As the metallic-mean increases, the size of honeycomb domains bounded by the parallelograms also increases, and the whole tiling converges to the honeycomb lattice. Conversely, the metallic-mean IC modulation is introduced to the honeycomb crystals in terms of the metallic-mean tilings. The domain walls composed of parallelograms in the honeycomb crystal are regarded as ingredients of a quasicrystal adding superspace dimensions. Significantly, we show that the metallic-mean tiling scheme is applicable to a polymer system~\cite{Izumi2015} and colloidal systems~\cite{Engel2011,Schoberth2016} in soft-matter self-assemblies. 
\bigskip

\begin{figure*}[htb]
{\small
\begin{center}
\includegraphics[width=\linewidth]{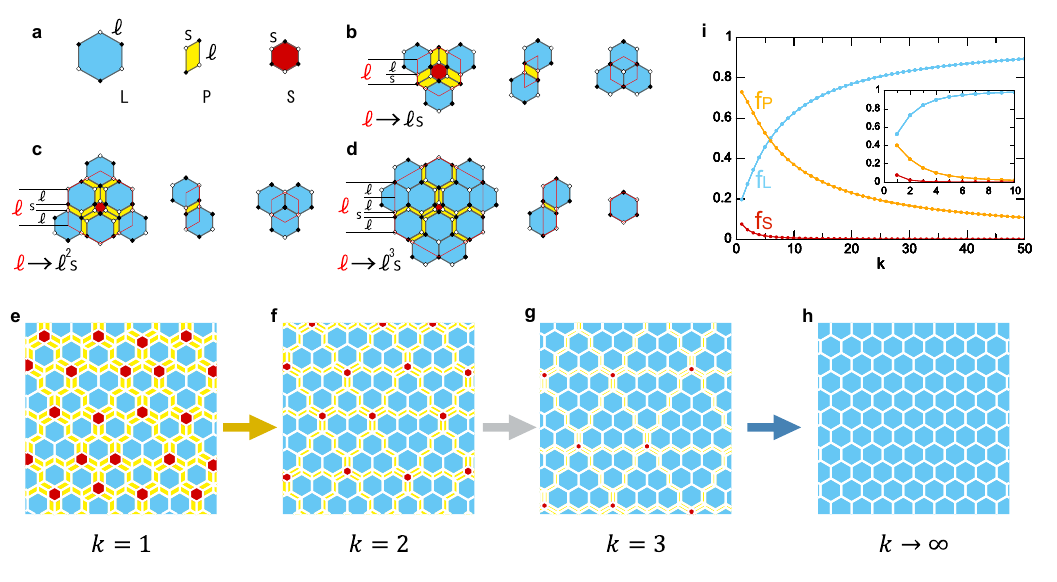}
\caption{{\bf Metallic-mean tilings.} 
{\bf a}, Large hexagons (L), parallelograms (P) and small hexagons (S) with edge lengths $\ell$ and $s$.
Vertices are decorated with open and solid circles alternatively 
to introduce the matching rule of the tiling.
{\bf b}-{\bf d}, Substitution rules for the golden-mean tiling ($k=1$)({\bf b}),
 silver-mean tiling ($k=2$)({\bf c}), and bronze-mean tiling ($k=3$)({\bf d}).
{\bf e}, Golden-mean tiling. {\bf f}, Silver-mean tiling. {\bf g}, Bronze-mean tiling.
{\bf h}, Honeycomb lattice.
{\bf i}, Frequencies of L, P, and S tiles as a function of $k$, and the corresponding fraction for each area (inset). The convergence to the periodic honeycomb lattice is assessed.
}
\label{fig2: deflation}
\end{center}
}
\end{figure*}

\noindent
{\bf Metallic-mean tilings}\label{sec:tiling}\\
\noindent
We construct the metallic-mean approximants of the honeycomb lattice, which are composed of large hexagons (L), parallelograms (P) and small hexagons (S) shown in Fig.~\ref{fig2: deflation}{\bf a}. The ratio between the long ($\ell$) and short ($s$) lengths is given by the metallic-mean $\tau_k(=\ell/s)$. Consequently, the ratio of areas for the three tiles is given by $3\tau_k^2 : \tau_k : 3$. We elaborate the substitution rules for these tiles as a natural extension of those for the hexagonal golden-mean tiling~\cite{Sam_big} (Fig.~\ref{fig2: deflation}{\bf b}). The substitution rules for $k=2$ and $k=3$ are illustrated in Figs.~\ref{fig2: deflation}{\bf c} and \ref{fig2: deflation}{\bf d}, respectively. Notice that the matching rule of the tilings are introduced by solid and open circles. When the deflation rule is applied to an L tile, an S tile is generated at the center of the original L tile, thereby, six zig-zag chains of P tiles emanate from the central S tile, which is clearly found in the case with $k=3$. The rest region is filled by L tiles. Upon one deflation process, a P tile is changed to one P tile and L tiles, and an S tile is changed to one L tile. Hence, one can construct the substitution rules of three tiles for any $k$, which are subjected to the substitution rule for the generalized Fibonacci sequence: $\ell\rightarrow \ell^k s$ and $s\rightarrow \ell$. See also Supplementary Fig.~1 showing how these rules are extended to the cases of $k=4$ and $k=5$.

Two-dimensional space is covered without gaps after iterative deflation processes, as shown in Figs.~\ref{fig2: deflation}{\bf e-g} and Supplementary Fig.~2 for $k=1-5$. Because of the deflation process, self-similarity is an inherent property of the metallic-mean tilings: 
Supplementary Fig.~3 exemplifies exact self-similarity for $k=2$ and $k=3$. We find that a finite number of adjacent L tiles are bounded by the P tiles, which can be regarded as an isolated ``honeycomb domain". For examples, in the case with $k=2$, 
the domains are composed of one, three, or six L tiles, as shown in Fig.~\ref{fig2: deflation}{\bf f}. We confirm that each honeycomb domain bounded by the P tiles is composed of $a_{k-1}, a_k$, or $a_{k+1}$ adjacent L tiles in the $k$th metallic-mean tiling, where $a_k=k(k+1)/2$, see Supplementary Note 3. Therefore, increasing $k$, the number of the L tiles in each honeycomb domain quadratically increases. On the other hand, the S and P tiles are located around the corners and edges of the honeycomb domains,
and thereby their numbers should be $O(1)$ and $O(k)$, respectively. These suggest that the L tiles become majority 
in the large $k$ case and the single honeycomb domain is realized in the limit $k\rightarrow\infty$, as shown in Fig.~\ref{fig2: deflation}{\bf h}. Using a deflation matrix described in Method, it is easy to evaluate the frequencies of tiles ($f_{\rm L}, f_{\rm P},$ and $f_{\rm S}$) and the ratio of the corresponding areas ($S_{\rm L}, S_{\rm P},$ and $S_{\rm S}$) rendered in Fig.~\ref{fig2: deflation}{\bf i} and its inset. For $k=5$, more than ninety percent of the two-dimensional space is occupied by L tiles. See Method in details.

\begin{figure}[htbp]
{\small
\centering
\includegraphics[width=\linewidth]{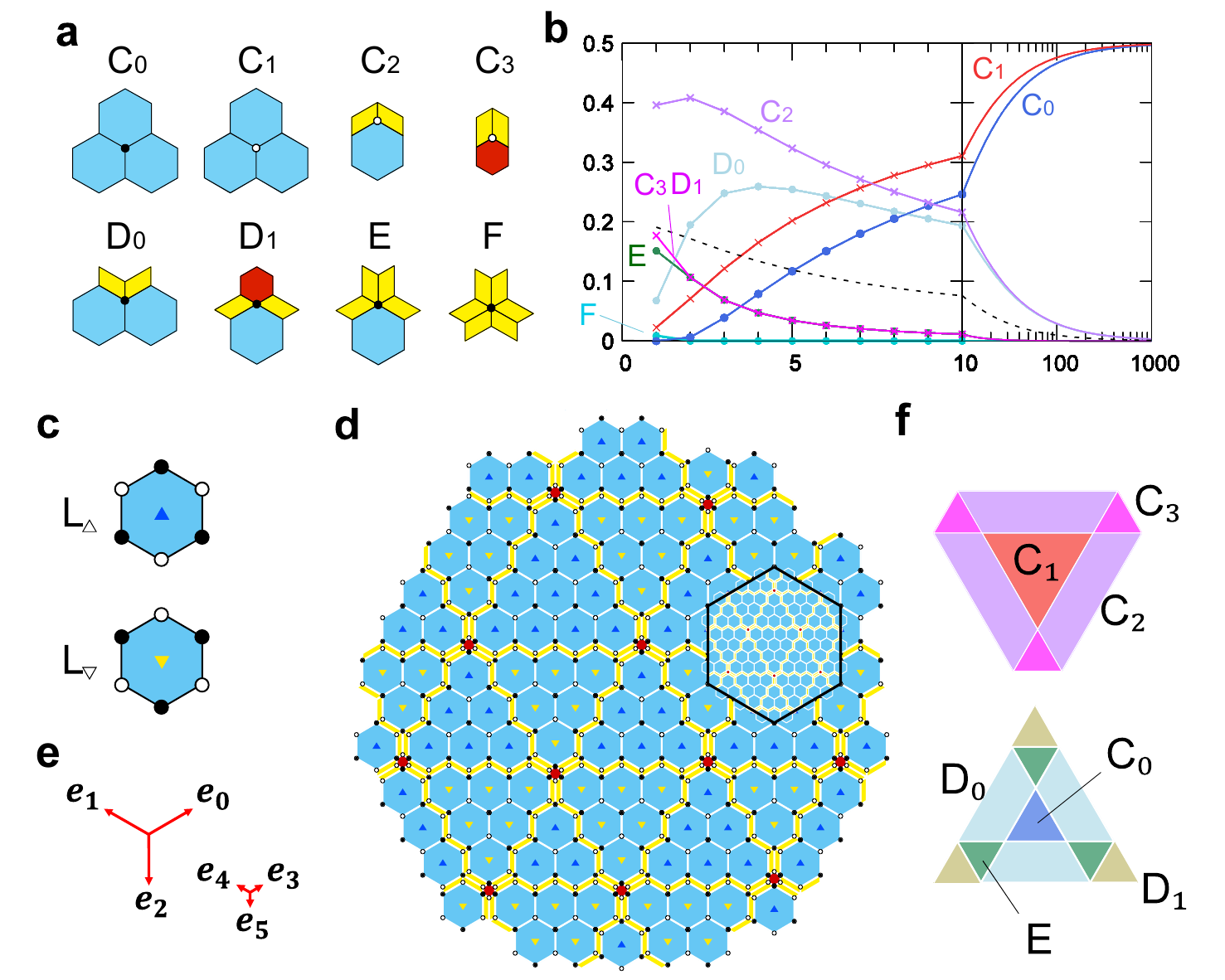}
\caption{{\bf Vertex types.} Vertices are alternatively decorated with the open and solid circles to define A and B sublattices, respectively. {\bf a}, Eight types of vertices. {\bf b}, 
Frequencies of the vertex types. Dashed line represents the sublattice imbalance $\Delta$. {\bf c}, Two kinds of the L tiles, L$_\triangle$ and L$_\triangledown$. {\bf d}, Honeycomb domain structures for the hexagonal bronze-mean tilings. L$_\triangle$ tiles form up-triangular domains, and L$_\triangledown$ tiles form down-triangular domains.
{\bf e}, Projected basis vectors ${\bf e}_i$ ($i=0, \cdots, 5$) from fundamental translation vectors in six dimensions. {\bf f}, Windows in the perpendicular space. Each area shows the vertex types.}
\label{fig3: vertex}
}
\end{figure}

The tiling has eight unique types of vertices as shown in Fig.~\ref{fig3: vertex}{\bf a} classified by their coordination numbers and their circumstances. The frequency of each type can be exactly computed, and the explicit formulae for any $k$ are presented in Supplementary Note 2. Figure~\ref{fig3: vertex}{\bf b} shows the frequencies of the vertex types as a function of $k$. As expected, the frequencies of the C$_0$ and C$_1$ vertices monotonically increase and approach $1/2$ implying the convergence to the honeycomb lattice.

The metallic-mean tilings are bipartite since they are composed of hexagons and parallelograms. As shown in Fig.~\ref{fig3: vertex}{\bf a}, the vertex types C$_1$-C$_3$ belong to the A sublattice and the others belong to the B sublattice, as depicted by open and solid circles, respectively. We find that the sublattice imbalance in the system given as $\Delta = f_A - f_B=1/(3+\sqrt{k^2 + 4})$, where $f_A(=f_{\rm C_1} + f_{\rm C_2} + f_{\rm C_3})$ and $f_B(=f_{\rm C_0} + f_{\rm D_0} + f_{\rm D_1}+ f_{\rm E} + f_{\rm F})$ are the fractions of the A and B sublattices, respectively. This distinct property is in contrast to those for the bipartite Penrose, Ammann-Beenker, and Socolar dodecagonal tilings where each type of vertices equally belongs to both sublattices~\cite{Koga_Penrose}.

As shown in Fig.~\ref{fig3: vertex}{\bf c}, we can distinguish two kinds of L tiles denoted by L$_\triangle$ and L$_\triangledown$, 
introducing up and down triangles located at their centers so that three corners of each triangle point the filled circles on the vertices of the L tile.
In Fig.~\ref{fig3: vertex}{\bf d}, we find the following properties: (1) Two kinds of honeycomb domains composed of L$_\triangle$ or L$_\triangledown$ tiles are alternatively arranged. (2) The shape of L$_\triangle$ domains is up-triangular and
that of L$_\triangledown$ domains is down-triangular. (3) Domain walls are composed of consecutive zigzag P tiles. (4) Three domain walls should meet at a point of S hexagons. (5) Intervals between domain walls are not periodic, but metallic-mean incommensurately modulated, as shown in Method and Supplementary Fig.~6.
\bigskip

\noindent {\bf Superspace representation}\\
\noindent
To provide a theoretical basis for the metallic-mean tilings, we construct their higher-dimensional description. In this powerful method, the superspace is divided into the physical space and its complement, known as the perpendicular space. A tiling is viewed as a projection of a hypercubic crystal in the superspace onto the two-dimensional physical space. The projections onto the perpendicular space are densely filled in specific areas, as illustrated in Fig.~\ref{fig3: vertex}{\bf f}, which are referred to as windows. These windows are derived from sections perpendicular to the threefold axis of a rhombohedron (octahedron) that is the projection of the hypercubic unit cell, showcasing hexagonal and triangular shapes in Extended Data Fig.~\ref{fig5}. The figure also highlights the regions associated with the eight vertex types, as detailed in the Method section and Supplementary Fig.~8.
\bigskip

\begin{figure*}[htbp]
{\small
\centering
\includegraphics[width=\linewidth]{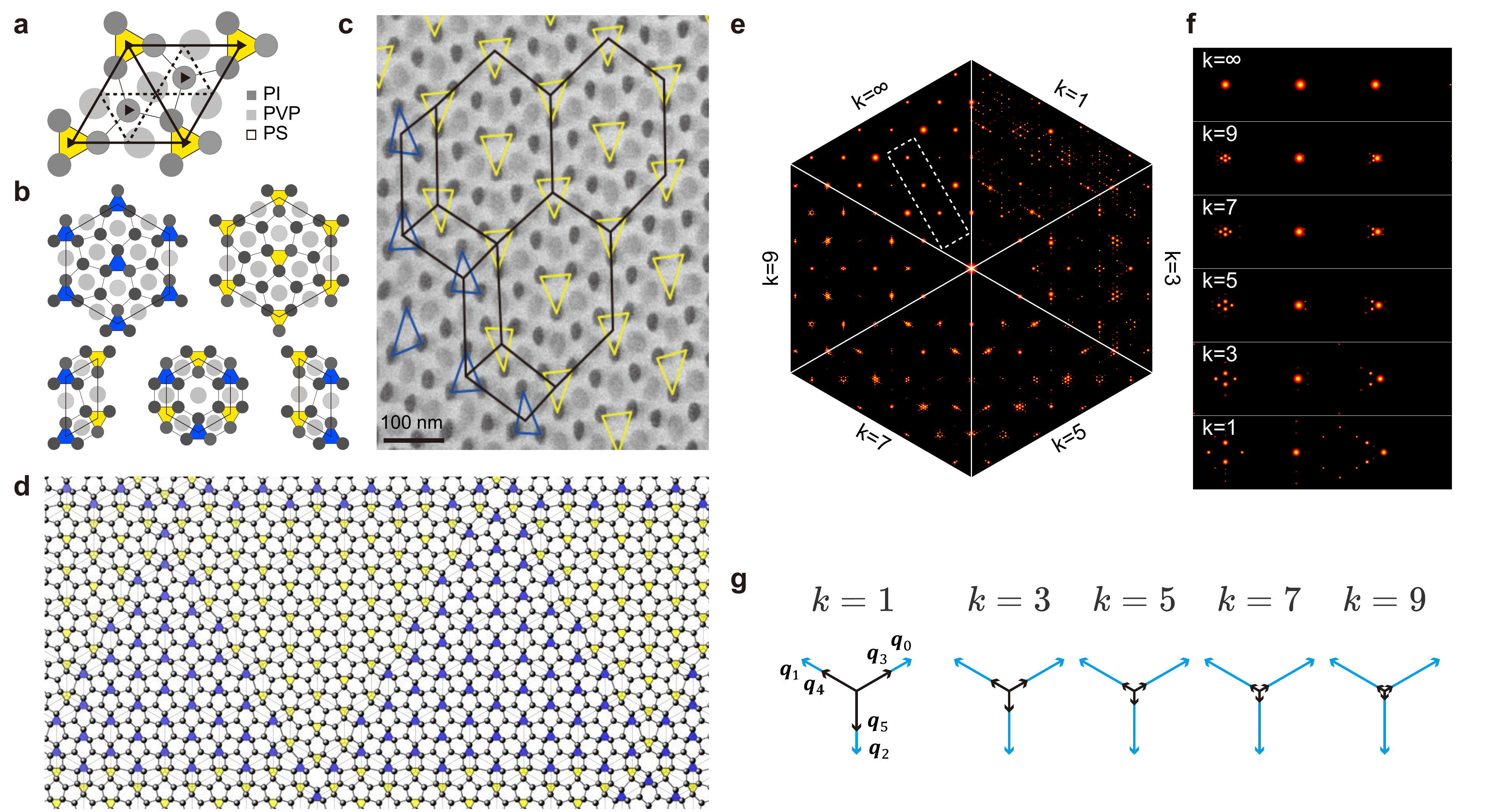}
\caption{
{\bf Application to soft matter.} 
{\bf a}, Diagram of the P31m plane group. 
{\bf b}, Schematic decoration of L, P, and S tiles by ABC triblock terpolymer/homopolymer blend ISP-III/S. 
Dark gray circles indicate polyisoprene (PI), light gray circles indicate poly(2-vinylpyridine) (PVP), and the other matrix region is polystyrene (PS). 
{\bf c}, TEM image from the ABC triblock terpolymer/homopolymer ISP-III/S.
{\bf d}, Ideal particle decoration for a colloidal system generated by the 5-th metallic-mean tiling. Up and down triangles form blue and yellow triangular domains reproducing colloidal simulations. 
{\bf e}, Each sector shows the structure factor for the decorated $k$-th metallic-mean tiling when $k=1, 3, 5, 7, 9$, and $\infty$.
{\bf f}, Magnified views of slices of the structure factor indicated by a dashed rectangle in {\bf e}.
In the vicinity of main peaks, the aperiodic modulation yields satellite peaks characterizing IC structures.
{\bf g}, Six reciprocal vectors ${\bf q}_i\;(i=0, 1, \cdots, 5)$ for $k=1$, 3, 5, 7, and 9.
}
\label{fig4}
}
\end{figure*}

\noindent {\bf Application to soft matter}\\
\noindent
The metallic-mean tilings are physical entities in two soft-matter systems. 
We consider self-assembled crystalline structures obtained in soft materials with the P31m plane group, as illustrated in Fig.~\ref{fig4}{\bf a}, which belongs to the two-dimensional hexagonal Bravais lattice but lacks hexagonal rotational axes. Further crystallographic description is given in Supplementary Notes 8 and 9 for colloidal particles and polymer blends, respectively.

The first application of the metallic-mean tiling is a polymer system reported by Izumi~{\it et.~al.}, who found a complex ordered structure in an ABC triblock terpolymer/homopolymer blend system~\cite{Izumi2015}. Sample preparation is provided in Method.
Figure~\ref{fig4}{\bf b} illustrates the decoration of L, P, and S tiles by three kinds of polymers. In the previous study, regular large domains consisting of only L tiles were observed. It is noticed that the triangle inside a hexagon has two directions, up and down. In the present study, we searched samples again and found P tiles in a TEM picture rendered in Fig.~\ref{fig4}{\bf c}. In Fig.~\ref{fig4}{\bf c}, a regular region of an extended L$_\triangledown$ area in the center and a domain wall represented by a row of zigzag P tiles on the left-hand side. We can interpret the rows of P tiles within the L sea as twin boundaries, which mark a transition between different crystal orientations, L$_\triangle$ and L$_\triangledown$. It's worth emphasizing that a row of P tiles physically changes the crystal orientations, demonstrating the tangible properties of P tiles beyond mathematical concepts. We note that the decoration of S tile (Fig.~\ref{fig4}{\bf b}) is hypothetical and it has not been observed in the samples.

The second application of the metallic-mean tiling is a colloidal particle simulation in two dimensions conducted by Engel~\cite{Engel2011}. 
It utilizes a Lennard-Jones--Gauss (LJG) potential~\cite{Engel2007} that has two distinct length scales. In Method, we have reproduced his result. We find that the LJG particles occupy the same positions as the dark gray circles in Figs.~\ref{fig4}{\bf a}-{\bf c}. 
Moreover, in Fig.~2 of the Engel's paper and his Supplementary Figure S1 in particular, it was shown to form twin-boundary superstructures on a scale much larger than the potential range: the size of superstructures depends on the temperature reversibly; the lower the temperature, the larger the size. One finds that regular L$_\triangle$ or L$_\triangledown$ domains form triangle shapes of several sizes, which property is also characteristic of the metallic-mean tiling. Additionally, it was observed that twin boundaries only intersect at triple junctions, which situation mimics the metallic-mean tilings, where triple rows consisting of the P tiles meet at the location of an S hexagon, though the correspondence between the P tiles and domain walls in the LJG system is not always exact. 
Nonetheless, the metallic-mean scheme mimics Engel's modulated superstructures with changing scale ratios or $k$ values.

In Fig.~\ref{fig4}{\bf d}, an ideal decoration model for the particle system generated by the higher-dimensional quasicrystal theory with the 5-th metallic-mean modulation (Supplementary Note 12). In these cases, as shown in Figs.~\ref{fig4}{\bf e} and \ref{fig4}{\bf f}, the structure factor $S({\bf q})=\left|\frac{1}{N}\sum_i e^{i{\bf q}\cdot{\bf r}_i}\right|^2$ theoretically calculated in terms of the superspace representation dramatically reproduces the numerical FFT for the diffraction images shown in Fig.~3 of the Engel's paper. As clearly shown in the magnified views (Fig.~\ref{fig4}{\bf f}), the prominent peaks
appear at almost the same positions, while the aperiodic modulation of the metallic-mean tiling yields the satellite peaks in the vicinity of the main peaks, which is the characteristic property of IC structures.

\bigskip

\noindent
{\bf Discussion} \\
Our previous study has covered the multiples-of-3 metallic-means, through the hexagonal aperiodic approximants of the triangular lattice~\cite{Nakakura2019}. The present work broadens the scope of aperiodic approximants. Firstly, our tiling serves as the approximant of the honeycomb lattice. Secondly, it enables an inflation ratio of {\it any} metallic-mean, thereby enhancing the applicability.

In fact, we have applied the tiling concept to explore real materials, such as polymer and colloidal systems.
Our analysis successfully identifies large hexagons as regular structures and parallelograms as twin boundaries. 
It is noted that similar IC triangular domain structures were discovered in quartz and aluminum phosphate long time ago~\cite{Tendeloo1976,Landuyt1985}, known as Dauphin\'{e} twins in trigonal quartz. We surmise that there is a similar mechanism behind the formation.

We emphasize that the decorated perpendicular space windows in 6D generate the 2D IC structures, whose method has been developed in the field of QC studies. It is striking that the satellite peaks can be calculated not by direct real-space Fourier transform, but by perpendicular-space Fourier transform of the windows. By comparing these peaks with those observed in two-dimensionally IC modulated structures, we establish a foundation for analysis of IC structures in terms of the QC methodology. 

One of the origins of the P31m plane group demonstrated here is the aggregation tendency of pentagons. Regular pentagons cannot tile the entire plane without gaps, as shown by D\"{u}rer-Kepler-Penrose, however, there are pentagon-related tilings if we abort five-fold symmetry. In Supplementary Note 10, we demonstrate the accommodation of pentagons within both a square and a hexagon. Using 4-fold symmetry, the Cairo pentagonal tiling and its dual, {\it i.e.}, the $3^2.4.3.4$ Archimedean tiling with the P4gm have been considered~\cite{Takano2005}. The latter Archimedean tiling is associated with the $\sigma$ phase found in complex metallic and soft-matter phases, which is recognized as a periodic approximant of dodecagonal QCs~\cite{Zeng2004,Talapin2009,Iacovella2011,Xiao2012,Zhang2012,Foerster2013}. It is noteworthy that P31m plane group structure is a 3-fold variant of the Cairo tiling and the $\sigma$ phase.

Our study highlights the effectiveness of aperiodic approximants in inducing modulations within self-assembled soft-matter systems employing the P31m plane group. Specifically, we utilized the rows of P tiles as domain boundaries in the honeycomb lattice, thereby bridging metallic-mean hexagonal QCs and IC modulated honeycomb lattices. The dynamic movement of domain walls while maintaining triple junctions can be explained by the phason flips of L, S, and P tiles, as illustrated in Extended Data Fig.~\ref{fig7: phasons} and Supplementary Note 4. In this context, the colloidal system appears to be a phason-random tiling version of the metallic-mean tiling system. 
Lastly, applying the deterministic growth rules, known as OSDS rules~\cite{Onoda1988}, reveals that dead surfaces consist of these domain walls. Overall, our research offers a fresh perspective, providing novel insights into the realm of both aperiodic crystals and their broader implications for domain wall structures across various fields.

\bibliographystyle{naturemag}
\bibliography{./refs}

\clearpage
\setcounter{figure}{0}
\renewcommand{\figurename}{Extended Data Fig.~$\!\!$}

\noindent
{\bf \large{Methods}}\\
\noindent
{\bf Deflation matrix of the metallic-mean tiling}\\
The metallic-mean tilings are regarded as the aperiodic approximants of the honeycomb lattice. To discuss quantitatively how the metallic-mean tilings approach the honeycomb lattice with increasing $k$, we construct the deflation matrix. At each deflation process, the increase of the numbers of L, P, and S tiles is explicitly given by ${\bf v}_{n+1}=M {\bf v}_n$ with ${\bf v}_n=(N_{\rm L}^{(n)} \, N_{\rm P}^{(n)} \, N_{\rm S}^{(n)})^T$ and
\begin{equation}
M=\left(
\begin{array}{ccc}
k^2 & \displaystyle\frac{k}{3} & 1 \\
6 k & 1 & 0 \\
1 & 0 & 0
\end{array}
\right),
\end{equation}
where $N_\alpha^{(n)}$ is the number of the tile $\alpha$, which stands for L, P, or S at iteration $n$. The maximum eigenvalue of the matrix $M$ is $\tau_k^2$, and the corresponding eigenvector is given as $(\tau_k^2\; 6\tau_k\; 1)^T$. We evaluate the frequencies for these tiles in the large $k$ limit approach $f_{\rm L}=\tau_k^2/Z$, $f_{\rm P}=6\tau_k/Z$, $f_{\rm S}=1/Z$, where 
$Z=\tau_k^2+6\tau_k+1$. The $k$-dependent frequencies for three tiles are shown in Fig.~\ref{fig2: deflation}{\bf i}.
Increasing $k$, the frequency of the L tiles monotonically increases and approaches unity.

\bigskip

\noindent
{\bf Domain boundaries}\\
\noindent
The domain boundaries composed of consecutive zig-zag P tiles intersect at small hexagons and pass through the opposite edge of the small hexagons with keeping alternating directions of P tiles. If we ignore these slithering configuration of P tiles, there are three sets of parallel domain walls, as displayed in Fig.~\ref{fig3: vertex}{\bf d}. Focusing on a set of parallel domain walls, we observe two types of intervals between the domain walls denoted by ${\cal S}_S$ and
${\cal S}_L$, as shown in Supplementary Fig.~6 for silver- and bronze-mean tilings. There are intriguing properties for the intervals. First, for the $k$-th metallic-mean tilings, the interval ${\cal S}_S$ and ${\cal S}_L$ consists of $k$ and $k+1$ consecutive L tiles. Second, upon the deflation, we find the substitution rules: 
${\cal S}_L \rightarrow {\cal S}_L^\frac{1}{2} {\cal S}_S^k {\cal S}_L^\frac{1}{2}$, and 
${\cal S}_S \rightarrow {\cal S}_L^\frac{1}{2} {\cal S}_S^{k-1} {\cal S}_L^\frac{1}{2}$.
The numbers of intervals $N_{{\cal S}_S}^{(n)}$ and 
$N_{{\cal S}_L}^{(n)}$ of the $n$-th generation satisfy
\begin{equation}
\left(\begin{array}{l}
N_{{\cal S}_S}^{(n+1)} \\
N_{{\cal S}_L}^{(n+1)}
\end{array}\right)=\left(\begin{array}{lc}
k -1 & k \\
1 & 1
\end{array}\right)\left(\begin{array}{l}
	N_{{\cal S}_S}^{(n)} \\
	N_{{\cal S}_L}^{(n)}
\end{array}\right),
\end{equation}
where the maximum eigenvalue of the matrix is given by the metallic-mean $ \tau_k$. The eigenvector of the matrix is given by $(\tau_k-1/\tau_k, 1+1/\tau_k,)^T$, indicating
$N_{{\cal S}_S}^{(n)} / N_{{\cal S}_L}^{(n)} \rightarrow\tau_k$
as $n\rightarrow\infty$, where the sequence is filled with ${\cal S}_S$ intervals for large $k$ values. Therefore, we conclude that the intervals between domain walls are metallic-mean modulated.
\bigskip

\noindent
{\bf Superspace representation}\\
\noindent
We outline main steps of the construction of the metallic-mean tiling by projection of a higher-dimensional hyperlattice onto the physical space.
Let $\ell$ and $s$ be the lengths of the long and short edges of the tiling. We here assume that the ratio $\eta=s/\ell$ is a variable to apply the tiling to soft-matter systems, while the ratio in the perpendicular space is set to be $1/{\tau_k}$ to keep the arrangement of the metallic-mean tiling. When $\eta =1/{\tau_k}$, the tiling is the exact self-similar metallic-mean tiling generated by the deflation rules.

\begin{figure}[htb]
{\small
\centering
\includegraphics[width=\linewidth]{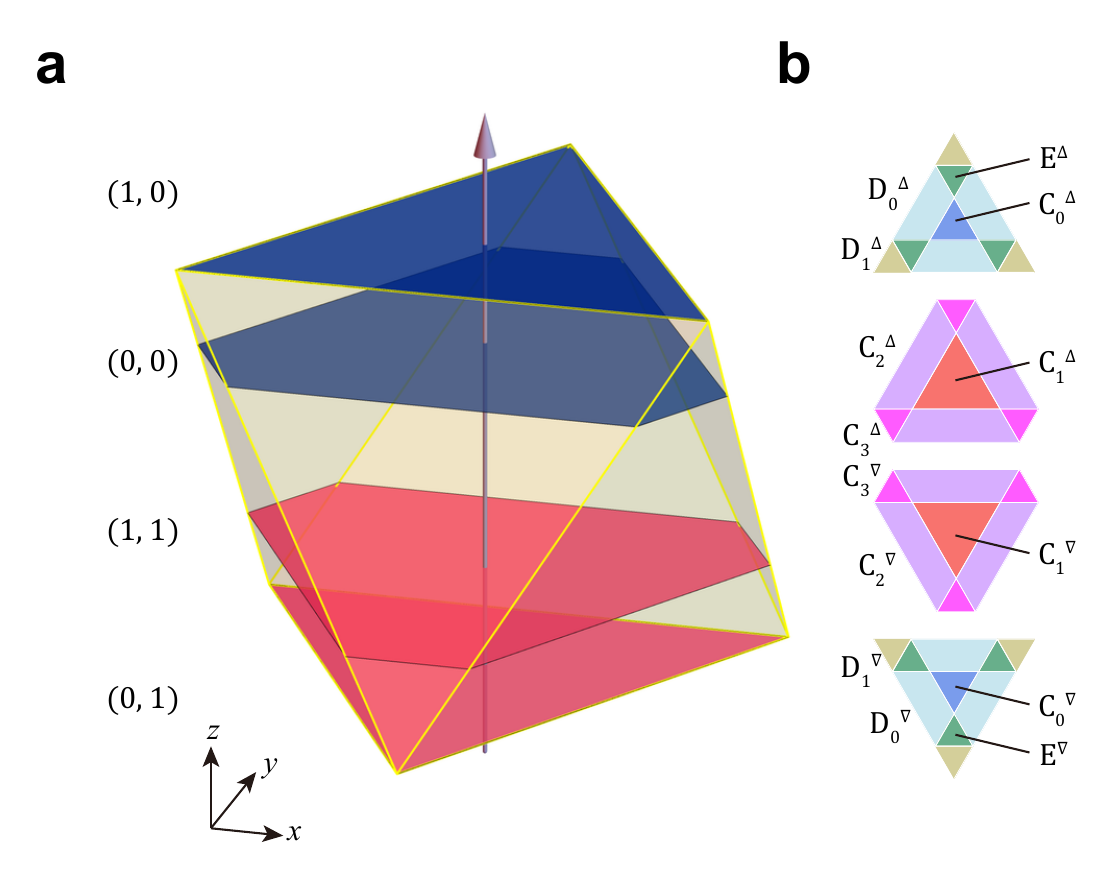}
\caption{
	{\bf Superspace perspective.} {\bf a,} the perpendicular space for the bronze-mean tiling. 
{\bf b,} Four windows on the right-hand side are obtained from a regular octahedron (middle part of a rhombohedron) of edge length $\sqrt{3}(1+\tau_k^{-1})$. 
The top $(1, 0)$ and bottom $(0, 1)$ windows are equilateral triangular faces of the solid, and
hexagonal windows indicated by $(0, 0)$ and $(1, 1)$ are the sections of the octahedron. 
In the solid, blue and red colors correspond to honeycomb domains with L$_\triangle$ and L$_\triangledown$, respectively.
In each window, each color corresponds to the vertex type rendered in Fig.~\ref{fig3: vertex}{\bf b}.
}
\label{fig5} 
}
\end{figure}

Each vertex site in the tiling is described by a six-dimensional lattice point $\vec{n} = (n_0, n_1, \cdots , n_5)^T$, labeled with integers $n_m$. Let the six-dimensional lattice point $\vec{r}^{\,h}$ in the six-dimensional space ${\cal S}^h$ as $\vec{r}^{\,h}=R\vec{n}$:
\small
\begin{eqnarray}
	 R &=& 
	 \begin{pmatrix}
	   \ell c_6 & -\ell c_6 & 0&
	   s c_6 & -s c_6 & 0\\
	   \ell s_6 & \ell s_6 & -\ell&
	   s s_6 & s s_6 & -s\\
		\tau_k^{-1} c_6 & -\tau_k^{-1}c_6 & 0&
	   - c_6 & c_6 & 0\\
		\tau_k^{-1} s_6 & \tau_k^{-1} s_6 & -\tau_k^{-1}&
	   - s_6 & - s_6 & 1\\
	  \sqrt{2} \tau_k^{-1} & \sqrt{2} \tau_k^{-1} & \sqrt{2} \tau_k^{-1} & 0 & 0 & 0\\
	   0 & 0 & 0 & -\sqrt{2} & -\sqrt{2}  & -\sqrt{2} \\
	 \end{pmatrix},
\end{eqnarray}
where $R$ is the mapping matrix and $c_6= \cos(\pi /6)$, $s_6 = \sin(\pi/6)$. 
Namely, the matrix is represented by the six-dimensional basis vectors $\vec{e}_i^{\, h}\,(i=0,1,\cdots,5)$: $(\vec{e}_i^{\, h})_j=R_{ji}$. The vertex site ${\bf r}$ in the physical space ${\cal S}$ is given by the first two components of the vector:
${\bf r}=\left((\vec{r}^{\,h})_0, (\vec{r}^{\,h})_1 \right)=\sum_{m=0}^5 n_m {\bf e}_m$, where the projected vectors of the form ${\bf e}_m=(R_{0m}, R_{1m})$ with lengths $\ell$ and $s$ are displayed in Fig.~\ref{fig3: vertex}{\bf e}. The remaining four-dimensional perpendicular space is split into two-dimensional spaces $\tilde{\cal S}$ and ${\cal S}^\perp$, and the corresponding coordinates $\tilde{\bf r}$ and ${\bf r}^\perp$ are given as
$\tilde{\bf r}=\left((\vec{r}^{\,h})_2, (\vec{r}^{\,h})_3 \right)=\sum_{m=0}^5 n_m \tilde{\bf e}_m$, 
${\bf r}^\perp=\left((\vec{r}^{\,h})_4, (\vec{r}^{\,h})_5 \right)=\sum_{m=0}^5 n_m {\bf e}^\perp_m$,
where $\tilde{\bf e}_m=(R_{2m}, R_{3m})$ and ${\bf e}^\perp_m=(R_{4m}, R_{5m})$.

Note that $\tilde{\bf r}$ points are densely filled on four planes with
${\bf r}^\perp=$ $\{(\sqrt{2}{\tau^{-1}_k},0)$, $(0,0)$, $(\sqrt{2}{\tau^{-1}_k},-\sqrt{2})$, $(0,-\sqrt{2})\}$
denoted by \{(1,0), (0,0), (1,1), (0,1)\}, having polygonal windows shown in Extended Data Fig.~\ref{fig5}. Notice that the windows are faces and sections for a regular octahedron. This octahedron is the middle part of a rhombohedron of edge length $\sqrt{3}(1+\tau_k^{-1})$, which is the projection of the hypercubic unit cell. In Extended Data Figure~\ref{fig5}, $\tilde{\bf r}$ is plotted in the $(x, y)$-directions, while for ${\bf r}^\perp$ both $(\vec{r}^{\,h})_4$ and $(\vec{r}^{\,h})_5$ are projected onto the $z$ component. We find that in the limit $k\rightarrow \infty$, the upper and lower hexagons get closer to the top and bottom faces, respectively, and finally they become the equilateral triangles. The explicit sizes of hexagonal windows are presented in Supplementary Fig.~8. 

The six-dimensional reciprocal lattice vectors $\vec{q}_i^{\,h}$ are defined 
to have the following property $\vec{e}_i^{\, h} \cdot \vec{q}_j^{\, h} = 2\pi \delta_{ij}$ with $\delta_{ij}$ is the Kronecker delta. It is easy to find 
$(\vec{q}_j^{\, h})_i=Q_{ij}$,
where $RQ^T = 2\pi \delta_{ij}$ and 
\small
\begin{eqnarray}
Q &=& C
	\begin{pmatrix}
	c_6 & -c_6 & 0&
	   {\tau^{-1}_k}c_6 & -{\tau^{-1}_k} c_6 & 0\\
	s_6 & s_6 & -1&
	   {\tau^{-1}_k} s_6 & {\tau^{-1}_k} s_6 & -{\tau^{-1}_k}\\
	s c_6 & -s c_6 & 0&
	   -\ell c_6 & \ell c_6 & 0\\
	s s_6 & s s_6 & -s&
	   -\ell s_6 & -\ell s_6 & \ell \\
	\psi & \psi & \psi & 0 & 0 & 0\\
	0 & 0 & 0 & -\psi \tau^{-1}_k & -\psi \tau^{-1}_k  & -\psi \tau^{-1}_k \\
	 \end{pmatrix}
\end{eqnarray}
with $C = 4\pi/[3(\ell+s\tau_k^{-1})]$ and $\psi=(\ell\tau_k+s)/(2\sqrt{2})$. ${\bf q}=\left((\vec{q}_i^{\,h})_0, (\vec{q}_i^{\,h})_1 \right)=\sum_{m=0}^5 n_m {\bf q}_m$, where the projected vectors ${\bf q}_m=(Q_{0m}, Q_{1m})$ with lengths $1/\ell$ and $1/(\ell\tau_k)$ are displayed in Fig.~\ref{fig4}{\bf g}. The remaining four-dimensional perpendicular space is split into two-dimensional reciprocal spaces and the corresponding reciprocal vectors $\tilde{\bf q}$ and ${\bf q}^\perp$ are given as
$\tilde{\bf q}=\left((\vec{q}_i^{\,h})_2, (\vec{q}_i^{\,h})_3 \right)=\sum_{m=0}^5 n_m \tilde{\bf q}_m$, 
${\bf q}^\perp=\left((\vec{q}_i^{\,h})_4, (\vec{q}_i^{\,h})_5 \right)=\sum_{m=0}^5 n_m {\bf q}^\perp_m$,
where $\tilde{\bf q}_m=(Q_{2m}, Q_{3m})$ and ${\bf q}^\perp_m=(Q_{4m}, Q_{5m})$. The detailed procedure is given in Supplementary Note 5. 

When computing the Fourier transforms, we rely on the following identity for any pair of vectors in the superspace lattice $\vec{r}^{\,h}$ and in the corresponding reciprocal lattice $\vec{q}^{\,h}$: $1 = \exp(i\vec{q}^{\,h}\cdot\vec{r}^{\,h}) = \exp(i\bf{q}\cdot\bf{x}) \exp(i\bf{\tilde{q}}\cdot\bf{\tilde{x}}) \exp(i\bf{q}^\perp\cdot\bf{x}^\perp).$ If particle's positions are described by $\delta$-functions so that the density reads $f({\bf r}) = \sum^N_{j=1}\delta({\bf r} - {\bf r}_j)$, then the Fourier transform of the density is calculated as
\begin{eqnarray*}
\int d{\bf r} e^{-i\bf{q}\cdot\bf{r}}f(\bf{r})
&=&\sum_{j=1}^Ne^{-i\bf{q}\cdot\bf{x}_j}= \sum_{j=1}^Ne^{i\bf\tilde{q}\cdot\bf\tilde{x}_j}e^{i\bf{q}^\perp\cdot\bf{x}^\perp_j},
\end{eqnarray*}
in the last step, we resorted to the above identity. 

To construct decorated tilings (Fig.~\ref{fig4}{\bf d}) for soft-matter systems, we set $\eta=s/\ell=0.6249$. In this case, we employ sections of a rhombohedron as extensional windows. Detailed procedures are presented in Supplementary Notes 8 and 12. 
\bigskip

\noindent
{\bf Polymer details}\\
\noindent
An ISP (I:~polyisoprene, S:~polystyrene, P:~poly(2-vinylpyridine)) triblock terpolymer sample was prepared by a sequential monomer addition technique of an anionic polymerization from cumyl-potassium as an initiator in tetrahydrofuran (THF), while styrene homopolymer was synthesized anionically with {\it sec}-butyllithium in benzene. The average molecular weight of the terpolymer is 161k and the composition is $\phi_{\rm I}/\phi_{\rm S}/\phi_{\rm P}=0.25/0.53/0.22$, whereas that of the styrene homopolymer is 9k. 
The overall composition of the blend sample is $\phi_{\rm I}/\phi_{\rm S}/\phi_{\rm P}=0.17/0.68/0.15$, where polystyrene block/styrene homopolymer ratio of $w_{\rm S}({\rm b)}/w_{\rm S}({\rm h})=1.4$. The sample film was obtained by casting for two weeks from a dilute solution of THF followed by heating at 150$^\circ$C for two days. The specimens for morphological observation were cut by an ultramicrotome of Leica model Ultracut UCT into ultrathin sections of about 100 nm thickness and stained with OsO$_4$ for the TEM observation. Further details are provided in the previous reference~\cite{Izumi2015}.
\bigskip

\begin{figure}[htb]
{\small
\centering
\includegraphics[width=0.8\linewidth]{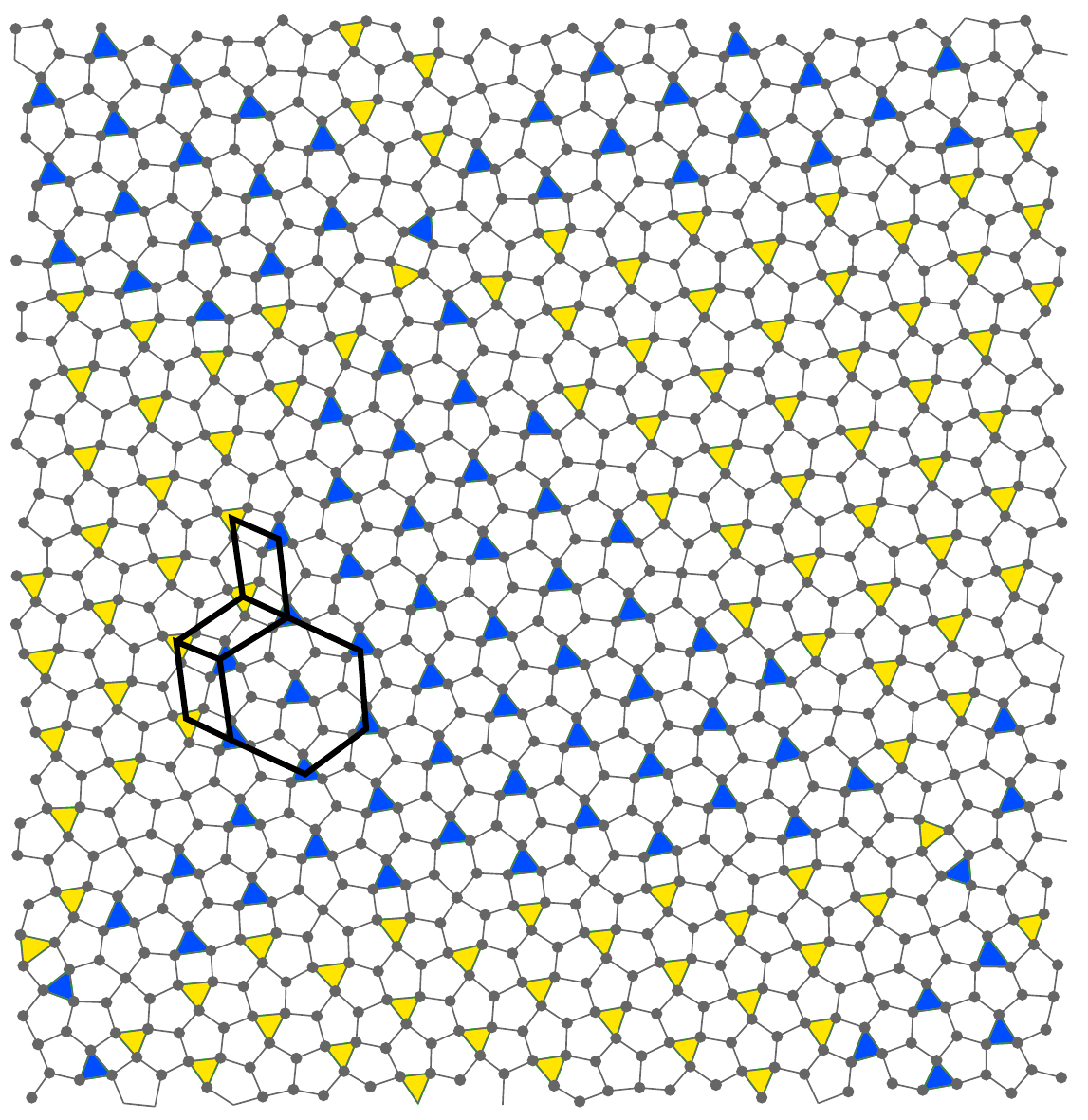}
\caption{
	{\bf Colloidal simulation.} Monte Carlo simulation for the Lennard-Jones-Gauss potential at $T=0.270$.}\label{fig6}
}
\end{figure}

\noindent
{\bf Simulations of colloidal particles}\\
\noindent We used {\it NPT} Monte Carlo simulations of $N=10000$ colloidal particles interacting with the Lennard-Jones-Gauss potential~\cite{Engel2007,Engel2011} given by \begin{equation} V(r)=\frac{1}{r^{12}}-\frac{2}{r^{6}}-\epsilon\exp\left(-\frac{(r-r_0)^2}{2\sigma^2}\right), \end{equation} with parameters $\sigma^2=0.042$, $\epsilon=1.8$, $r_0=1.42$ at $T=0.270$, $P=0.0$. There are slight differences between simulations (Extended Data Fig.~\ref{fig6}) and the metallic-mean tiling model (Fig.~\ref{fig4}): (1) Dynamically, P tiles are not always perfect. (2) There are five particles in an S tile in simulations, while six particles in the latter. The effect of these is negligible in the structure factor. Further data including diffraction images is provided in Supplementary Note 11.

\bigskip
\noindent
{\bf Phasons}\\
\noindent
Domain walls dynamically move with keeping triple junctions can be explained by the phason flips of L, S and P tiles, as shown in Extended Data Fig.~\ref{fig7: phasons}. In this sense, the colloidal system appears to be a phason-random tiling version of the metallic-mean tiling system. The existence and the conservation of S tiles in the phason flips is the key of triple junctions of domain walls at moderate thermal excitations. See also Supplementary Note 4. 
\bigskip
\begin{figure}[htb]
{\small
\centering
\includegraphics[width=\linewidth]{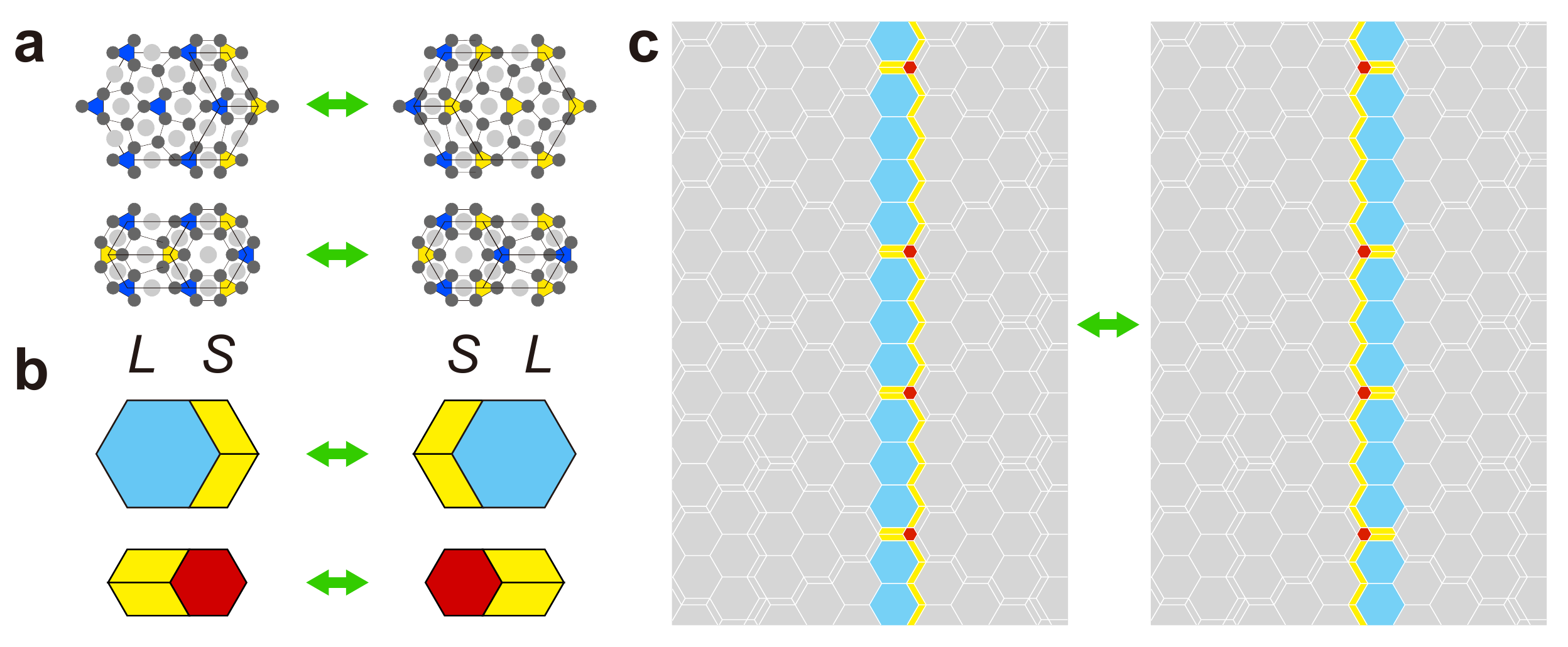}
\caption{
	{\bf Phason flips.} {\bf a}, Schematic phason move in self-assembled pattern from ABC triblock terpolymer/homopolymer blend. {\bf b}, Two types of phason flips in the metallic-mean tiling.
{\bf c}, Move of a twin boundary by a row of phason flips.}\label{fig7: phasons}
}
\end{figure}
\bigskip

\noindent
{\bf Data availability}\\
\noindent
The whole datasets are available from the corresponding author on reasonable request.

\bigskip

\noindent
{\bf Code availability}\\
\noindent
The codes used to construct the tilings and the projection windows are available from the corresponding author on reasonable request.

\bigskip

\noindent
{\bf Acknowledgements}\\
\noindent
The authors thank M. Engel, M. Mihalkovic, and P. Ziherl for valuable discussions.
Parts of the numerical calculations are performed
in the supercomputing systems in ISSP, the University of Tokyo.
This work was supported by Grant-in-Aid for Scientific Research from
JSPS, KAKENHI Grant Nos.
JP22K03525, JP21H01025, JP19H05821 (A.K.).\\

\noindent
{\bf Author contributions}\\
\noindent
T.M. \& A.K. proposed the tiling theory. T.D. developed the application to soft matter. 
A.T. \& Y.M. conducted polymer experiments. T.M., A.K. \& T.D. wrote the manuscript. \\

\noindent
{\bf Additional information} 

\noindent
{\bf Supplementary Information} accompanies this paper at {\tt http://XXX}.

\noindent
{\bf Competing interests:} The authors declare no competing interests.

\noindent
{\bf Reprints and permission} information is available online at {\tt http://npg.nature.com/
reprintsandpermissions/}

\noindent
{\bf Publisher's note:} Springer Nature remains neutral with regard to jurisdictional claims in published maps and institutional affiliations. 

\bigskip

\noindent
Correspondence and requests for materials should be addressed to
A. K. (email: koga@phys.titech.ac.jp) and
T. D. (email: dotera@phys.kindai.ac.jp).

\end{document}


\date{\today}

\title{Aperiodic approximants bridging quasicrystals and modulated structures}
\
\vspace*{2mm}
{\sl Supplementary Information}
\vspace*{1mm}
\author{Toranosuke Matsubara}
\author{Akihisa Koga}
\affiliation{Department of Physics, Tokyo Institute of Technology, Meguro, Tokyo 152-8551, Japan}

\author{Atsushi Takano}
\affiliation{Department of Molecular and Macromolecular Chemistry, 
Nagoya University, Nagoya, Aichi 464-8603, Japan}
\author{Yushu Matsushita}
\affiliation{Toyota Physical and Chemical Research Institute, Nagakute, Aichi 480-1192, Japan}

\author{Tomonari Dotera}
\affiliation{Department of Physics, Kindai University,
Higashi-Osaka, Osaka 577-8502, Japan}

\thispagestyle{empty}

\vspace{3cm}
\maketitle

\tableofcontents

\newpage

\setcounter{page}{1}
\section{Substitution rules for the hexagonal metallic-mean tilings}
\label{app: tilings}
The hexagonal metallic-mean tiling we propose in the main text is composed 
of large hexagons (L), parallelograms (P), and small hexagons (S).
The length ratio is given by the metallic mean $\tau_k$,
with $\tau_k=(k+\sqrt{k^2+4})/2$. 
Extending the rule for the hexagonal golden-mean tiling~\cite{Sam_big},
we propose the substitution rules for the metallic-mean tilings,
which are explicitly shown in Supplementary Fig.~\ref{fig:deflation}.
\begin{figure}[htb]
  \begin{center}
    \includegraphics[width=0.9\linewidth]{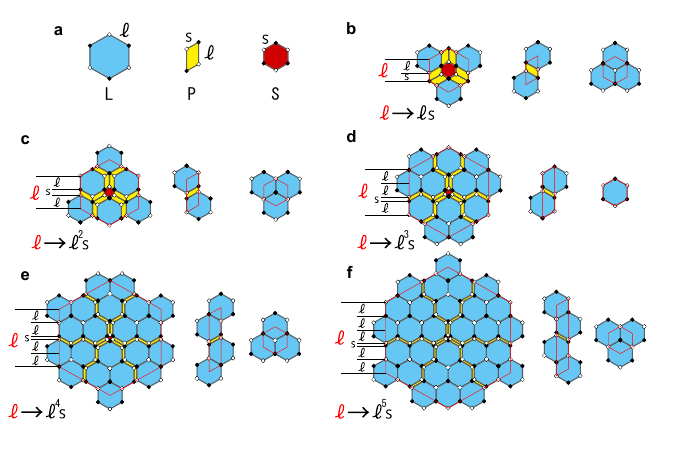}
    \caption{{\bf a} Large hexagon, parallelogram, and small hexagon. 
    {\bf b}-{\bf f} Substitution rules for the golden-mean, silver-mean, bronze-mean,
    4th metallic-mean, and 5th metallic-mean tilings.}
    \label{fig:deflation}
  \end{center}
\end{figure}
When the deflation rule is applied to an L tile, 
an S tile is generated at the center of the original L tile.
Furthermore, six chains sharing a short edge of $k$ P tiles are adjacent to the S tile
and are located along six directions $(\cos \pi i/3, \sin \pi i/3)\; 
(i=0, 1,\cdots, 5)$.
The rest region is filled by $k^2$ L tiles.
As for a P tile, $k/3$ L tiles are generated along the longer edge of the original P tile, 
and one P tile is generated around the center.
Note that there are two kinds of the P tiles: 
a P tile shown in Supplementary Fig.~\ref{fig:deflation}{\bf a} and its reflected tile (\={P} tile).
We find that the P (\=P) tile appears for odd (even) $k$ case when the substitution rule is applied to a P tile.
An S tile is replaced to one L tile under one deflation operation.
We find the triple periodicity in $k$ for the deflation process of the S tile. 
These allow us to generalize the substitution rules for any metallic-mean tilings.

We obtain the hexagonal metallic-mean tilings, applying 
the substitution rule to a certain tile iteratively.
The metallic-mean tilings for $k=1, 2, 3, 4, 5$, and the honeycomb lattice, 
which can be regarded as the tiling with $k\rightarrow\infty$, 
are shown in Supplementary Fig.~\ref{fig: tilings}.
\begin{figure}[htb]
  \begin{center}
    \includegraphics[width=0.8\linewidth]{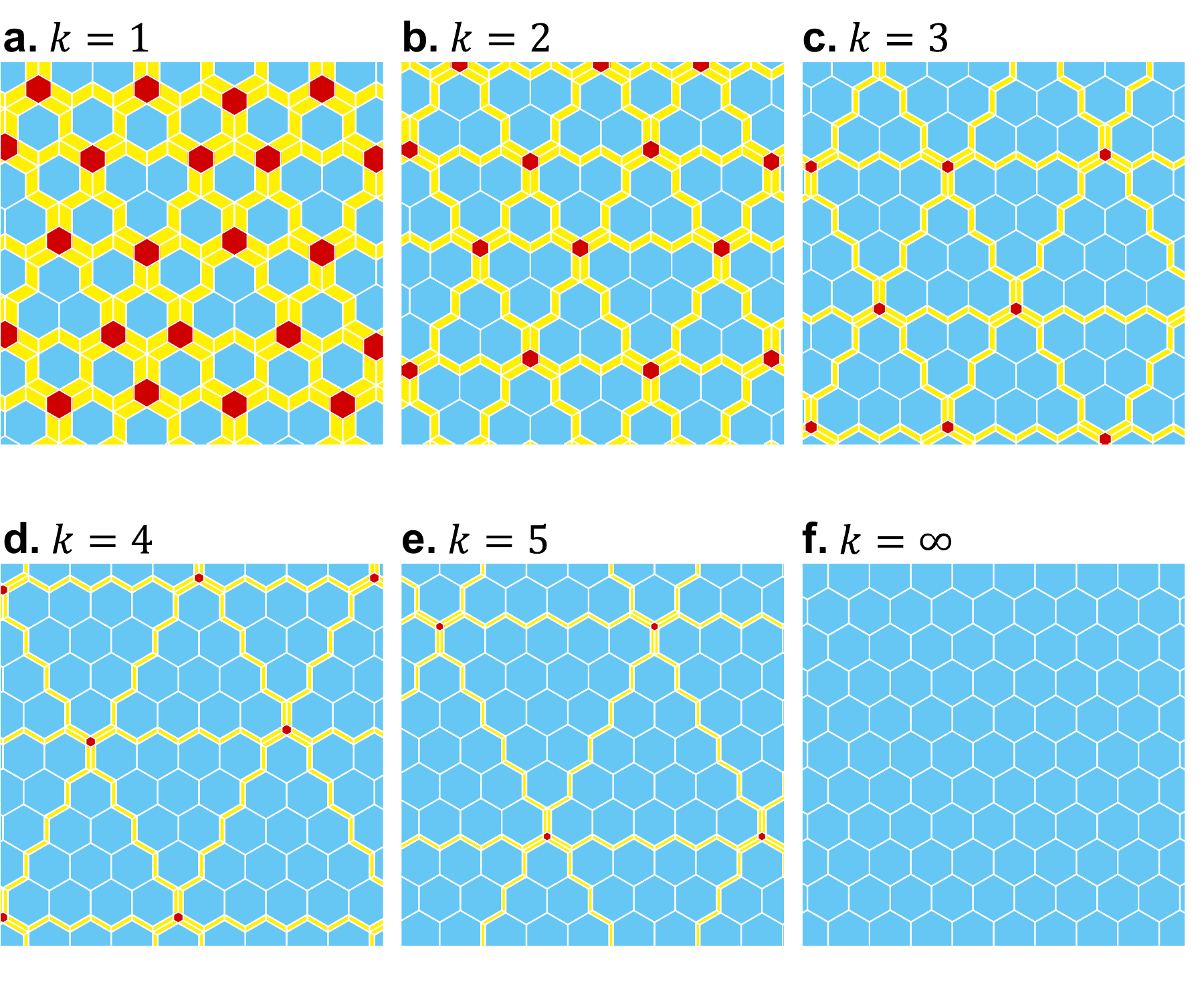}
    \caption{Hexagonal metallic-mean tilings. {\bf a} Golden-mean tiling $(k=1)$~\cite{Sam_big}, 
    {\bf b} silver-mean tiling $(k=2)$ and 
    {\bf c} bronze-mean tiling $(k=3)$. {\bf d} and {\bf e} represent metallic-mean tilings with $k=4$ and $k=5$. 
    {\bf f} represents the honeycomb lattice with $k\rightarrow\infty$.}
    \label{fig: tilings}
  \end{center}
\end{figure}
When one deflation operation is applied to the tilings,
the number of tiles increases and the self-similar structure appears,
which are shown in Supplementary Figs.~\ref{fig: inflation}{\bf a} and \ref{fig: inflation}{\bf b}. 
\begin{figure}[htb]
  \begin{center}
    \includegraphics[width=\linewidth]{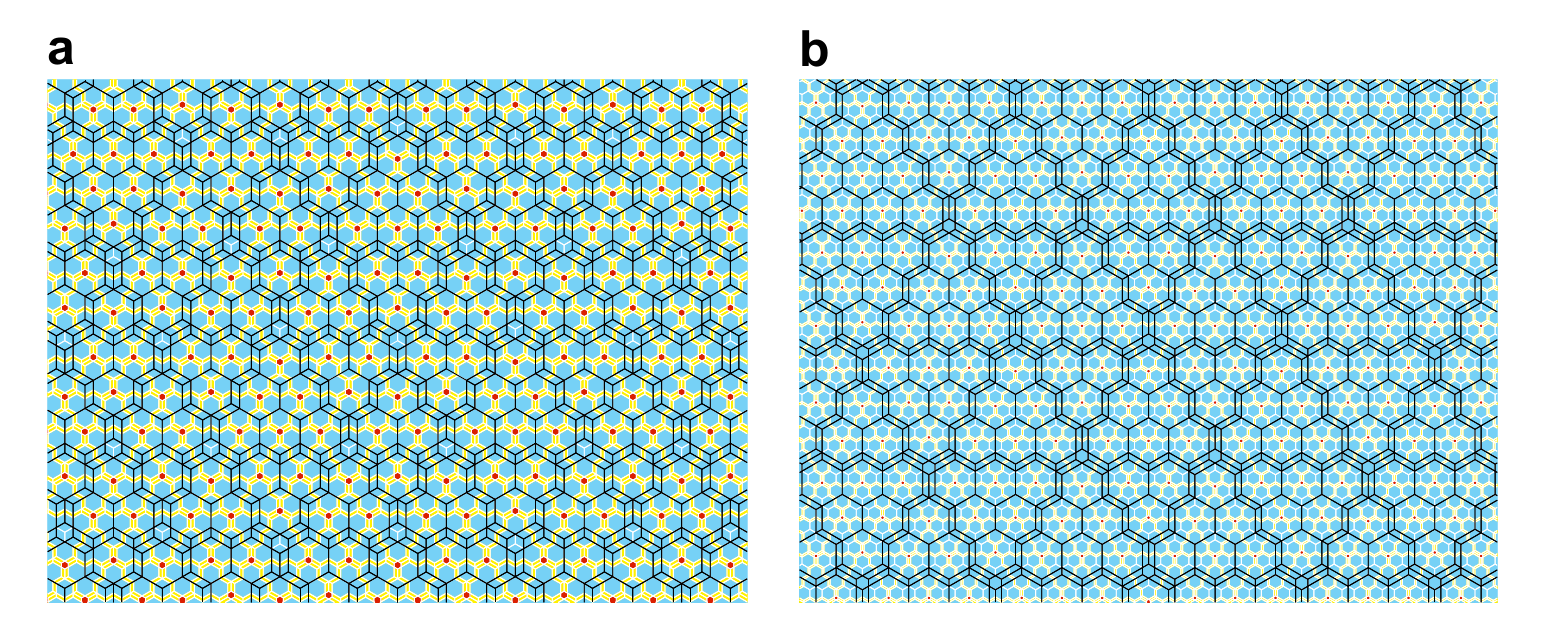}
    \caption{Self-similarity of hexagonal metallic-mean tilings. 
    {\bf a} Silver-mean tiling. {\bf b} Bronze-mean tiling.
      The colored tilings are obtained by applying the deflation operation to the tilings shown as the black lines.
    }
    \label{fig: inflation}
  \end{center}
\end{figure}

\clearpage
\section{Fractions of the vertices}\label{app: fraction}
We derive the fractions of vertices for $k \neq 1$
(the fractions for $k=1$ have been given in Ref~\cite{Sam_big}).
In the case of $k\neq 1$, we could not find the F vertex shared by six P tiles ($f_F=0$).
This is because vertices shared by two adjacent P tiles are always shared by the L or S tile
according to the substitution rules for $k\neq 1$,
as shown in Supplementary Figs.~\ref{fig:deflation}{\bf c}-\ref{fig:deflation}{\bf f}.
When one evaluates the fractions for certain graphs
such as vertices and domains,
it is convenient to consider the ratio between numbers of tiles and vertices
for the hexagonal metallic-mean tiling in the thermodynamic limit.
Supplementary Figure~\ref{fig:deflation}{\bf a} clearly shows that
the net numbers of sites in L, P, and S tiles are
two, one, and two, respectively.
Therefore, we obtain the ratio $r_k$ as
\begin{eqnarray}
  r_k=2f_L+f_P+2f_S= \frac{P_k}{\tau_k^2 + 6\tau_k + 1},
\end{eqnarray}
where $ P_k = 2\tau_k^2 + 6\tau_k + 2.$
We first focus on the bipartite structure.
The sublattice structures for the L, P, and S tiles are shown as
the open and solid circles in Supplementary Fig.~\ref{fig:deflation}{\bf a},
where these are referred to as A and B sublattices.
By counting the net numbers of the site belonging to each sublattice
in L, P, and S tiles,
we obtain its fractions as,
\begin{eqnarray}
f_{\rm A} &=& \frac{1}{r_k}\left(f_L+\frac{2}{3}f_P + f_S\right) = \frac{1}{2}+\frac{\tau_k}{P_k},\\
f_{\rm B} &=& \frac{1}{r_k}\left(f_L+\frac{1}{3}f_P + f_S\right) = \frac{1}{2}-\frac{\tau_k}{P_k}.
\end{eqnarray}
This naturally leads to the sublattice imbalance
in the hexagonal metallic-mean tilings,
\begin{eqnarray}
  \Delta=f_A - f_B=\frac{1}{3+\sqrt{k^2 + 4}}.
\end{eqnarray}
Since the sublattice A (B) is composed of
C$_1$, C$_2$, and C$_3$ (C$_0$, D$_0$, D$_1$, and E) vertices,
we obtain the following equations,
\begin{eqnarray}
f_{\rm A} &=& f_{\rm C_1} + f_{\rm C_2} + f_{\rm C_3},\\
f_{\rm B} &=& f_{\rm C_0} + f_{\rm D_0} + f_{\rm D_1}+ f_{\rm E}.
\end{eqnarray}
In the tilings, two adjacent tiles share the edge,
which is connected between the neighboring sites in A and B sublattices.
Therefore, we obtain the equations for the total number of
longer and shorter edges,
\begin{eqnarray}
3 f_{\rm C_1} + 2f_{\rm C_2} +f_{\rm C_3} &=& 
3f_{\rm C_0} + 3f_{\rm D_0} + 2f_{\rm D_1}+ 3f_{\rm E},\\
f_{\rm C_2} + 2f_{\rm C_3} &=& f_{\rm D_0} + 2f_{\rm D_1}+ 2f_{\rm E},
\end{eqnarray}
where the left (right) hand side of the equations represents the total number of edges, which is expressed by the numbers of vertices belonging to the A (B) sublattice.
According to the substitution rule,
the C$_3$, D$_1$, and E vertices always appear around the S tile for $k \neq 1$.
Therefore, these fractions are then given as
\begin{eqnarray}
f_{\rm C_3} = f_{\rm D_1} = f_{\rm E} =\frac{3f_{\rm S}}{r_k}
=\frac{3}{P_k}.
\end{eqnarray}

From these equations,
we obtain the exact fractions of vertices in the hexagonal metallic-mean tilings as
\begin{eqnarray}
  f_{\rm C_0} &=&\left\{
  \begin{array}{ll}
    \displaystyle 0& (k=1)\\
    \displaystyle\frac{1}{2} - \frac{7}{4\tau_k} + \frac{1}{2P_k}(27-7k) & (k\neq 1)
  \end{array}
  \right.,\label{eq: f_C0} \\
  f_{\rm C_1} &=& \frac{1}{2} - \frac{5}{4\tau_k} + \frac{5}{2P_k}(3-k),\\
  f_{\rm C_2} &=& \frac{3}{2\tau_k} - \frac{3}{P_k}(4-k),\\
  f_{\rm C_3} &=& \frac{3}{P_k},\\
  f_{\rm D_0} &=& \left\{ 
  \begin{array}{ll}
  \displaystyle\frac{3}{4\tau_1^5} & (k=1) \\
  \displaystyle\frac{3}{2\tau_k} - \frac{3}{P_k}(6-k) & (k \neq 1)
  \end{array}
  \right.,\\
  f_{\rm D_1} &=& \frac{3}{P_k},\\
  f_{\rm E} &=& \left\{
  \begin{array}{ll}
  \displaystyle\frac{3\sqrt{5}}{4\tau_1^5} & (k=1) \\
  \displaystyle\frac{3}{P_k} & (k \neq 1)
  \end{array}\right.,\\
  f_{\rm F}&=& \left\{
  \begin{array}{ll}
    \displaystyle  \frac{1}{4\tau_1^7} & (k=1)\\
    \displaystyle 0 & (k\neq 1)
  \end{array}\right. . \label{eq: f_F} 
  \end{eqnarray}

The average of the coordination number is given by
  \begin{eqnarray}
    z_k&=&3\sum_i f_{{\rm C}_i}+4\sum_if_{{\rm D}_i}
  +5f_{\rm E}+6f_{\rm F}\nonumber\\
  &=&\displaystyle 3+\frac{3}{2\tau_k}+\frac{3(k-3)}{P_k}.
  \end{eqnarray}
In the hexagonal metallic-mean tilings,
the average of the coordination number depends on $k$.
$z_k\rightarrow 3$ 
when the system approaches the honeycomb lattice $k\rightarrow \infty$.

\clearpage
\section{Honeycomb domain}\label{app: domain}
\begin{figure}[htb]
  \begin{center}
    \includegraphics[width=\linewidth]{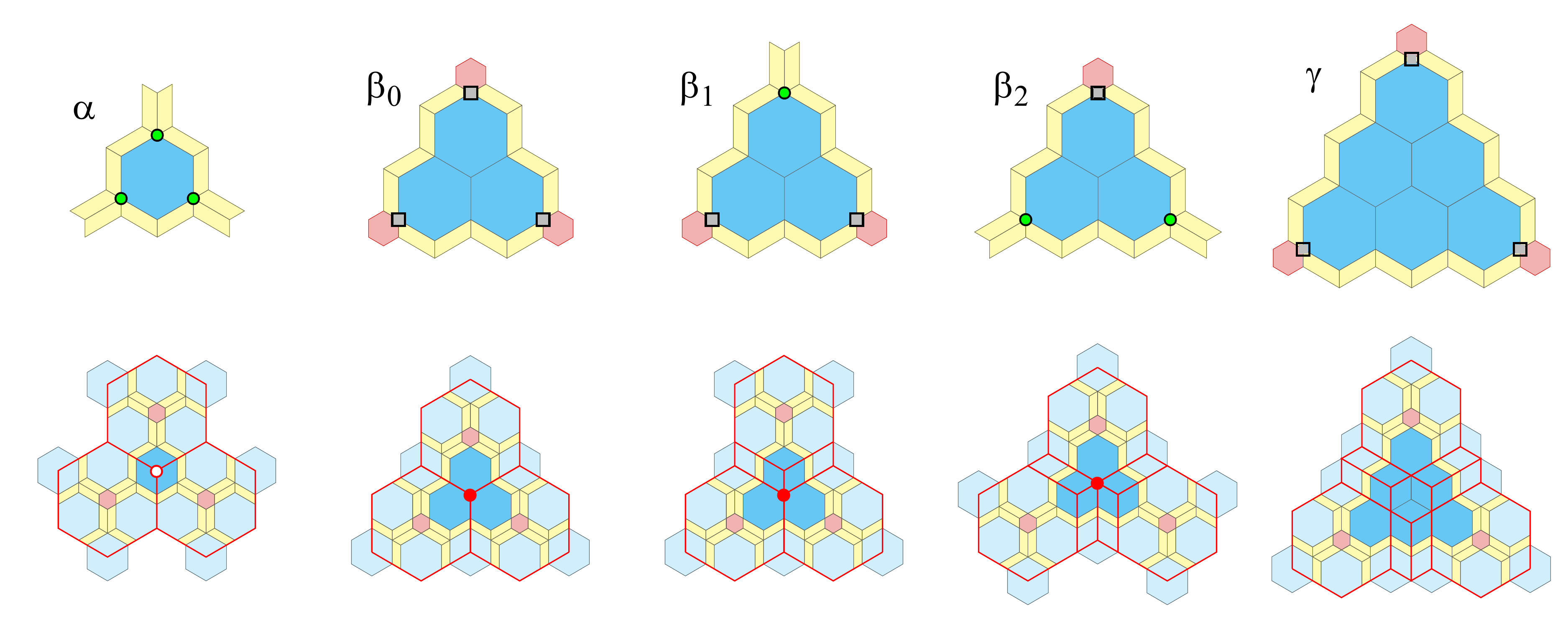}
    \caption{Upper panels: $\alpha, \beta_0, \beta_1, \beta_2$,
      and $\gamma$ domains
      in the hexagonal silver-mean tiling,
      which are bounded by some P and S tiles.
      Circles and squares indicate E and D$_1$ vertices at
      the corners of the honeycomb domains.
      Lower panels: the red lines represent
      the results of the inflation operation applied to
      the tiles around the honeycomb domains shown in the upper panels.
      Red open (solid) circles at the vertices indicate the A (B) sublattice
      in the inflated tiling.
    }
    \label{fig: domain}
  \end{center}
\end{figure}
Here, we focus on the honeycomb domain in the hexagonal metallic-mean tilings with $k\neq 1$,
which is composed of finite number of the L tiles
and is bounded by the P and S tiles.
As seen in Supplementary Fig.~\ref{fig: tilings}, in the hexagonal metallic-mean tiling with $k\neq 1$,
there exist three kinds of domains composed of $a_{k-1}, a_k$ or $a_{k+1}$ L tiles,
where $a_k=k(k+1)/2$.
These are referred to as $\alpha$, $\beta$ and $\gamma$ domains.
Supplementary Figure~\ref{fig: domain} shows $\alpha$, $\beta$, and $\gamma$ domains in the silver-mean tilings, as an example.
We find that in the $\alpha$ domain,
the E vertex shared by one L tile and four P tiles, which is shown as the circle,
is located at each corner site.
In the $\gamma$ domain,
the D$_1$ vertex shared by one L tile, two P tiles and one S tile, which is shown as the square,
is located at each corner site.
On the other hand, the $\beta$ domains can be divided into the $\beta_i (i=0, 1, 2)$ domains,
where $i$ E vertices and $(3-i)$ D$_1$ vertices 
are located at three corner sites, as shown in Supplementary Fig.~\ref{fig: domain}. %
The absence of the $\beta_3$ domains will be proved below.

To examine the fraction of each domain,
we consider the substitution rule for the tiles.
In the lower panels of Supplementary Fig.~\ref{fig: domain},
we show the tiling structure obtained by the inflation operation as the red lines.
We find that
the C$_1$, C$_0$, D$_0$, E vertices, and S tiles generated by an inflation operation
are located at the center of the $\alpha$, $\beta_0$, $\beta_1$, $\beta_2$,
and $\gamma$ domains. 
Therefore, we obtain the following equations as
\begin{eqnarray}
f_{\alpha} &=& \frac{r_k f_{\rm C_1}}{\tau_k^2},\\
f_{\beta_0} &=& \frac{r_k f_{\rm C_0}}{\tau_k^2},\\
f_{\beta_1} &=& \frac{r_k f_{\rm D_0}}{\tau_k^2}, \label{eq: beta_1}\\
f_{\beta_2} &=& \frac{r_k f_{\rm E}}{\tau_k^2},\label{eq: beta_2}\\
f_{\gamma} &=& \frac{ f_{\rm S}}{\tau_k^2},\label{eq: gamma}
\end{eqnarray}
where $f_X$ is the ratio of the number of $X(=\alpha, \beta_i, \gamma)$ domains
to the total number of tiles.
Since $f_{\rm L} = a_{k-1} f_\alpha + a_k\sum_{i=0}^2 f_{\beta_i} + a_{k+1}f_\gamma$,
we prove that each L tile belongs to $\alpha$, $\beta_i$\; $(i=0, 1, 2)$ 
or $\gamma$ domain, and $\beta_3$ domains never appear in the hexagonal metallic-mean tiling with $k\neq 1$.
As for the golden-mean tiling with $k=1$,
$\alpha$ and $\beta_0$ domains do not appear,
but $\beta_3$ domains appear
due to the existence of the F vertices.
The fractions of the $\beta_1$, $\beta_2$, and $\gamma$ domains are given
by Eqs.~(\ref{eq: beta_1}), (\ref{eq: beta_2}), and (\ref{eq: gamma}), and
the fraction of the $\beta_3$ domains is given as
\begin{eqnarray}
  f_{\beta_3} = \frac{r_1 f_{\rm F}}{\tau_1^2}=\frac{5+3\tau_1}{31\tau_1^9}.
\end{eqnarray}

We show in Supplementary Fig.~\ref{fig: domainfraction} the fraction of the L tiles which belong to 
each domain.
\begin{figure}[htb]
  \begin{center}
    \includegraphics[width=0.85\linewidth]{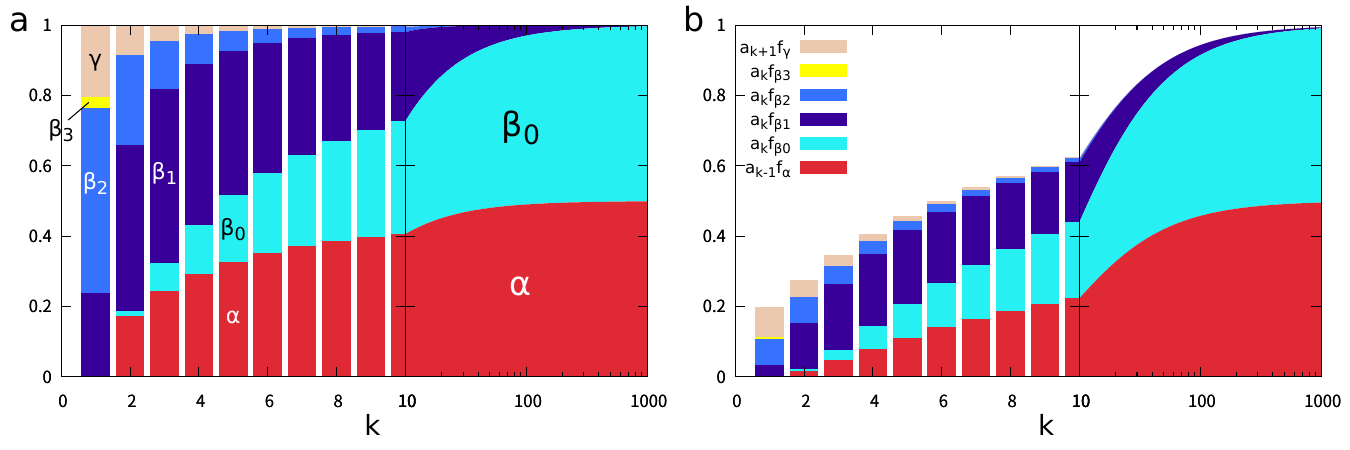}
    \caption{{\bf a}
      Fractions of $\alpha, \beta_i$ $(i=0,1,2,3)$, and $\gamma$ domains
      for the total number of the domains
      are shown as the cumulative bar chart.
      {\bf b}
      Fraction of the L tiles belonging to each domain are shown as
      the cumulative bar chart.
    }
    \label{fig: domainfraction}
  \end{center}
\end{figure}
When $k$ is small, the $\beta_1$, $\beta_2$, and $\gamma$ domains are majority 
in the tilings.
On the other hand, when the system approaches the honeycomb lattice,
$\alpha$ and $\beta_0$ domains become dominant in the system. 
This originates from the fact that, in the large $k$ case, 
the vertices are almost composed of the C$_1$ and C$_0$ vertices, and 
thereby $\alpha$ and $\beta_0$ domains, which are generated
by applying the deflation operation to the above vertices, become dominant.

\clearpage
\section{Domain boundaries}

Honeycomb domains are separated by the P and S tiles and are distributed like triangular structure, 
as shown in Supplementary Fig.~\ref{fig: tilings}.
This may allow us to regard the hexagonal metallic-mean tilings as
the honeycomb lattice modulated by the one-dimensional domain boundaries along three directions
with the angles $\theta=0$ and $\pm \pi/3$.
Here, we clarify how the domain boundaries are 
distributed in the metallic-mean tilings with $k\neq 1$.

\begin{figure*}[htb]
  \begin{center}
    \includegraphics[width=0.8\linewidth]{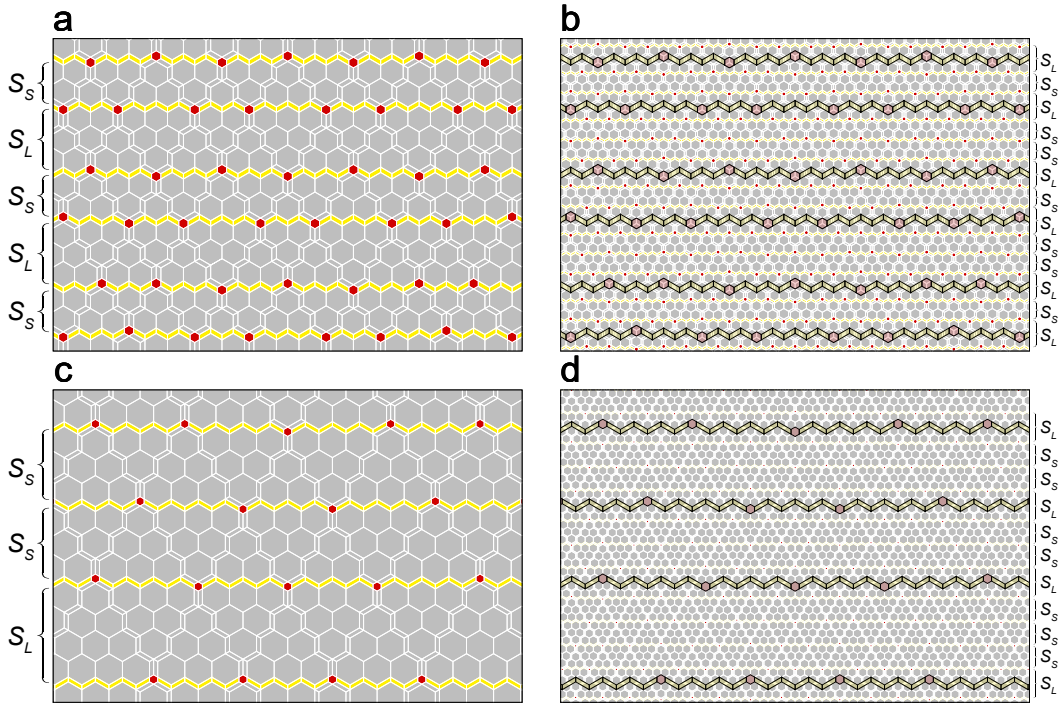}
    \caption{
      {\bf a} ({\bf c}) The colored tiles represent the domain boundaries
      along the horizontal direction for the hexagonal silver-mean (bronze-mean) tiling with $k=2$ ($k=3$).
      {\bf b} ({\bf d}) The tilings are obtained by applying the deflation
      operation to the ones shown in {\bf a} ({\bf c}).
      The transparent tiles represent the original domain boundaries along the horizontal direction.
    }
    \label{fig: wall}
  \end{center}
\end{figure*}

Supplementary Figures~\ref{fig: wall}{\bf a} and \ref{fig: wall}{\bf c} show the domain boundaries along the horizontal direction
as colored tiles for $k=2$ and $k=3$.
Each domain boundary is composed of the P and S tiles,
and adjacent P tiles always share their shorter edges.
We find that each P tile is adjacent to its reflected P tile (\=P tile) or an S tile.
Therefore, the zigzag chains of P tiles appear in the domain boundary.
We note that three domain boundaries with distinct directions do not share any P tiles, 
but always cross at a certain S tile.

Now, we discuss how the domain boundaries along a certain direction are arranged in the metallic-mean tilings.
We find in Supplementary Figs.~\ref{fig: wall}{\bf a} and \ref{fig: wall}{\bf c} 
that the spaces between the domain boundaries
are classified by two groups ${\cal S}_S$ and ${\cal S}_L$.
In the smaller space ${\cal S}_S$, there exist the $\alpha$ and $\beta$ honeycomb domains,
and in the other space ${\cal S}_L$, $\beta$ and $\gamma$ domains appear.
To clarify the distributions of these spaces, we use the substitution rule of the tiles. 
Supplementary Figures~\ref{fig: wall}{\bf b} and \ref{fig: wall}{\bf d} show
the deflations of the tilings shown in Supplementary Figs.~\ref{fig: wall}{\bf a} and \ref{fig: wall}{\bf c},
respectively.
When the deflation operation is applied to the tiling,
$k$ and $k+1$ domain boundaries are equally-spaced generated in the original ${\cal S}_S$ and ${\cal S}_L$ spaces,
respectively.
Namely, $k-1$ and $k$ ${\cal S}_S$ are generated.
On the other hand, under one deflation operation, 
one ${\cal S}_L$ is generated at each domain boundary. 
Therefore, the numbers of ${\cal S}_{S}$ and ${\cal S}_{L}$ at iteration $n$ ($N_{{\cal S}_S}$ and $N_{{\cal S}_L}$) 
satisfy 
\begin{equation}
  \left(\begin{array}{l}
    N_{{\cal S}_S}^{(n+1)} \\
    N_{{\cal S}_L}^{(n+1)}
  \end{array}\right)=\left(\begin{array}{cc}
  k-1 & k \\
  1 & 1
  \end{array}\right)\left(\begin{array}{l}
    N_{{\cal S}_S}^{(n)} \\
    N_{{\cal S}_L}^{(n)}
  \end{array}\right),
  \end{equation}
where the maximum eigenvalue is given by $\tau_k$
and the corresponding eigenvector $(k \tau_k, 1+\tau_k)^T$.
This means that the self-similarity inherent in the metallic ratio $\tau_k$ 
appears in the tiles, honeycomb domains, and spaces between adjacent domain boundaries.

\clearpage
\section{Higher-dimensional representation}\label{6d}
\subsection*{Perpendicular space}
In this section, 
we consider the perpendicular space,
introducing the six-dimensional representations of the vertices.
First, we describe the vertex site in the two-dimensional physical space ${\cal S}$
by six vectors ${\bf e}_m$ and six integer indices $\vec{n}=(n_0, n_1, n_2, n_3, n_4, n_5)^T$,
as
\begin{eqnarray}
{\bf r} = (x, y) = \sum_{m=0}^5 n_m {\bf e}_m,\label{eq:2dmap}
\end{eqnarray}
with
\begin{equation}
  {\bf e}_m =\left\{
\begin{array}{ll}
  \Big(\ell \cos(m\phi + \theta) , \ell \sin(m\phi + \theta)\Big) & m=0,1,2\\
  \Big(s \cos(m\phi + \theta) , s \sin(m\phi + \theta) \Big) & m=3,4,5\\
\end{array}
  \right.,
\end{equation}
where $\phi=2\pi/3$ and $\theta$ is constant.
$\ell$ and $s$ are the lengths for longer and shorter edges of the tiles.
The vectors ${\bf e}_m$ are schematically shown in Supplementary Fig.~\ref{fig: vec}.
\begin{figure}[htb]
  \begin{center}
    \includegraphics[width=0.55\linewidth]{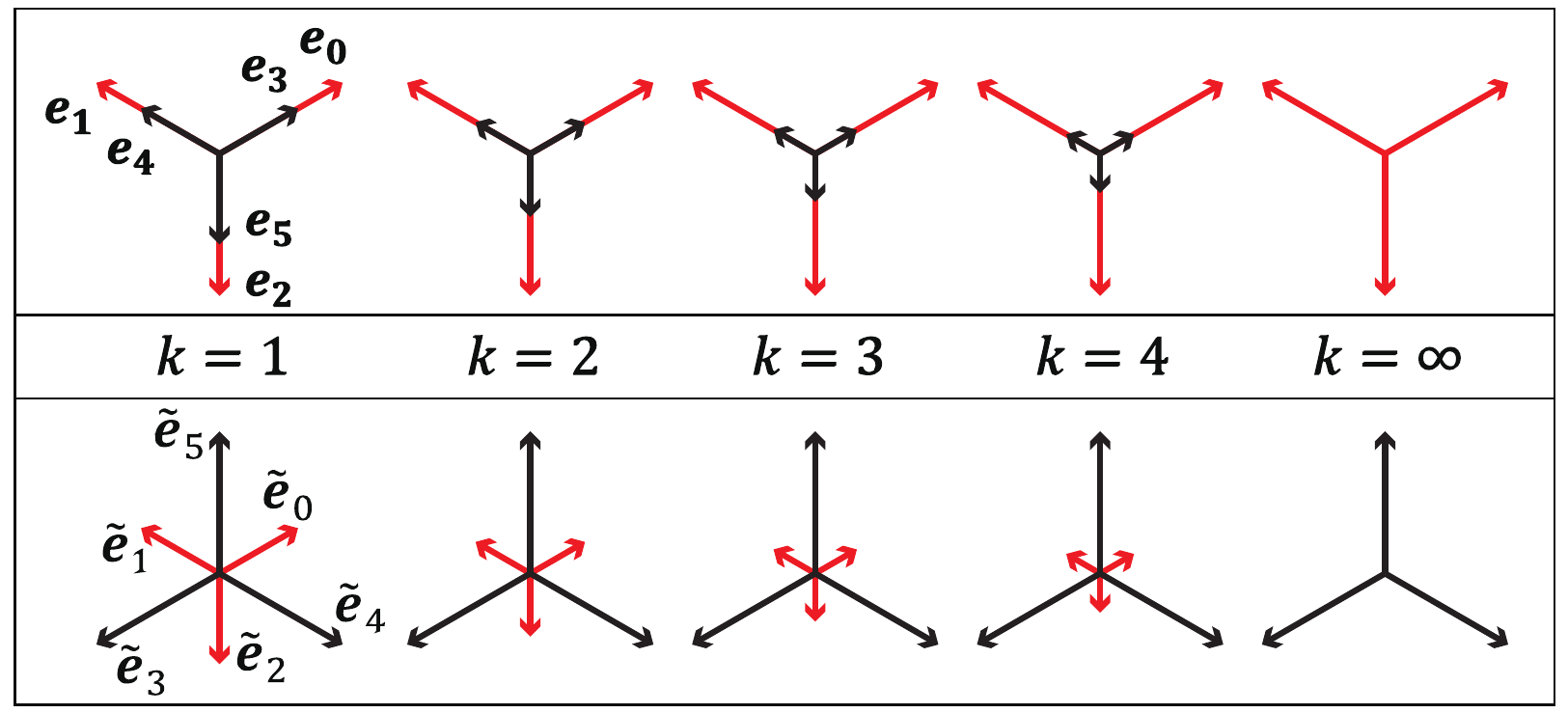}
    \caption{Real space basis ${\bf e}_0, \cdots , {\bf e}_5$ 
    and perpendicular space basis ${\bf \tilde{e}}_0, \cdots , {\bf  \tilde{e}}_5$.
    }
    \label{fig: vec}
  \end{center}
\end{figure}

In Eq.~(\ref{eq:2dmap}), the vertex site ${\bf r}$ can be regarded
as the projection from a six-dimensional lattice point,
where the vectors ${\bf e}_m$ are the projections from the six-dimensional basis vectors.
Thereby one can define the projections onto the other four-dimensional space (perpendicular space).
For a unified understanding of the projection,
it is convenient to introduce the six-dimensional space ${\cal S}^h$ including
the physical and perpendicular spaces.
Then, $\vec{n}$ is mapped to
the six-dimensional lattice point $\vec{r}^{\,h}$ in ${\cal S}^h$ as,
\begin{eqnarray}
 \vec{r}^{\, h}&=&M\vec{n},\\
  M &=& 
  \begin{pmatrix}
    \ell \cos\theta & \ell \cos(\phi + \theta) & \ell \cos(2\phi + \theta)&
    s \cos\theta & s \cos(\phi + \theta) & s \cos(2\phi + \theta)\\
    \ell \sin\theta & \ell \sin(\phi + \theta) & \ell \sin(2\phi + \theta)&
    s \sin\theta & s \sin(\phi + \theta) & s \sin(2\phi + \theta)\\
     \tau_k^{-1}\cos\theta & \tau_k^{-1}\cos(\phi + \theta) & \tau_k^{-1}\cos(2\phi + \theta)&
    - \cos\theta & - \cos(\phi + \theta) & - \cos(2\phi + \theta)\\
     \tau_k^{-1}\sin\theta & \tau_k^{-1}\sin(\phi + \theta) & \tau_k^{-1}\sin(2\phi + \theta)&
    - \sin\theta & - \sin(\phi + \theta) & - \sin(2\phi + \theta)\\
   \sqrt{2} \tau_k^{-1} & \sqrt{2} \tau_k^{-1} & \sqrt{2} \tau_k^{-1} & 0 & 0 & 0\\
    0 & 0 & 0 & -\sqrt{2} & -\sqrt{2}  & -\sqrt{2} \\
  \end{pmatrix},\nonumber \\ 
  & &\label{eq:M}
\end{eqnarray}
where $M$ is the mapping matrix.
Then, one can discuss vertex properties in both physical and perpendicular space.
Namely, in the physical space ${\cal S}$, the vertex site ${\bf r}$ is given 
by the first two components of the vector as
\begin{eqnarray}
  {\bf r}=\Big((\vec{r}^{\, h})_0, (\vec{r}^{\, h})_1 \Big)=\sum_{m=0}^5 n_m {\bf e}_m.
\end{eqnarray}
The four-dimensional perpendicular space is split into two-dimensional spaces 
$\tilde{\cal S}$ and ${\cal S}^\perp$, and
the corresponding coordinates $\tilde{\bf r}$ and ${\bf r}^\perp$ are given as
\begin{eqnarray}
  \tilde{\bf r}&=&\Big((\vec{r}^{\, h})_2, (\vec{r}^{\, h})_3 \Big)=\sum_{m=0}^5 n_m \tilde{\bf e}_m,\\
  {\bf r}^\perp&=&\Big((\vec{r}^{\, h})_4, (\vec{r}^{\, h})_5 \Big)=\sum_{m=0}^5 n_m {\bf e}^\perp_m,
\end{eqnarray}
where $\tilde{\bf e}_m=(M_{2m},M_{3m})$ and ${\bf e}^\perp_m=(M_{4m},M_{5m})$.
We find that ${\bf r}^\perp = (x^\perp , y^\perp)$ takes
four values $x^\perp = 0, \sqrt{2} \tau_k^{-1}$ and
$y^\perp = -\sqrt{2},0$
in the hexagonal metallic-mean tilings.
In each ${\bf r}^\perp$ plane, the $\tilde{\bf r}$ points densely
cover a certain window.
We find that the window in planes
${\bf r}^\perp  = (0,0), (\sqrt{2}\tau_k^{-1},-\sqrt{2})\;[(0,-\sqrt{2}), (\sqrt{2}\tau_k^{-1},0)]$
has a hexagonal (triangular) structure.
\begin{figure}[htb]
  \begin{center}
    \includegraphics[width=0.85\linewidth]{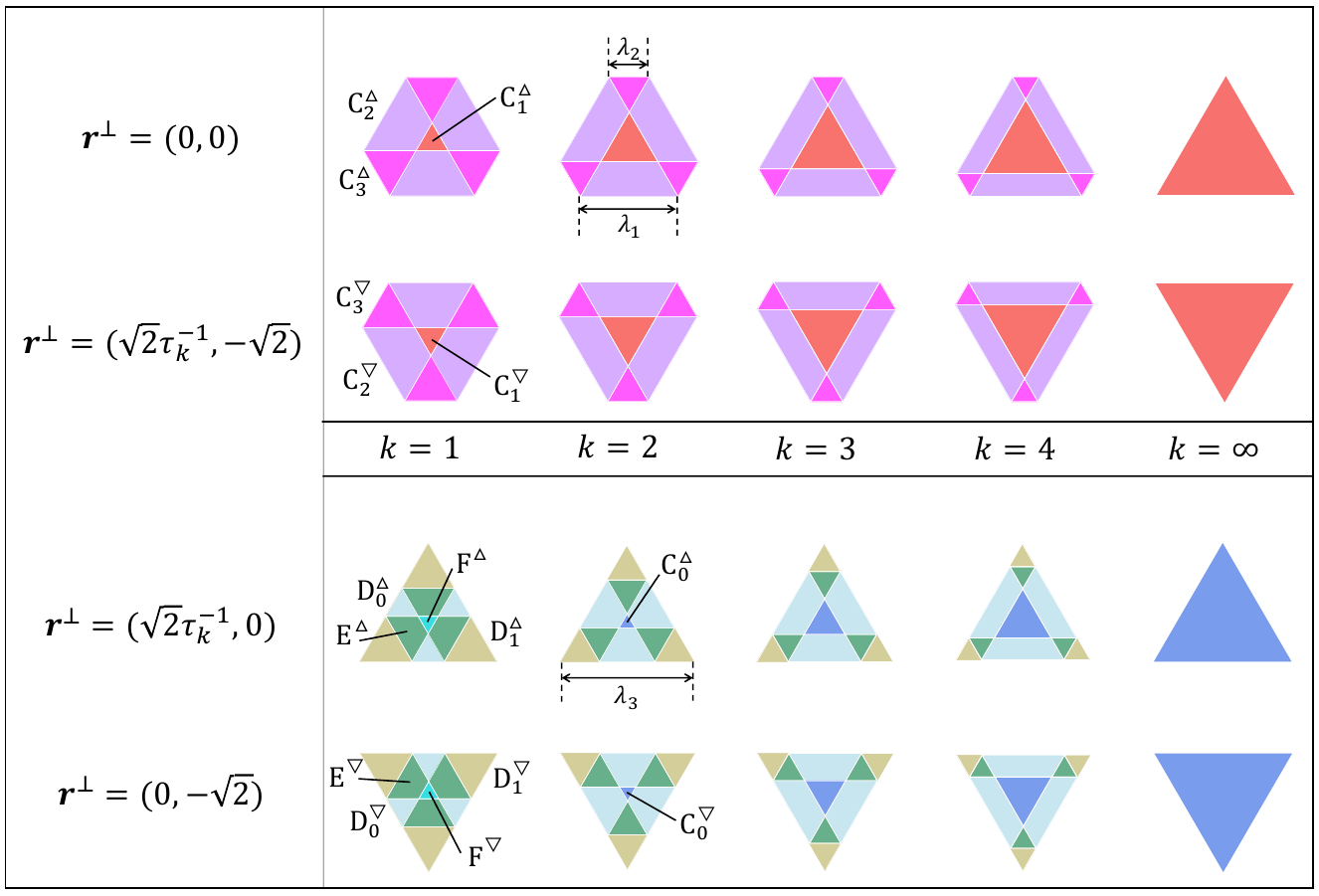}
    \caption{Perpendicular spaces
      [${\bf r}^\perp =(0,0), (\sqrt{2}\tau_k^{-1},-\sqrt{2}), (\sqrt{2}\tau_k^{-1},0)$, and $(0,-\sqrt{2})$]
        of the hexagonal metallic-mean tilings for $k=1, 2, 3, 4$, and $\infty$.
        Each area bounded by the solid lines is the region of one of eight types of vertices with $\triangle$ and $\triangledown$.
        $\lambda_1$ and $\lambda_2$ ($\lambda_3$) are the characteristic lengths
    of the windows with ${\bf r}^\perp =(0,0)$ and $(\sqrt{2}\tau_k^{-1},-\sqrt{2})$ [$(\sqrt{2}\tau_k^{-1},0)$ and $(0,-\sqrt{2})$].}
    \label{fig: PS}
  \end{center}
\end{figure}
Supplementary Figure~\ref{fig: PS} shows the perpendicular spaces
for the hexagonal metallic-mean tilings with $k=1, 2, 3, 4$, and $\infty$.
The characteristic lengths $\lambda_1$, $\lambda_2$ and $\lambda_3$
will be obtained in Supplementary Note~\ref{sec: phason}.
We find that eight types of vertices are mapped into specific regions.
This implies that the perpendicular spaces reflect the local environments for the lattice sites.
Namely, the areas of each vertex region in perpendicular spaces are proportional
to its fraction in the physical space.
The region of the F (C$_0$) vertices appears only in the case of $k=1$ ($k \neq 1$).
This means the absence of C$_0$ (F) vertices
in the hexagonal metallic-mean tilings with $k=1$ ($k\neq 1$),
which is consistent with the results discussed in the main text and Supplementary Note~\ref{app: fraction}.
The areas of the C$_0$, C$_1$ vertices (the others) monotonically increase (decrease)
with increasing $k$.
The planes ${\bf r}^\perp=(0,0)$ and $(\sqrt{2}\tau_k^{-1},-\sqrt{2})$ [$(0,-\sqrt{2})$ and $(\sqrt{2}\tau_k^{-1},0)$] are fully occupied
only by the C$_0$ and C$_1$ vertices in the limit $k \rightarrow \infty$
since $f_{\rm C_0}, f_{\rm C_1} \rightarrow 1/2$.
We also find that the vertices in the A (B) sublattice are mapped to the planes
with $(0,0)$ and $(\sqrt{2}\tau_k^{-1},-\sqrt{2})$ [$(0,-\sqrt{2})$ and $(\sqrt{2}\tau_k^{-1},0)$].
This can be explained by the following.
The sublattice index for each vertex is uniquely determined, as discussed above.
Since upon moving from one site to its neighbor only one of the $n_m$'s changes by $\pm 1$,
the site with an even (odd) number $(\tau_k x^\perp + y^\perp)/\sqrt{2}$ corresponds to the A (B) sublattice.
Correspondingly, the areas for both sublattices are different from each other,
meaning the existence of the sublattice imbalance in the system.

We wish to note that the perpendicular space analysis clarifies
whether each vertex belongs to the honeycomb domain with $\triangle$ or $\triangledown$.
As seen in Supplementary Fig.~\ref{fig: domain-tiling},
the vertices in the space
${\bf r}^\perp =  (\sqrt{2}\tau^{-1}_k,-\sqrt{2})$ and $(0,-\sqrt{2})$ [(0,0) and $(\sqrt{2}\tau^{-1}_k,0)$]
belong to the honeycomb domain with $\triangle$ ($\triangledown$).
This can be explained by the following.
Each honeycomb domain is composed of only L tiles.
When one moves within the domain, 
the longer length appears in the physical space, changing $x^\perp$.
Therefore, we find that vertices in a certain honeycomb domain take the common value of $y^\perp$.
Furthermore, any honeycomb domains with $\alpha(=\triangle$ or $\triangledown$) are adjacent 
to the honeycomb domains with $\bar{\alpha}$
across the single boundary composed of the zig-zag P tiles, as discussed in the main text.
When one moves from one honeycomb domain to its neighboring honeycomb domain, 
one shorter length appears in the physical space, changing $y^\perp$ by $\pm \sqrt{2}$. 
Therefore,
we find that vertices of the adjacent domains take the different values of $y^\perp$ each other,
and understand that vertices,
which are mapped to the perpendicular spaces
${\bf r}^\perp =  (\sqrt{2}\tau^{-1}_k,-\sqrt{2})$ and $(0,-\sqrt{2})$ [(0,0) and $(\sqrt{2}\tau^{-1}_k,0)$],
belong to the honeycomb domains with $\triangle$ ($\triangledown$).
By these reasons, the perpendicular analysis clarifies 
not only the sublattice structure but also the honeycomb domain structure.
One can define the vertex belonging to the honeycomb domain 
with $\alpha(=\triangle, \triangledown)$ as X$^\alpha$, where 
$\rm X(=C_0, C_1, \cdots, F)$.
Namely, the fractions for $X^\triangle$ and $X^\triangledown$ are identical 
since the corresponding areas are identical in the perpendicular space.

\begin{figure}[htb]
  \begin{center}
    \includegraphics[width=0.7\linewidth]{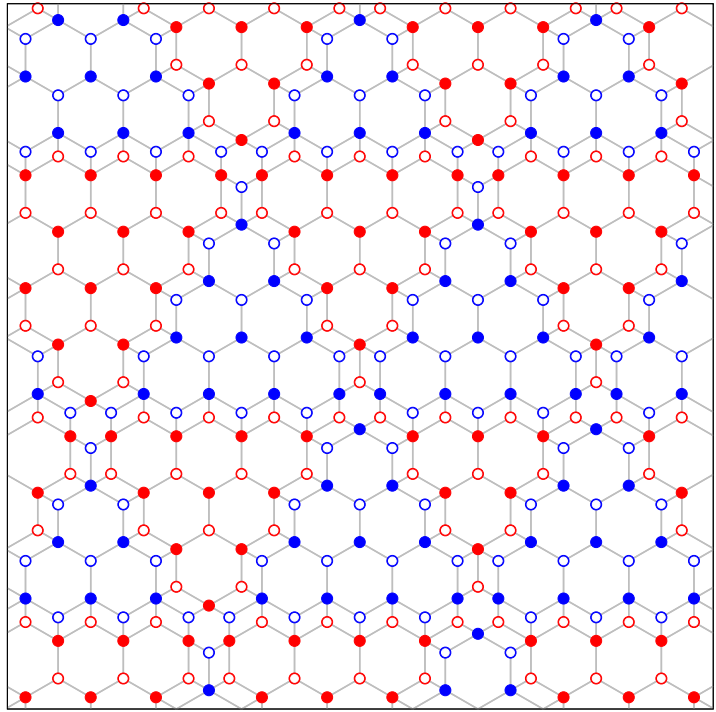}
    \caption{Hexagonal bronze-mean tiling
    with vertices classified according to their coordinates in the perpendicular space ${\bf r}^\perp$.
    Red open, blue open, red filled, and blue filled circles represent the vertices,
    which are mapped to the perpendicular space with
    ${\bf r}^\perp =(0,0), (\sqrt{2}\tau^{-1}_k,-\sqrt{2}), (\sqrt{2}\tau^{-1}_k,0)$ and $(0,-\sqrt{2})$,
    respectively.
    The blue (red) symbols belong to honeycomb domains with $\triangle$ ($\triangledown$).
    }
    \label{fig: domain-tiling}
  \end{center}
\end{figure}

\subsection*{Cut-and-project scheme}
In the six-dimensional representations,
we have introduced
two lengths $\ell$ and $s$, and the metallic mean $\tau_k$.
Note that, in Eq.~(\ref{eq:M}), the length scale appears only in the physical space and 
the metallic mean appears only in the perpendicular space.
This may mean that the metallic mean is important in the perpendicular space while
plays no role in the physical space. 
This allows us to generate the hexagonal metallic-mean tilings 
with arbitrary lengths $\ell$ and $s$, in contrast to the tilings generated by means of the substitution rule. 
This is known as the cut-and-project scheme, which should be useful to clarify how relevant the tiling is for the
atomic position in the MC and MD simulations discussed in the main text.

The six-dimensional lattice points $\vec{r}^{\, h}$ with indices $\vec{n}$
relevant for the generalized hexagonal metallic-mean tiling
satisfies the condition that
their projections onto the perpendicular space
are located inside the windows, which are shown in Supplementary Fig.~\ref{fig: PS}.
Then, projecting these six-dimensional coordinates onto the physical space,
we can obtain the generalized hexagonal tilings.
Golden-mean, silver-mean, and bronze-mean tilings with $\ell/s=1/2$ and $\ell/s=2$ 
generated by means of the cut-and-project scheme are shown
in Supplementary Fig.~\ref{fig: variable}.
We wish to note that
the self-similarity in the tilings is inherent in the case $\ell /s = \tau_k$
and the substitution rule cannot be defined in generic case with arbitrary $\ell$ and $s$.
\begin{figure}[htb]
  \begin{center}
    \includegraphics[width=0.8\linewidth]{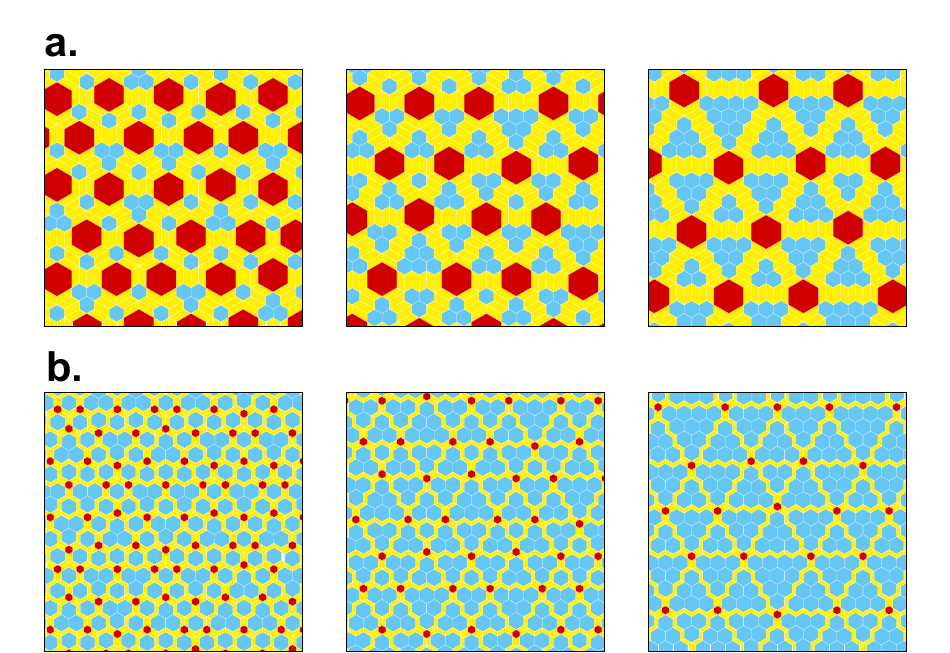}
    \caption{
    Hexagonal golden-, silver- and bronze-mean tilings
    generated by the cut-and-project scheme
    with {\bf a} $\ell/s = 1/2$ and {\bf b} $\ell/s = 2$.}
    \label{fig: variable}
  \end{center}
\end{figure}

\subsection*{Reciprocal vectors}

Each vertex site in the tilings is described by the six integer indices $\vec{n}$ 
and that mapped to the six-dimensional space ${\cal S}^h$ is represented as
\begin{eqnarray}
  \vec{r}^{\, h}=M\vec{n}=\sum_i n_i \vec{e}_i^{\, h},
\end{eqnarray}
where $\vec{e}_i^{\, h}\,(i=0,1,\cdots,5)$ are the six kinds of six-dimensional basis vectors
with $(\vec{e}_i^{\, h})_j[=M_{ij}]$.
The six kinds of six-dimensional reciprocal vectors $\vec{q}_i^{\, h}$ are obtained
by imposing the condition $\vec{e}_i^{\, h} \cdot \vec{q}_j^{\, h} = 2\pi \delta_{ij}$ with $\delta_{ij}$ is Kronecker delta.
These are explicitly given as,
\begin{eqnarray}
  \vec{q}_m^{\, h} &=&C\left(
  \begin{array}{c}
   \cos(m \phi + \theta) \\ \sin(m \phi + \theta) \\ s \cos(m \phi + \theta) \\ s \sin(m \phi + \theta) \\ (\ell\tau_k+s)/(2\sqrt{2}) \\ 0
  \end{array}
  \right) \qquad\quad {\rm for}\quad m=0, 1, 2,\; \\
  \vec{q}_m^{\, h} &=& C \left(
  \begin{array}{c}
  \tau^{-1}_k \cos(m \phi + \theta) \\ \tau^{-1}_k \sin(m \phi + \theta) \\  -\ell \cos(m \phi + \theta) \\  -\ell \sin(m \phi + \theta) \\ 0 \\ -(\ell+s\tau^{-1}_k)/(2\sqrt{2})
  \end{array}
  \right) \qquad {\rm for}\quad m=3, 4, 5,
\end{eqnarray}
where $C = 4\pi/[3(\ell+s\tau_k^{-1})]$.
The reciprocal vectors projected onto the physical space are given as,
\begin{eqnarray}
  {\bf q}_m= \Big((\vec{q}_m^{\, h})_0,(\vec{q}_m^{\, h})_1\Big).
\end{eqnarray}
The six reciprocal vectors are composed of three long vectors ${\bf q}_0,{\bf q}_1,{\bf q}_2$ and 
three short vectors ${\bf q}_3,{\bf q}_4,{\bf q}_5$. 
The ratio of their lengths is given by $\tau_k$.
The longer length $|{\bf q}_0|=|{\bf q}_1|=|{\bf q}_2|=4\pi/[3(\ell+s\tau_k^{-1})]$ 
monotonically increases with increasing $k$
and approaches the constant $4\pi/(3\ell)$ in larger $k$.
On the other hand, the shorter length
$|{\bf q}_3|=|{\bf q}_4|=|{\bf q}_5|=4\pi/[3(\ell\tau_k+s)]$ monotonically decreases
and vanishes in the limit $k\rightarrow\infty$. 
This is consistent with the fact that, in the limit $k\rightarrow\infty$,
three of the reciprocal vectors are reduced to those for the honeycomb lattice and 
the metallic-mean tiling can be regarded as the aperiodic approximant of the honeycomb lattice.
We wish to note that the net number of the reciprocal vectors
is two due to satisfying ${\bf q}_0=-({\bf q}_1 + {\bf q}_2)$ in the limit $k\rightarrow\infty$.
This is consistent with the fact that the honeycomb lattice is periodic.
\begin{figure}[htb]
  \begin{center}
    \includegraphics[width=0.55\linewidth]{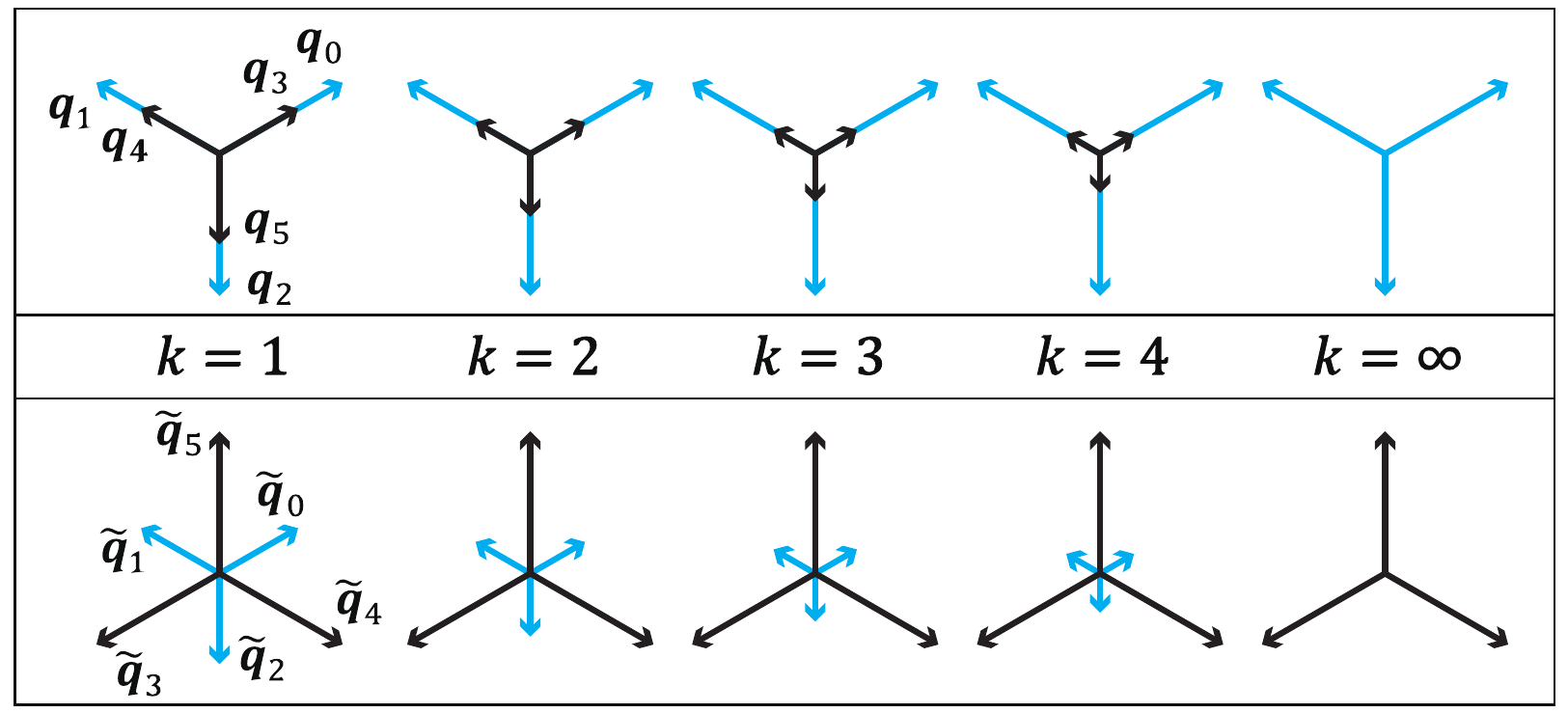}
    \caption{Reciprocal lattice basis in the real space ${\bf q}_0, \cdots , {\bf q}_5$ 
    and those in the perpendicular space ${\bf \tilde{q}}_0, \cdots , {\bf  \tilde{q}}_5$.
    }
    \label{fig: qvec}
  \end{center}
\end{figure}

\clearpage
\section{Phason flips}\label{sec: phason}
We examine phason flips in the hexagonal metallic-mean tilings.
As seen in Supplementary Note~\ref{6d},
the vertices in the metallic-mean tilings can be mapped 
inside of the windows in the perpendicular space.
When the windows in the perpendicular space slightly slide,
positions for some vertices become located outside of the windows and
those for some vertices become located inside.
This slightly changes the vertex structure in the physical space, 
that is, some vertices can be regarded to move, so called, phason flips.
To discuss how the phason flips occur in the metallic-mean tiling,
we show the bronze-mean tiling and the tiling with the slightly shifted windows in
Supplementary Figs.~\ref{fig: phason}{\bf a} and \ref{fig: phason}{\bf b}.
For clarify, the difference between these two tilings is shown in color.
We find the phason flip along a certain domain boundary.
Furthermore, we find that this change can be described
by three kinds of local flips.
Note that a single local flip never appears
due to the matching rule of the tilings.
One of local flips is the position change around the C$_2$ vertex
sharing one L tile and two P tiles, as shown in Supplementary Fig.~\ref{fig: phason}{\bf c},
where the C$_2$ vertex and the D$_0$, D$_1$ and E vertices connected to it by the longer edges
change their positions.
The local flip around the C$_3$ vertex also occurs at the same time, 
as shown in Supplementary Fig.~\ref{fig: phason}{\bf d},
where the C$_3$ vertex and the D$_1$ vertices connected to it by the shorter edges
change their positions.
The other flip appears at the intersection of two domain boundaries,
as shown in Supplementary Fig.~\ref{fig: phason}{\bf e}.
In the case,
some D$_0$, D$_1$ and E vertices change their positions.
When one focuses on the E vertex,
the change in the physical space
is characterized by ${\bf e}_2 - {\bf e}_0$,
as shown in Supplementary Fig.~\ref{fig: phason-E}{\bf a}.
The corresponding move in the perpendicular space
appears between the corners of the triangular window for E vertices,
which is characterized by $\tilde{\bf e}_2 - \tilde{\bf e}_0$
schematically shown in Supplementary Fig.~\ref{fig: phason-E}{\bf b}.
Therefore, an edge length of the window of E vertex in the perpendicular space
(if $k=1$, a longer edge length of the trapezoid)
is %
$|\tilde{\bf e}_2 - \tilde{\bf e}_0|=\sqrt{3}\tau^{-1}_k$.
We immediately obtain the characteristic lengths of the windows
in the perpendicular space as,
\begin{eqnarray}
  \lambda_1 &=& \sqrt{3},\\
  \lambda_2 &=& \sqrt{3}\tau^{-1}_k,\\
  \lambda_3 &=& \sqrt{3}(1+\tau^{-1}_k).
 \end{eqnarray}
We note that the above phason flip also changes
the areas of the honeycomb domains.
This should be observed as thermal fluctuations of Monte Carlo and MD simulations~\cite{Engel2011} 
(See the main text for the details).

\begin{figure*}[htb]
  \begin{center}
    \includegraphics[width=0.8\linewidth]{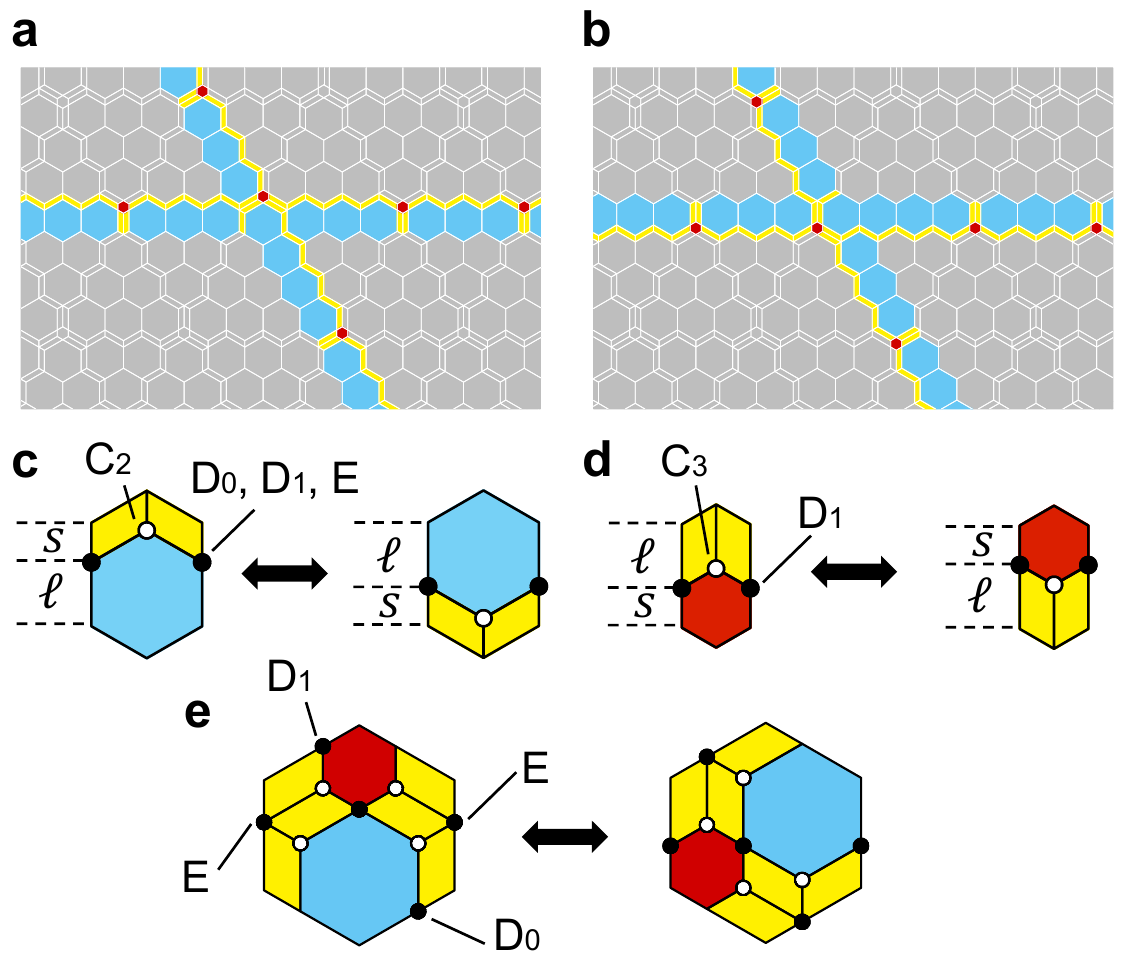}
    \caption{Schematic pictures for the phason flips
      in the hexagonal bronze-mean tiling.      
    {\bf a} and {\bf b} show the bronze-mean tiling and the tiling with the slightly shifted windows, respectively.
    The difference is explicitly shown in color.
    {\bf c} ({\bf d}) The local flips around the C$_2$ (C$_3$) vertex,
    {\bf e} The local flip at the intersection of two domain boundaries.
    In {\bf c},{\bf d} and {\bf e}, circles represent the vertices
    which move due to the phason flip.
    }
    \label{fig: phason}
  \end{center}
\end{figure*}

\begin{figure}[htb]
  \begin{center} 
    \includegraphics[width=0.5\linewidth]{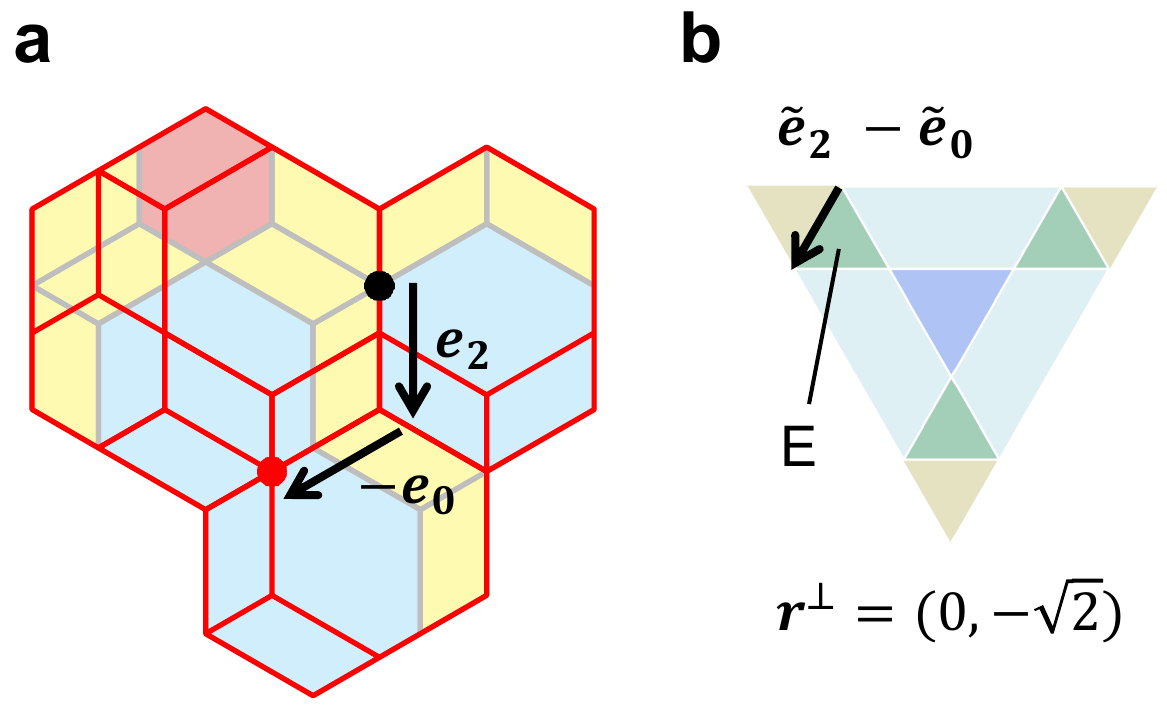}
    \caption{
    {\bf a} The position change of the E vertex due to the phason flip.
    The black (red) circle indicates the position of the E vertex before (after) the flip.
    {\bf b} shows the corresponding position change in the perpendicular space
    ${\bf r^\perp}=(0,-\sqrt{2})$.
    }
    \label{fig: phason-E}
  \end{center}
\end{figure}

\clearpage
\section{Lattice structure factors}\label{app: fourier}
\begin{figure}[htb]
  \begin{center}
    \includegraphics[width=\linewidth]{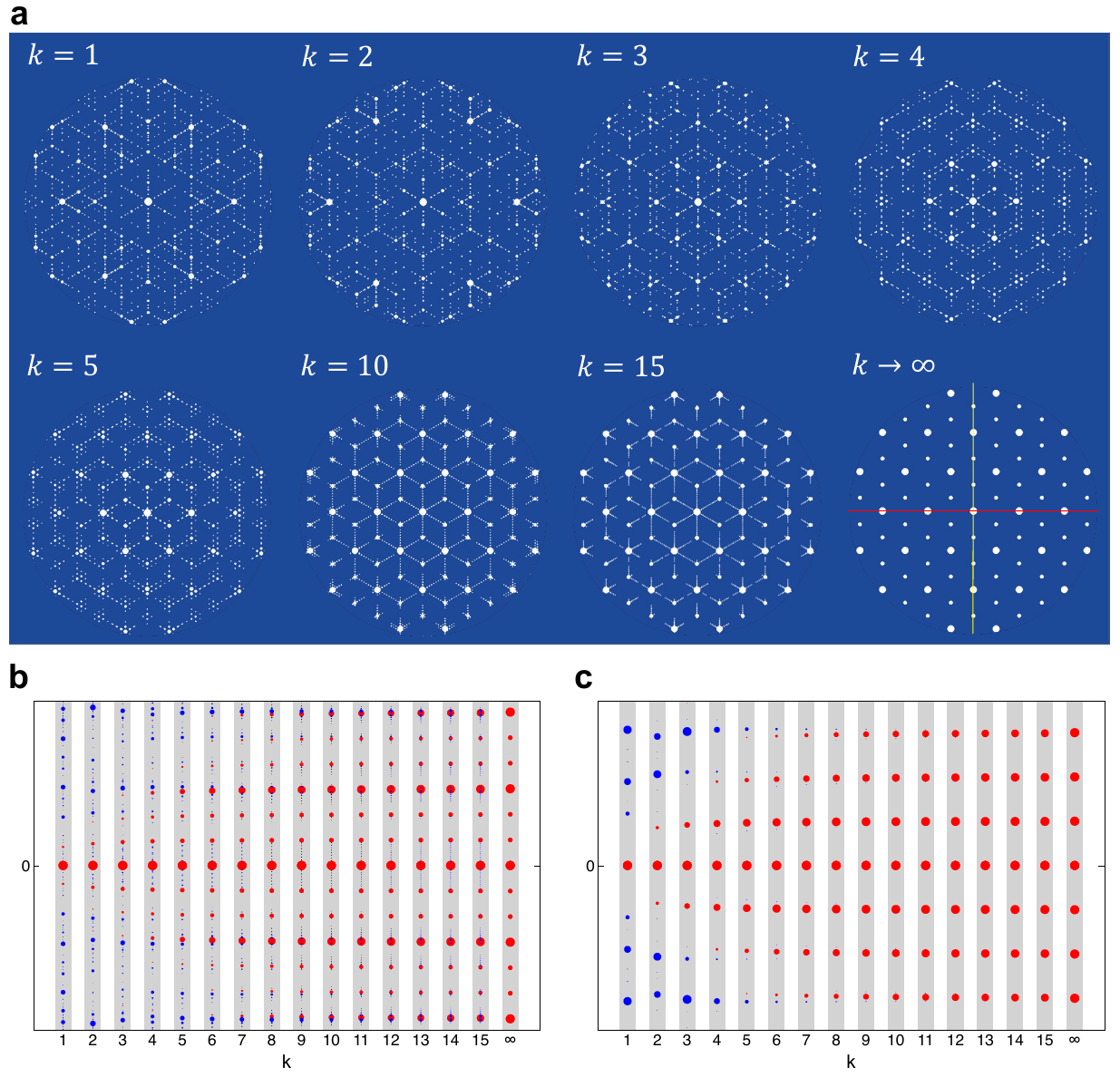}
    \caption{{\bf a} The lattice structure factors for each $k$.
    {\bf b} ({\bf c}) The crosssection of the lattice structure factors on the mirror axis shown as the yellow (red) line in {\bf a}.
    The red circles show peak positions spanned by only ${\bf q}_0, {\bf q}_1$ and ${\bf q}_2$, and the blue circles show the others.}
    \label{fig: fourier}
  \end{center}
\end{figure}

We study the lattice structure factors of the hexagonal metallic-mean tilings in detail.
Supplementary Figure~\ref{fig: fourier}{\bf a} shows the lattice structure factors for each $k$,
where the circular area corresponds to the weight of the peak.
It is clearly found that the structure factor have the sixfold rotational symmetry,
meaning that the metallic-mean tilings have a hexagonal symmetry.

To clarify how the peak structure is changed with varying $k$, 
we show in Supplementary Fig.~\ref{fig: fourier}{\bf b} ({\bf c})
the crosssection of the lattice structure factors on two mirror axes,
which are explicitly shown as the yellow (red) line
in Supplementary Fig.~\ref{fig: fourier}{\bf a}. 
When $k \rightarrow \infty$, the peaks are periodically distributed
since the system is reduced to the honeycomb lattice.
These positions are spanned by ${\bf q}_0, {\bf q}_1$ and ${\bf q}_2$, and 
the weights of the peaks take two values, 1 and 0.25,
as shown in Supplementary Figs.~\ref{fig: fourier}{\bf b} and \ref{fig: fourier}{\bf c}. 
Decreasing $k$, main peaks remain with larger weights and 
satellite peaks with smaller weights are induced slightly away from the main peaks.
The main peak positions are mainly spanned by ${\bf q}_0, {\bf q}_1$ and ${\bf q}_2$.
On the other hand, the other peak positions
are represented by six reciprocal vectors as
\begin{eqnarray}
  {\bf Q}=\sum_{m=0}^5 \tilde{n}_m {\bf q}_m,
\end{eqnarray}
where %
$\tilde{n}_m$ is an integer.
These are consistent with the fact that
the metallic-mean tilings are quasiperiodic.
Furthermore, the existence of the satellite peaks originates from 
the incommensurate modulated structure formed by the domain boundary
composed of P tiles in the real space.
When $k\le 4$, the main peaks with large weights are represented by
six reciprocal vectors, which are shown as the blue circles
in Supplementary Figs.~\ref{fig: fourier}{\bf b} and \ref{fig: fourier}{\bf c}.
This means that the vertex structures for small $k$ are
no longer described by the modulated honeycomb lattice.

Here, we calculate the lattice structure factors.
The density of the vertex sites in the tiling is given
by $\rho ({\bf r}) = \sum_j \delta ({\bf r}_j - {\bf r})$,
where $\delta$ is the delta function and ${\bf r}_i$ represents the position of the $i$th vertex site.
The summation is over all vertex sites in the tiling.
The lattice structure factors are give by $|\rho({\bf q})|^2$,
where %
\begin{eqnarray}
  \rho({\bf q}) = \int \rho ({\bf r}) \exp(-i{\bf r} \cdot {\bf q}) d{\bf r}
           = \sum_j  \exp(-i{\bf r}_j \cdot {\bf q}).
\end{eqnarray}
It is convenient to make use of the higher-dimensional representation discussed in Supplementary Note~\ref{6d}.
Since the six-dimensional basis and reciprocal vectors $\vec{e}_m^{\, h}$ and $\vec{q}_m^{\, h}$ 
satisfy the orthogonal relation
$\vec{e}_m^{\, h} \cdot \vec{q}_n^{\, h} = 2\pi \delta_{mn}$,
we obtain 
\begin{eqnarray}
   \exp(i\vec{e}_m^{\, h} \cdot \vec{q}_n^{\, h})
= \exp(i{\bf e}_m \cdot {\bf q}_n) \exp(i\tilde{\bf e}_m \cdot \tilde{\bf q}_n) \exp(i{\bf e}^\perp_m \cdot {\bf q}^\perp_n)=1,
\end{eqnarray}
where ${\bf r}_m({\bf q}_m)$, $\tilde{\bf r}_m (\tilde{\bf q}_m)$, and ${\bf r}_m^\perp ({\bf q}_m^\perp)$ are the two-dimensional vectors in the spaces
${\cal S}, {\cal \tilde{S}}$, and ${\cal S}^\perp$, respectively.
Therefore, we obtain
\begin{eqnarray}
  \rho({\bf q})
= \sum_j  \exp(i\tilde{\bf r}_j \cdot \tilde{\bf q}) \exp(i{\bf r}^\perp_j \cdot {\bf q}^\perp),\label{rhoq}
\end{eqnarray}
where the $j$th vertex site ${\bf r}_j$ is mapped to $\tilde{\bf r}_j$ and ${\bf r}_j^\perp$ 
in the perpendicular spaces ${\cal\tilde{S}}$ and ${\cal S}^\perp$.
Each vertex site is mapped to the point in one of four domains with
${\bf r}^\perp =(0,0), (\sqrt{2}\tau_k^{-1},-\sqrt{2}), (\sqrt{2}\tau_k^{-1},0)$, and $(0,-\sqrt{2})$.
Therefore,
the equation (\ref{rhoq}) can be divided into four, as
\begin{eqnarray}
    \rho({\bf q}) &=& {\cal F}_{00}(\tilde{\bf q})+\exp(i \sqrt{2} \tau^{-1}_k q^\perp_0) {\cal F}_{10}(\tilde{\bf q})\nonumber\\ 
    & &+\exp(-i\sqrt{2} q^\perp_1) {\cal F}_{01}(\tilde{\bf q})
    +\exp(i \sqrt{2}\tau^{-1}_k q^\perp_0)\exp(-i\sqrt{2} q^\perp_1) {\cal F}_{11}(\tilde{\bf q}), \\ 
  {\cal F}_{mn}(\tilde{\bf q})&=&\sum_{j \in (mn)^\perp}  \exp(i\tilde{\bf r}_j \cdot \tilde{\bf q}),
  \label{eq: Fourier}
\end{eqnarray}
where ${\cal F}_{mn}$ indicates the Fourier transform of the occupation domain in the perpendicular space with $(\tau_k x^\perp , -y^\perp)/\sqrt{2} = (m,n)$,
which are explicitly shown as the hexagons and triangles shown in Supplementary Fig.~\ref{fig: PS}.
Then, we can evaluate the lattice structure factor $|\rho(\bf q)|^2$.

\clearpage
\section{Crystallographic description of the P31m particle system}\label{sec: partices}
The particle system exhibits a regular structure with a plane group represented by P31m, as shown in Supplementary Fig.~\ref{fig:SIP31m}.
A solid bold line indicates a mirror plane, and a dashed line indicates a glide plane. A solid triangle symbol indicates a three-fold rotation axis. 
This group is characterized by its 3-fold symmetry. Let A, B, C, D, E, and F be the centers of particles.
As a result of this symmetry, we observe that the distances AB, BC, and BD are all equal.
To further analyze the system, we make the assumption that there are two types of equilateral polygons present: pentagons and triangles. By considering the equality AB = CF, we can determine the positions of all the particles in the system.
Within the unit cell, there are five particles, and their positions can be identified using the Wyckoff Positions 2b and 3c, as shown in the Supplementary Table \ref{table:SI1}. If the coordinates of particle A in Supplementary Fig.~\ref{fig:SIP31m}{\bf a} is set to be $(c, 0)$, we find 
\begin{eqnarray}
c=\frac{\sqrt{33}-3}{12}=0.2287, \quad
\overline{\rm AB}=\overline{\rm CF}=\sqrt{3}c=\frac{\sqrt{11}-\sqrt{3}}{4}=0.3961.
\end{eqnarray}
It is important to note that while the pentagon (BCFED) in the particle system is equilateral, it is not a regular pentagon, for it is elongated along the mirror plane.

\begin{figure}[h]
\begin{center}
\includegraphics[width=0.9\linewidth]{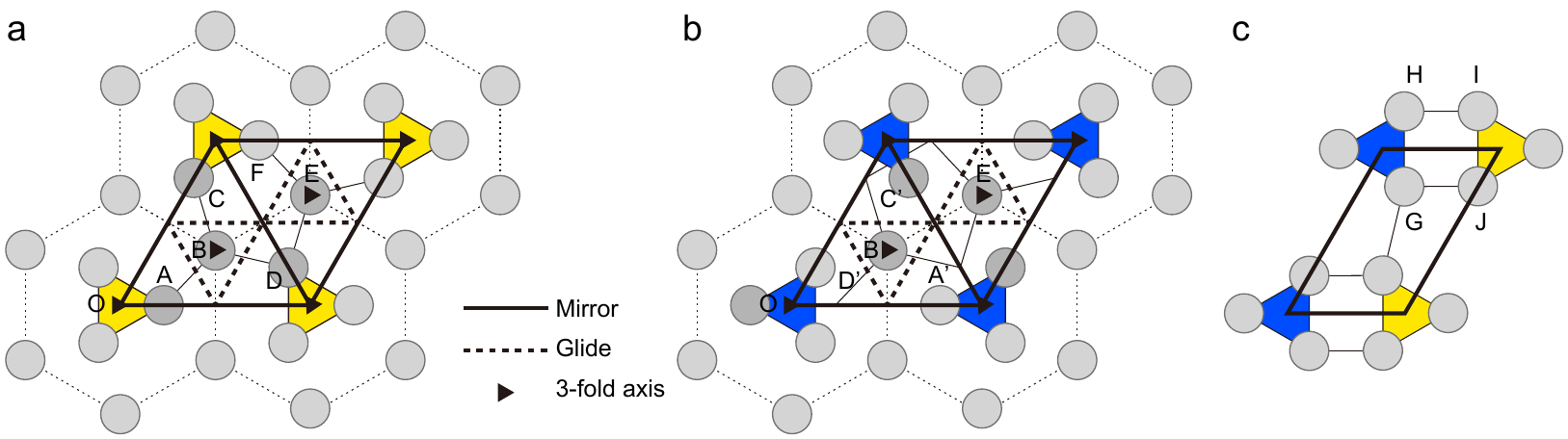}
\caption{{\bf a, b,} Plane group P31m for a particle system. {\bf c,} Decoration of a parallelogram. }
\label{fig:SIP31m}
\end{center}
\end{figure} 
\begin{table}[htp]
\begin{center}
\begin{tabular}{|c|c|c|c|}
\hline
Wyckoff Position & Site Sym. & Points & Coordinates\\
\hline
2b & 3.. & B, E &  
$\left(\frac{1}{2}, \frac{\sqrt{3}}{6}\right)$, 
$\left(1, \frac{\sqrt{3}}{3}\right)$\\
3c & ..m & A, C, D & $(c, 0)$, 
$\left(\frac{1}{2}(1-c), \frac{\sqrt{3}}{2}(1-c)\right)$, 
$\left(1-\frac{1}{2}c, \frac{\sqrt{3}}{2}c\right)$\\
3c & ..m & A', C', D' & $(1-c, 0)$, 
$\left(\frac{1+c}{2}, \frac{\sqrt{3}}{2}(1-c)\right)$,
$\left(\frac{1}{2}c, \frac{\sqrt{3}}{2}c\right)$ \\
\hline
\end{tabular}
\end{center}
\caption{Wyckoff Position for a particle system. $c=\frac{\sqrt{33}-3}{12}=0.2287.$}
\label{table:SI1}
\end{table}%

The basis vectors $\bf{a}_1$, $\bf{a}_2$ and 
the corresponding reciprocal basis vectors $\bf{b}_1$, $\bf{b}_2$ are given by
\begin{eqnarray*}
{\bf a}_1=a(1,0), \quad
{\bf a}_2=a\left(\frac{1}{2},\frac{\sqrt{3}}{2}\right), \quad
{\bf b}_1=\frac{2\pi}{a}\left(1,-\frac{\sqrt{3}}{3}\right), \quad
{\bf b}_2=\frac{2\pi}{a}\left(0,\frac{2\sqrt{3}}{3}\right),  
\end{eqnarray*}
where $a$ is the lattice constant (rhombus edge). For the reciprocal lattice vector $\bf{G}=h\bf{b}_1+ k\bf{b}_2$, 
the structure factor is evaluated as 
\begin{eqnarray}
F(h, k)=f\sum_{\alpha={\rm A,B,C,D,E}} \exp(2\pi i \bf{G}\cdot \bf{r}_\alpha).
\label{SFforCry}
\end{eqnarray}
The diffraction image calculated by this equation is shown in Supplementary Fig.~\ref{fig: simulationimage0}.\bigskip

\noindent {\bf Model for the metallic-mean tiling}: 
Supplementary Figure~\ref{fig:SIP31m}{\bf c} represents a decoration model for a parallelogram. If we assume that
the rectangle GHIJ is a square; {\it i.e.} $\overline{\rm AB}=\overline{\rm CF}=\overline{\rm GH}=\overline{\rm GJ}$, then the length ratio of the tile is 
\begin{eqnarray}
\frac{s}{\ell}=(1+\sqrt{3})c=0.6249.
\label{SFforCry}
\end{eqnarray}

\begin{figure*}[htb]
  \begin{center}
     \includegraphics[width=0.4\linewidth]{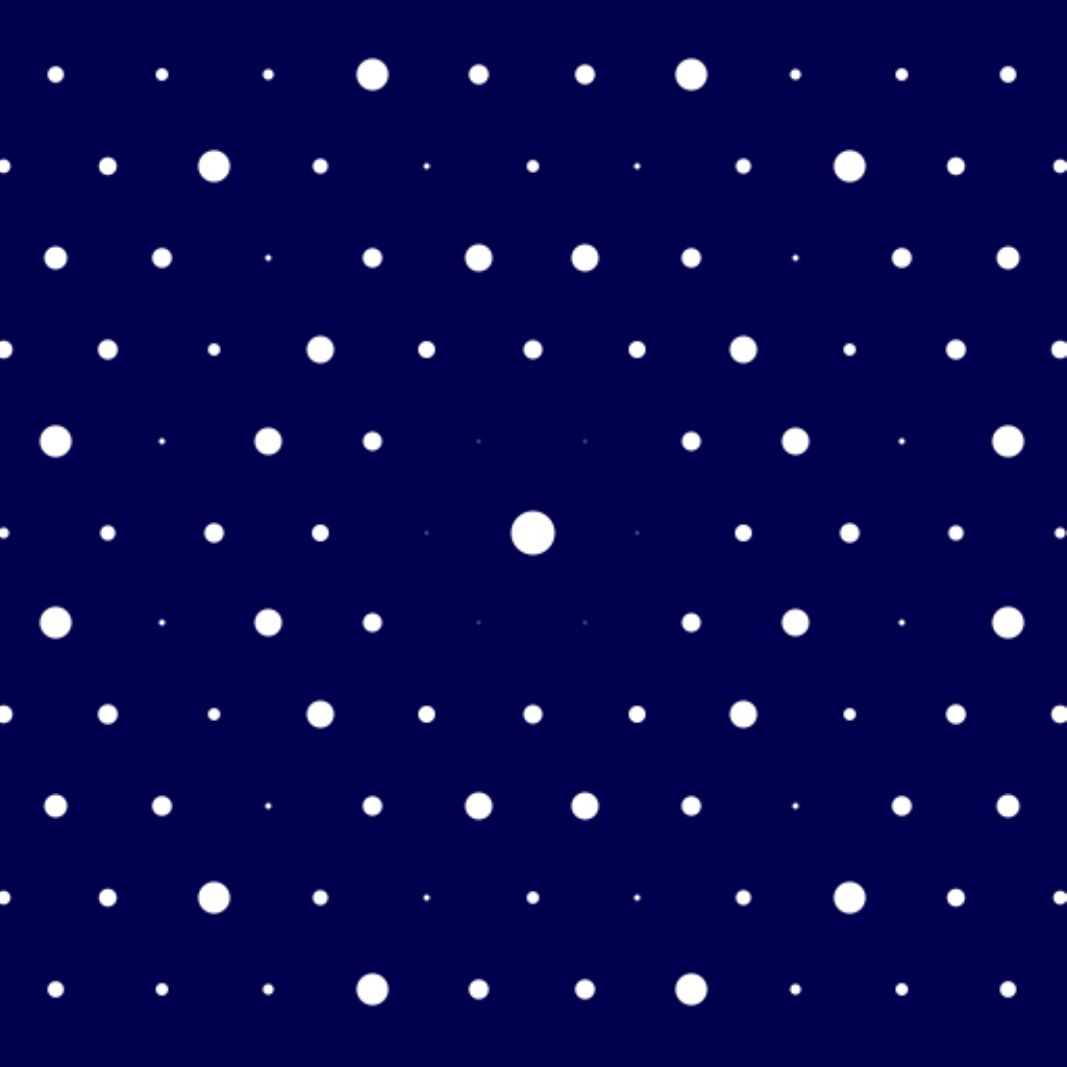}
    \caption{Diffraction image for P31m perfect crystal ($k\rightarrow \infty$) calculated by Eq.~(\ref{SFforCry}) rotated 90$^\circ$. The area is proportional to the intensity.
    }
    \label{fig: simulationimage0}
  \end{center}
\end{figure*}
\clearpage
\section{Crystallographic description of the P31m polymer blend system}
The same plane group is adopted by an ABC triblock terpolymer/homopolymer blend, where dark gray circles represent polyisoprene (PI), light gray circles represent poly(2-vinylpyridine) (PVP), and the matrix region is polystyrene (PS),
as shown in Supplementary Fig.~\ref{fig:SI2}.
We simply assume that domains of PI and PVP are all circles, and that the centers of polyisoprene domains occupy the same positions as the previous colloidal particle system. 
Let A, B, $\cdots$, I be the centers of circular domains, and 
let the distance $\overline{\rm OG}=d$ for the center positions of PVP. Although it is a crude treatment, we determine $d=\frac{3}{5}$ %
by minimizing the next (entropic) elastic free-energy function $S(d)$ inside a pentagon (BCFED):
\begin{eqnarray}
S(d) \propto \overline{\rm BI}^2+\overline{\rm CI}^2+\overline{\rm FI}^2+\overline{\rm EI}^2+\overline{\rm DI}^2=\overline{\rm AG}^2+2\overline{\rm BG}^2+2\overline{\rm DG}^2.
\end{eqnarray}
Wyckoff Positions for a polymer blend are shown in Supplementary Table~\ref{table:SI2}.

\begin{figure}[h]
\begin{center}
\includegraphics[width=0.6\linewidth]{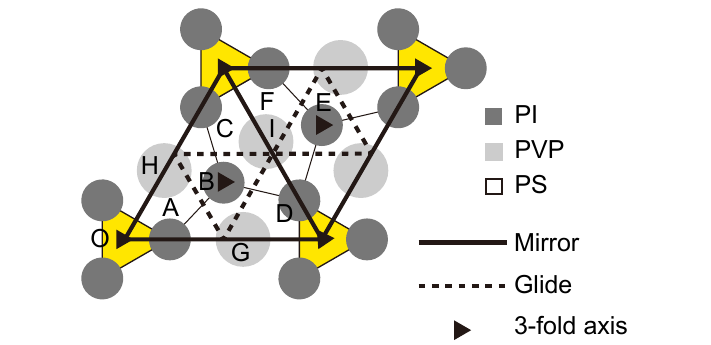}
\caption{Plane group P31m for a polymer blend.}
\label{fig:SI2}
\end{center}
\end{figure} 
\begin{table}[htp]
\begin{center}
\begin{tabular}{|c|c|c|c|c|}
\hline
Polymer & Wyckoff Position & Site Sym. & Points & Coordinates\\
\hline
PI & 2b & 3.. & B, E &  
$\left(\frac{1}{2}, \frac{\sqrt{3}}{6}\right)$, $\left(1, \frac{\sqrt{3}}{3}\right)$\\
PI & 3c & ..m & A, C, D & $(c, 0)$, $\left(\frac{1}{2}(1-c), \frac{\sqrt{3}}{2}(1-c)\right)$, $\left(1-\frac{1}{2}c, \frac{\sqrt{3}}{2}c\right)$\\
\hline
PVP & 3c & ..m & G, H, I & $\left(d, 0\right)$, $\left(\frac{1}{2}(1-d), \frac{\sqrt{3}}{2}(1-d)\right)$, $\left(1-\frac{1}{2}d, \frac{\sqrt{3}}{2}d\right)$\\
\hline
\end{tabular}
\end{center}
\caption{Wyckoff Position for a polymer blend. $c=\frac{\sqrt{33}-3}{12}=0.2287$ and $d=\frac{3}{5}$.}
\label{table:SI2}
\end{table}%

\clearpage
\section{Pentagon tilings}

\begin{figure}[h]
\begin{center}
\includegraphics[width=0.6\linewidth]{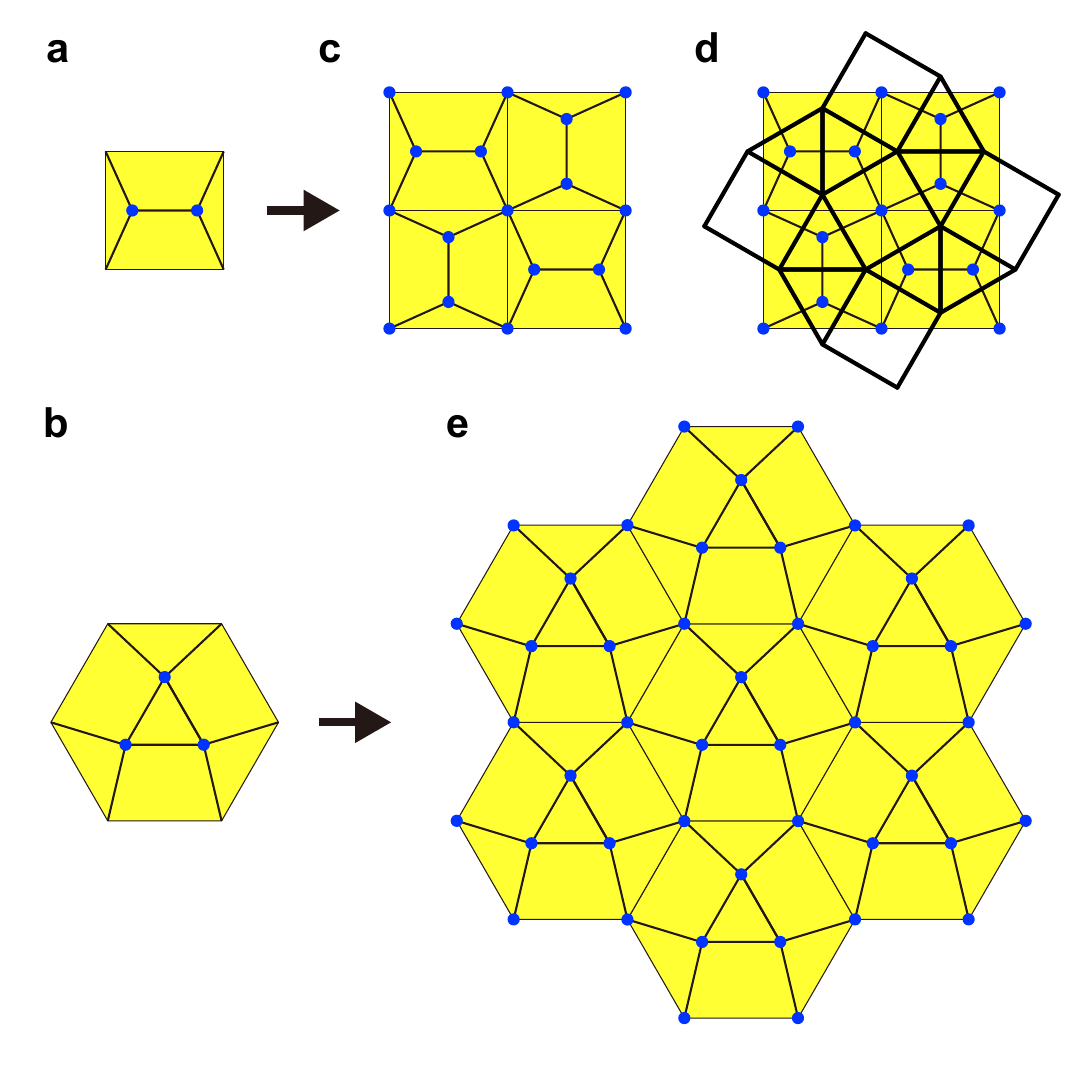}
\caption{Equilateral pentagon tilings. {\bf a}, elemental square, {\bf b} elemental hexagon, {\bf c}, 4-fold pentagonal Cairo tiling, {\bf d}, duality of the Cairo tiling and the $3^2.4.3.4$ Archimedean tiling, {\bf e,} 3-fold pentagon-triangle tiling.}
\label{fig:pentagon}
\end{center}
\end{figure}

We demonstrate the accommodation of pentagons within both a square and a hexagon. Firstly, we consider a square with two points placed inside (Supplementary Fig.~\ref{fig:pentagon}{\bf a}), and secondly, a hexagon with three points inside (Supplementary Fig.~\ref{fig:pentagon}{\bf b}). By connecting the points, we form pentagonal tilings. In Supplementary Fig.~\ref{fig:pentagon}{\bf c}, we present the equilateral Cairo pentagonal tiling, which serves as the dual of the $3^2.4.3.4$ Archimedean tiling with the P4gm plane group (Supplementary Fig.~\ref{fig:pentagon}{\bf d}). The $3^2.4.3.4$ Archimedean tiling is associated with the $\sigma$ phase found in complex metallic and soft-matter phases. It is recognized as a periodic approximant of dodecagonal quasicrystals. Additionally, Supplementary Figure~\ref{fig:pentagon}{\bf e} illustrates a 3-fold equilateral pentagon-triangle tiling with the P31m plane group discussed in the present paper. Despite their distinct symmetries, these two structures are closely related and are commonly observed in soft-matter systems.

\clearpage
\section{Colloidal system}
Engle used the Lennard-Jones-Gauss potential of the form\cite{Engel2007,Engel2011}:
\begin{equation}
V(r)=\frac{1}{r^{12}}-\frac{2}{r^{6}}-\epsilon\exp\left(-\frac{(r-r_0)^2}{2\sigma^2}\right),
\end{equation}
with parameters $\sigma^2=0.042$, $\epsilon=1.8$, $r_0=1.42$.
We have conducted Monte Carlo simulation with 
{\it NPT} ensemble at $T=0.29$ and $T=0.28$, $P=0.0$, and $N=19740$.
Snapshots of modulated structures quenched at $T=0.12$ 
are displayed in Supplementary Fig.~\ref{fig: simulationimage} ({\bf a}, $T=0.29$ and {\bf b}, $T=0.28$).
The diffraction images in Supplementary Fig.~\ref{fig: simulationimage} ({\bf c}, $T=0.29$ and {\bf d}, $T=0.28$) are obtained by the average of 200 Fourier transforms of quenched samples at $T=0.12$. 
In contrast to the diffraction image for the P31m perfect crystal, satellite peaks are observed.

\begin{figure*}[htb]
  \begin{center}
     \includegraphics[width=0.93\linewidth]{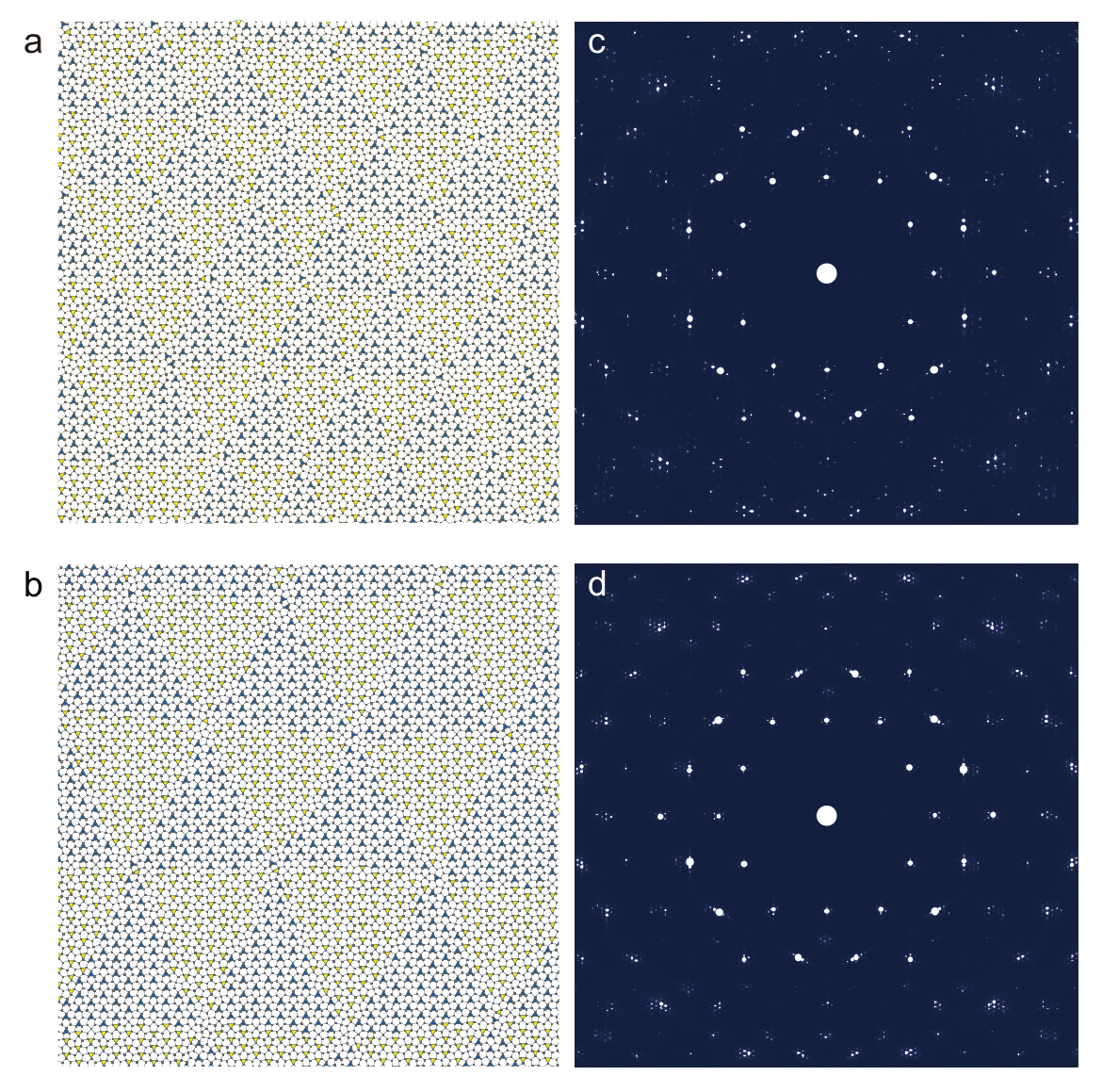}
    \caption{Colloidal simulation.      
    Modulated domain structure ({\bf a}, $T=0.29$ and {\bf b}, $T=0.28$) and  Diffraction image ({\bf c}, $T=0.29$ and {\bf d}, $T=0.28$). The area is proportional to the intensity.
    }
    \label{fig: simulationimage}
  \end{center}
\end{figure*}

\clearpage
\section{Atomic decorations}
In this section, we consider the decorations on the hexagonal metallic-mean tilings,
which should be relevant for the atomic structure of the MC and MD simulations~\cite{Engel2011}.
In a certain parameter regime, MC and MD simulations show the stable solution where
the atomic structure is described by large and small hexagons, and parallelograms,
as shown in Supplementary Fig.~\ref{fig: deco}.
The atomic positions are given in Supplementary Note~\ref{sec: partices}.
\begin{figure}[htb]
  \begin{center}
    \includegraphics[width=0.7\linewidth]{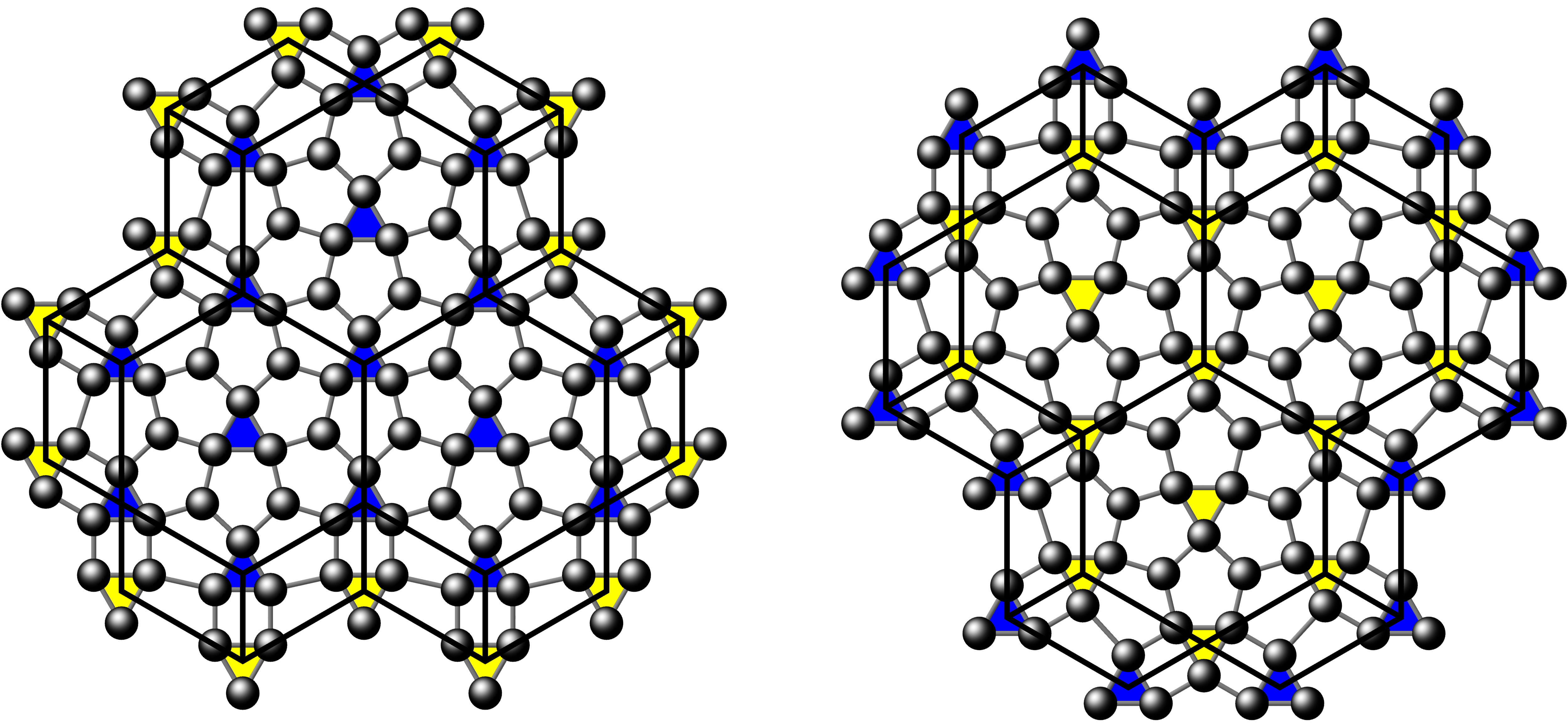}
    \caption{
    Atomic structures of the MC and MD simulations.
    The circles represent the locations of atoms.}
    \label{fig: deco}
  \end{center}
\end{figure}
We find that three atoms are located in the vicinity of each vertex in the tilings 
while the others are located around the center of the L tile.

To clarify how relevant the hexagonal metallic-mean tilings 
proposed in this study are for the atomic structure observed in MD and MC simulations,
we first introduce an auxiliary vertex C$_4^\triangle$ (C$_4^\triangledown$) 
placed at the center of the L$_\triangle$ (L$_\triangledown$) tile,
as shown in Supplementary Fig.~\ref{fig: C4}{\bf a}.
\begin{figure}[htb]
  \begin{center}
    \includegraphics[width=0.7\linewidth]{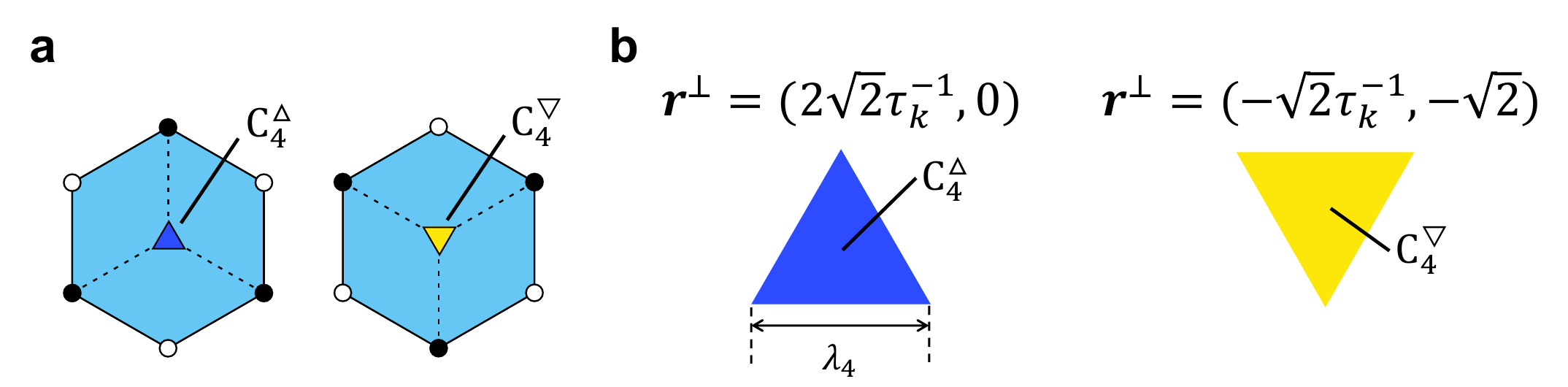}
    \caption{
    {\bf a} C$_4^\triangle$ and C$_4^\triangledown $vertices
    which are defined by
    dividing L$_\triangle$ and L$_\triangledown$ tiles into three rhombuses.
    {\bf b} Window of the C$_4^\triangle$ (C$_4^\triangledown$) vertex
    in the perpendicular space
    ${\bf r}^\perp =(2\sqrt{2}\tau^{-1}_k,0)$ [$(-\sqrt{2}\tau^{-1}_k,-\sqrt{2})$].
    $\lambda_4$ is the edge length of the triangles.
    }
    \label{fig: C4}
  \end{center}
\end{figure}
Since the number of the C$_4^\alpha$ ($\alpha=\triangle, \triangledown)$ vertices corresponds to that of the L$_\alpha$ tiles,
its fraction is given as,
\begin{equation}
  f_{\rm C_4^\alpha} = \frac{1}{4}\frac{\tau^2_k}{\tau^2_k + 3\tau_k + 1}.
\end{equation}
When the L$_\alpha$ tile is divided into three rhombuses, as shown in Supplementary Fig.~\ref{fig: C4}{\bf a},
each C$_4^\alpha$ vertex connects 
to the corner vertices in the B sublattice (the solid circles), and 
it can be regarded to belong to the A sublattice.
Since the distance between the C$_4^\alpha$ and corner vertices corresponds to the longer length $\ell$,
the set of the six integer indices for the C$_4^\alpha$ vertex is given by
\begin{eqnarray}
  \vec{n}_{{\rm C}_4^\alpha}=\vec{n}_{corner}+\varDelta\vec{n},
\end{eqnarray}
where $\vec{n}_{corner}$ is the set of the six integer indices for a certain corner vertex in the B sublattice and
$\varDelta \vec{n} = (n_0, n_1, n_2, 0, 0, 0)^T$ with integer $n_m$.
This means that the C$_4^\triangle$ and C$_4^\triangledown$ vertices
are clearly distinguished in the perpendicular space 
since $y^\perp$ $[=-\sqrt{2} \vec{n}_{{\rm C}_4^\alpha}\cdot (0,0,0,1,1,1)^T]$ for the C$_4^\alpha$ vertices
is the same as that for the vertices 
belonging to the honeycomb domain with $\alpha$.
Namely, the window for the C$_4^\triangle$ (C$_4^\triangledown$) vertex 
appears in the perpendicular spaces with ${\bf r}^\perp =(2\sqrt{2}\tau^{-1}_k,0)$ [$(-\sqrt{2}\tau^{-1}_k,-\sqrt{2})$],
as shown in Supplementary Fig.~\ref{fig: C4}{\bf b}. 
The edge length of the windows is given by $\lambda_4 = \sqrt{3}$,
by taking into account the fraction of the vertices.
Supplementary Figure~\ref{fig: C4-tiling} shows the hexagonal bronze-mean tiling with the C$_4$ vertices,
which is obtained by means of the cut-and-projection scheme.
We find that the C$_4^\alpha$ auxiliary vertices indeed belong to the honeycomb domain with $\alpha$.
To discuss the spatial structure of the vertices in the perpendicular space in detail, it is useful to consider the corresponding windows in three dimensions $(\tilde{x}, \tilde{y}, \tilde{z}=x^\perp+y^\perp)$, as discussed in the main text. Supplementary Figure~\ref{fig: rhombohedron} clarifies that these six windows are the crosssections of the trigonal trapezohedron with the edge length $\sqrt{3}(1+\tau^{-1}_k)$. We find that the vertices in the honeycomb domains with $\triangle$ ($\triangledown$) are located in the upper (lower) windows, which are shown as the blue (red) triangles and hexagon. This is consistent with the fact that certain two windows belonging to the same (different) honeycomb domains are bridged by the vectors ${\bf e}^h_0,{\bf e}^h_1,{\bf e}^h_2$ (${\bf e}^h_3,{\bf e}^h_4,{\bf e}^h_5$). The middle four windows can be regarded as the crosssections of the regular octahedron, which is shown in Extended Data Fig.~1 in the main text.

\begin{figure}[htb]
  \begin{center}
    \includegraphics[width=0.5\linewidth]{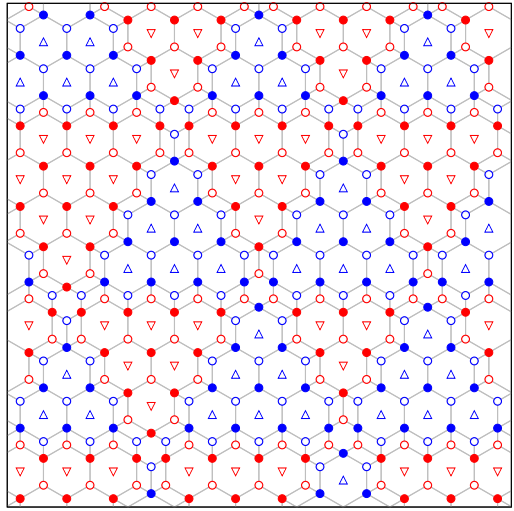}
    \caption{
      Hexagonal bronze-mean tiling with the C$_4$ vertices.
    Open (solid) circles at the vertices indicate the sublattice A (B).
    Open triangles represent the auxeliary C$_4$ vertices.
    Blue (red) symbols represent the vertices belonging to 
    the honeycomb domains with $\triangle$ ($\triangledown$).
    }
    \label{fig: C4-tiling}
  \end{center}
\end{figure}

\begin{figure}[htb]
  \begin{center}
    \includegraphics[width=0.6\linewidth]{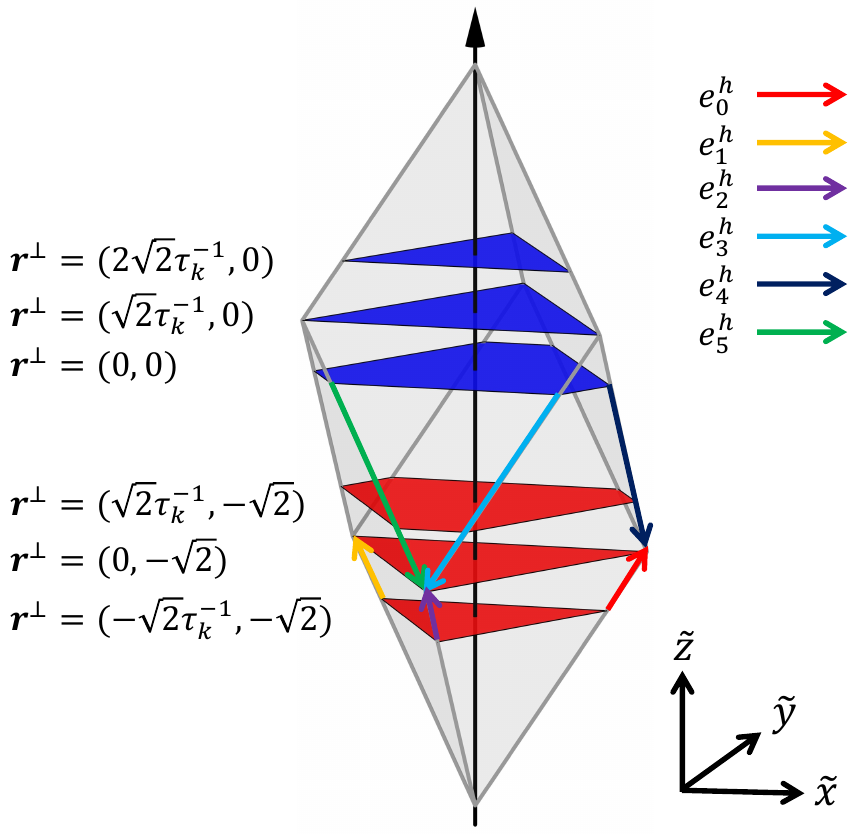}
    \caption{
    The perpendicular space
    [${\bf r}^\perp = (2\sqrt{2}\tau_k^{-1},0), (\sqrt{2}\tau_k^{-1},0),(0,0)$,
    $(\sqrt{2}\tau_k^{-1},-\sqrt{2}), (0,-\sqrt{2})$, and $(-\sqrt{2}\tau_k^{-1},-\sqrt{2})$.] of the hexagonal bronze-mean tiling with $k=3$ in three dimensions with $(\tilde{x}, \tilde{y}, \tilde{z}=x^\perp+y^\perp)$. The colored arrows indicate the projected basis vectors ${\bf e}^h_i$ and the blue (red) triangles and hexagon indicate the windows for the honeycomb domains with $\triangle$ ($\triangledown$).
    }
    \label{fig: rhombohedron}
  \end{center}
\end{figure}

Next, we consider the atomic decorations for the hexagonal metallic-mean tilings.
One can divide the atoms into three groups with $\alpha(=\triangle, \triangledown)$, 
as shown in Supplementary Fig.~\ref{fig: atom}.
The group u$_1^\alpha$ is composed of the atoms around the vertex in the hexagonal metallic-mean tiling and
the group u$_2^\alpha$ is composed of the atoms around the auxeliary vertex C$_4^\alpha$ (the center of L$_\alpha$ tiles).
The other group  v$^\alpha$ are located in the L$_\alpha$ tile, but far from the center.

Now, we would like to describe the atomic structures,
specifying the sets of the six indices for these atoms.
First, we focus on certain atoms in the groups u$_1^\triangle$ and u$_1^\triangledown$
around the vertex X$^\triangle$ and X$^\triangledown$,
which are pointed in Supplementary Figs.~\ref{fig: atom}{\bf a} and {\bf b}, respectively.
We find that the atom with $\alpha = \triangle$ ($\triangledown$)
is located in the backward (forward) direction of the vectors
${\bf e}_0$ %
from X$^\triangle$ (X$^\triangledown$).
Since the atoms also belong to the honeycomb domain with $\alpha$,
the sets of the six indices $\vec{n}_{\rm atom}$ for these atoms are given
as
\begin{eqnarray}
  &&\vec{n}_{\rm atom} = \vec{n}_{{\rm X}^\alpha} + \vec{n},\\
  &&\vec{n} = \left(\mp c,0,0,0,0,0\right)^T,
\end{eqnarray}
where
$\vec{n}_{{\rm X}^\alpha}$ is the set of the six integer indices for the X$^\alpha$ vertex, and
the above (under) sign corresponds to $\alpha = \triangle$ ($\triangledown$).
In the perpendicular space $\tilde{\cal S}$, the window of the corresponding atoms appears
$\tilde{\bf d}$ away from the window of X$^\alpha$,
where
$\tilde{\bf d}= \mp c\tilde{\bf e}_0$.
Examining the positions for the atoms of the u$_1^\alpha$, u$_2^\alpha$, and v$^\alpha$ groups,
we determine the sets of the six indices, which are explicitly shown in Supplementary Table~\ref{table: deco}.
Supplementary Figure~\ref{fig: PS_u}{\bf a} ({\bf b}) shows the windows
for the groups ${\rm u}^\triangle_1 , {\rm u}^\triangle_2$ (${\rm u}^\triangledown_1 , {\rm u}^\triangledown_2$)
in the perpendicular spaces.
Supplementary Figure~\ref{fig: PS_v}{\bf a} ({\bf b}) shows the windows
for the group ${\rm v}^\triangle$ (${\rm v}^\triangledown$)
in the perpendicular space with
${\bf r}^\perp = (2\sqrt{2}\tau^{-1}_k,0)$ $[(-\sqrt{2}\tau^{-1}_k,-\sqrt{2})]$.
We generate the hexagonal metallic-mean tilings
with the atomic decorations by the cut-and-project scheme,
which are schematically shown in Supplementary Fig.~\ref{fig: deco-tiling}.

\begin{figure}[htb]
  \begin{center}
    \includegraphics[width=0.6\linewidth]{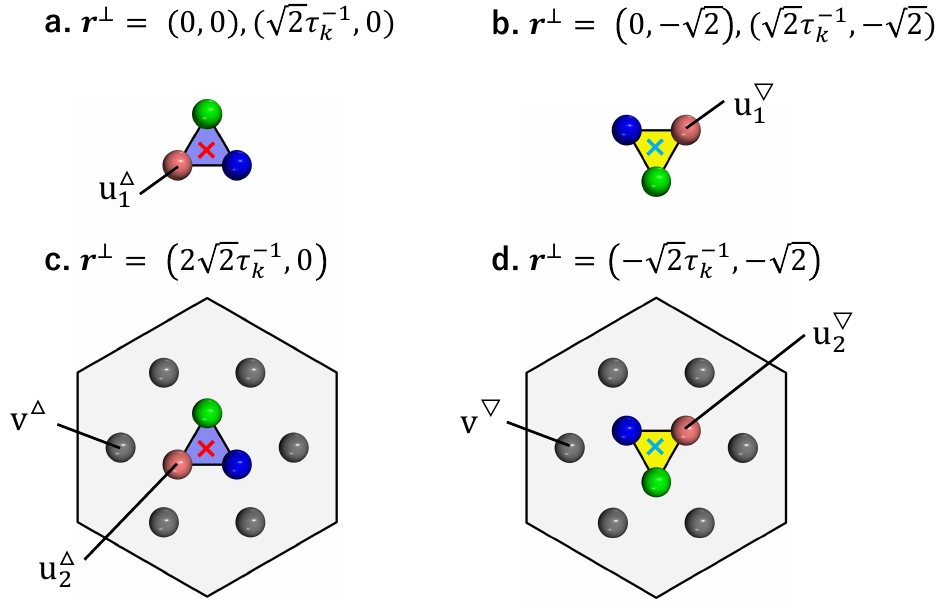}
    \caption{
      Positions of atoms in the decorated tiling.
      Blue, green, and red circles represent
      three kinds of the atoms in the u$_1$ and u$_2$ groups
      and gray circles represent the atoms in the v group.
    {\bf a} ({\bf b}) Red (blue) cross represents
    the vertex in the honeycomb domain with $\triangle$ ($\triangledown$), 
    which is mapped to the perpendicular space in
    ${\bf r}^\perp  = (0,0), (\sqrt{2}\tau^{-1}_k,0)$ [$(0,-\sqrt{2}), (\sqrt{2}\tau^{-1}_k,-\sqrt{2})$].
    {\bf c} ({\bf d}) Atomic decorations of the L$_\triangle$ (L$_\triangledown$) tile.
    Red (blue) cross represents
    the auxiliary C$_4^\triangle$ (C$_4^\triangledown$) vertex,
    which is mapped to the perpendicular space in
    ${\bf r}^\perp = (2\sqrt{2}\tau^{-1}_k,0)$ $[(-\sqrt{2}\tau^{-1}_k,-\sqrt{2})]$.
    }
    \label{fig: atom}
  \end{center}
\end{figure}

\begin{table}[htp]
  \begin{center}
  \begin{tabular}{|c|c|c|c|}
  \hline
  Group & ${\bf r}^\perp$ & Window & $3\vec{n}^T$ \\
  \hline
  u$_1^\triangle$ & $(-\sqrt{2}c\tau^{-1}_k,0),(\sqrt{2}\tau^{-1}_k-\sqrt{2}c\tau^{-1}_k,0)$ & \multirow{2}{*}{Hexagon} &
  \multirow{4}{*}{
    \begin{tabular}{l}
    $(\mp 3c,0,0,0,0,0)$\\
    $(0,\mp 3c,0,0,0,0)$\\
    $(0,0,\mp 3c,0,0,0)$   
    \end{tabular}
  }\\
  \cline{1-2}
  u$_1^\triangledown$ & $(\sqrt{2}c\tau^{-1}_k,-\sqrt{2}),(\sqrt{2}\tau^{-1}_k+\sqrt{2}c\tau^{-1}_k,-\sqrt{2})$ & &\\
  \cline{1-3}
  u$_2^\triangle$ &$(2\sqrt{2}\tau^{-1}_k-\sqrt{2}c\tau^{-1}_k,0)$ & \multirow{2}{*}{Triangle} &\\  
  \cline{1-2}
  u$_2^\triangledown$ &  $(-\sqrt{2}\tau^{-1}_k+ \sqrt{2}c\tau^{-1}_k,-\sqrt{2})$ & &\\  
  \hline
  \multirow{3}{*}{
    \begin{tabular}{c}
      v$^\triangle$
    \end{tabular}
  } &
  \multirow{3}{*}{
    \begin{tabular}{c}
      $(2\sqrt{2}\tau^{-1}_k,0)$ 
    \end{tabular}
  } &
  \multirow{7}{*}{Triangle} &
  \multirow{7}{*}{
    \begin{tabular}{l}
    $(-1,1,0,0,0,0)$\\     
    $(1,-1,0,0,0,0)$\\     
    $(-1,0,1,0,0,0)$\\     
    $(1,0,-1,0,0,0)$\\     
    $(0,-1,1,0,0,0)$\\     
    $(0,1,-1,0,0,0)$  
    \end{tabular}
  }\\
  &&&\\
  &&&\\
  \cline{1-2}
  \multirow{4}{*}{
    \begin{tabular}{c}
      v$^\triangledown$
    \end{tabular}
  } &
  \multirow{4}{*}{
    \begin{tabular}{c}
      $(-\sqrt{2}\tau^{-1}_k,-\sqrt{2})$
    \end{tabular}
  } & & \\
  &&&\\
  &&&\\
  &&&\\
  \hline
  \end{tabular}
  \end{center}
  \caption{
  Occupied plane ${\bf r}^\perp$ and corresponding window shape in the perpendicular space.
  Three sets of six indices for the u$_1$ and u$_2$ groups are given, where
  the above (under) sign corresponds to $\triangle$ ($\triangledown$).
  Six sets of six indices for the v group are also given.
  $c=(\sqrt{33}-3)/12=0.2287$.}
  \label{table: deco}
\end{table}

\begin{figure}[htb]
  \begin{center}
    \includegraphics[width=\linewidth]{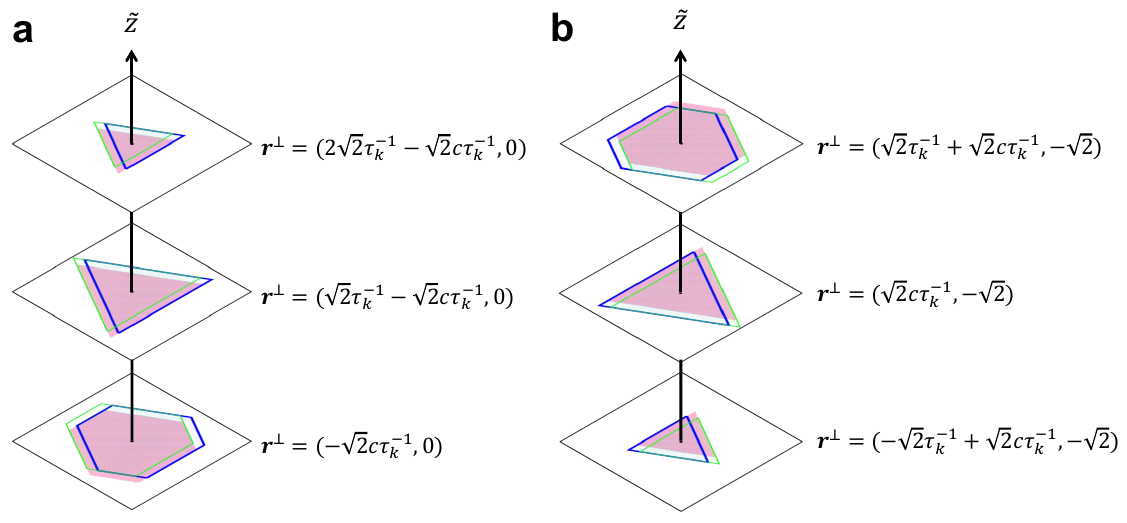}
    \caption{
      {\bf a} ({\bf b})
    Windows for the ${\rm u}^\triangle_1 , {\rm u}^\triangle_2$
    (${\rm u}^\triangledown_1 , {\rm u}^\triangledown_2$) groups
    in the perpendicular spaces %
    for the hexagonal golden-mean tiling with $k=1$.
    Each layer has three distinct windows, whose colors correspond to
    the colors fo the atom shown in Supplementary Fig.~\ref{fig: atom}.
    }
    \label{fig: PS_u}
  \end{center}
\end{figure}

\begin{figure}[htb]
  \begin{center}
    \includegraphics[width=0.9\linewidth]{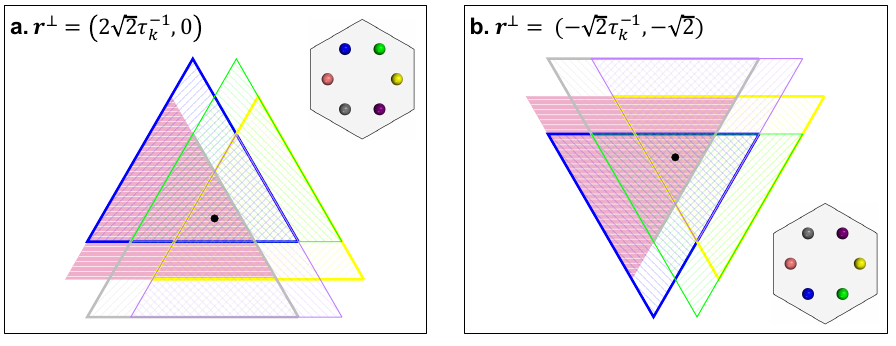}
    \caption{
    Windows for the ${\rm v}^\triangle$ and ${\rm v}^\triangledown$ groups
    in the perpendicular spaces with {\bf a}  ${\bf r}^\perp = (2\sqrt{2}\tau^{-1}_k,0)$
    and {\bf b} $(-\sqrt{2}\tau^{-1}_k,-\sqrt{2})$
    for the hexagonal golden-mean tiling with $k=1$, respectively.
    Insets show the positions of atoms in the v group.
    Each layer has six distinct triangular windows, whose colors correspond to
    the colors of the atoms shown in the inset.
    }
    \label{fig: PS_v}
  \end{center}
\end{figure}

\begin{figure}[htb]
  \begin{center}
    \includegraphics[width=0.8\linewidth]{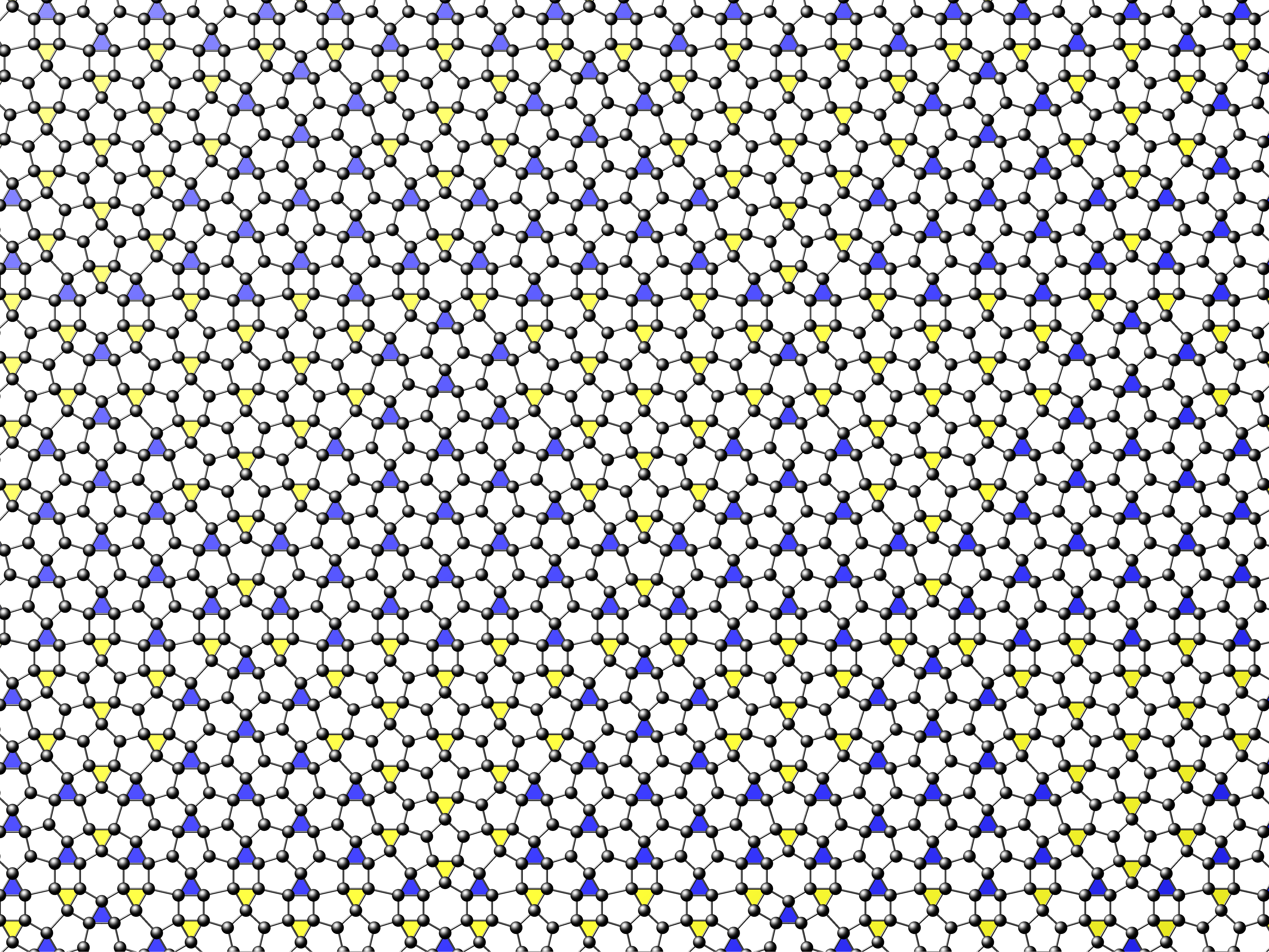}
    \caption{Decorated hexagonal silver-mean tilings. }
    \label{fig: deco-tiling}
  \end{center}
\end{figure}
\clearpage

We calculate the lattice structure factors of the decorated tilings,
extending the method shown in Supplementary Note~\ref{app: fourier}.
The atoms in the decorated tiling are divided into three groups $u_1$, $u_2$, and $v$,
as shown in Supplementary Fig.~\ref{fig: atom}.
The density of the atoms is then given as
\begin{eqnarray}
  \rho({\bf r})&=&\rho_{u_1}({\bf r})+\rho_{u_2}({\bf r})+\rho_{v}({\bf r}),\\
  \rho_{u_1}({\bf r})&=&\sum_{n=0}^2\left[
    \left( \sum_{j \in (0,0)^\perp}+\sum_{j \in (1,0)^\perp} \right)
    \delta \left({\bf r}_j - c {\bf e}_n - {\bf r}\right)\right] \nonumber\\
    & &+\sum_{n=0}^2\left[
      \left(\sum_{j \in (0,1)^\perp}+\sum_{j \in (1,1)^\perp} \right)
      \delta \left({\bf r}_j + {\bf e}_n  - {\bf r}\right)  \right],\\
  \rho_{u_2}({\bf r})&=&\sum_{n=0}^2\left[
        \sum_{j \in (2,0)^\perp}
        \delta \left({\bf r}_j - c {\bf e}_n  - {\bf r}\right)\right]
        +\sum_{n=0}^2\left[
          \sum_{j \in (-1,1)^\perp}
          \delta \left({\bf r}_j + c {\bf e}_n  - {\bf r}\right)  \right],\\
  \rho_{v}({\bf r})&=&\sum_{n=0}^5\left[
            \left( \sum_{j \in (2,0)^\perp}+\sum_{j \in (-1,1)^\perp} \right)
            \delta\left({\bf r}_j +R^n\left\{\frac{{\bf e}_1 -{\bf e}_0}{3}\right\}-{\bf r}\right)\right],
\end{eqnarray}
where ${\bf r}_i$ represents the position of the vertex site in the tiling and
the operator $R$ rotates the vectors by the angle $\pi/3$, as
\begin{equation}
  R=\left(\begin{array}{cc}
  \cos(\pi/3) & -\sin(\pi/3) \\
  \sin(\pi/3) & \cos(\pi/3)
  \end{array}\right).
\end{equation}
The operation is represented as $R {\bf e}_i= -{\bf e}_j$ $(i=0,1,2)$,
where $j = i-1 + 3\delta_{0i}$.
We have used the position of atoms, which are explicitly shown in Supplementary Table~\ref{table: deco}.
In the lattice structure factors, the peak positions are represented
as ${\bf q} = \sum_{m=0}^5 \tilde{n}_m {\bf q}_m$, where $\tilde{n}_m$ is an integer
and ${\bf q}_m$ is the reciprocal vector derived in Supplementary Note~\ref{6d}.
The Fourier transform of the density
$\rho({\bf q}) = \int \rho ({\bf r}) \exp(-i{\bf r} \cdot {\bf q}) d{\bf r}$
can be represented in terms of the Fourier transform of the occupation domains ${\cal F}_{mn}$ as,
\footnotesize
\begin{eqnarray}
  \rho({\bf q})
    &=&\exp(-i \sqrt{2}c\tau^{-1}_k q^\perp_0)
    \sum_{n=0}^2 \exp(-ic\tilde{\bf e}_n \cdot \tilde{\bf q})
    \Big[ {\cal F}_{00}(\tilde{\bf q})
    +\exp(i \sqrt{2}\tau^{-1}_k q^\perp_0){\cal F}_{10}(\tilde{\bf q})
    +\exp(i 2 \sqrt{2}\tau^{-1}_k q^\perp_0){\cal F}_{20}(\tilde{\bf q})\Big]\nonumber\\
    &+& \exp(i \sqrt{2}c\tau^{-1}_k q^\perp_0)
    \sum_{n=0}^2 \exp(ic\tilde{\bf e}_n \cdot \tilde{\bf q})\exp(-i\sqrt{2} q^\perp_1)
    \Big[{\cal F}_{01}(\tilde{\bf q})+ \exp(i \sqrt{2}\tau^{-1}_k q^\perp_0){\cal F}_{11}(\tilde{\bf q})+\exp(-i \sqrt{2}\tau^{-1}_k q^\perp_0){\cal F}_{-11}(\tilde{\bf q})\Big]\nonumber\\
    &+&\sum_{n=0}^5 \exp \left(iR^n\left\{\frac{\tilde{\bf e}_1 -\tilde{\bf e}_0}{3}\right\} \cdot \tilde{\bf q} \right)
    \Big[\exp(i 2 \sqrt{2}\tau^{-1}_k q^\perp_0) {\cal F}_{20}(\tilde{\bf q})
    + \exp(-i \sqrt{2}\tau^{-1}_k q^\perp_0)\exp(-i \sqrt{2}q^\perp_1){\cal F}_{-11}(\tilde{\bf q})\Big].
\end{eqnarray}
\normalsize
Then, we obtain the lattice structure factors $|\rho({\bf q})|^2$.

\begin{figure*}[htb]
  \begin{center}
     \includegraphics[width=0.93\linewidth]{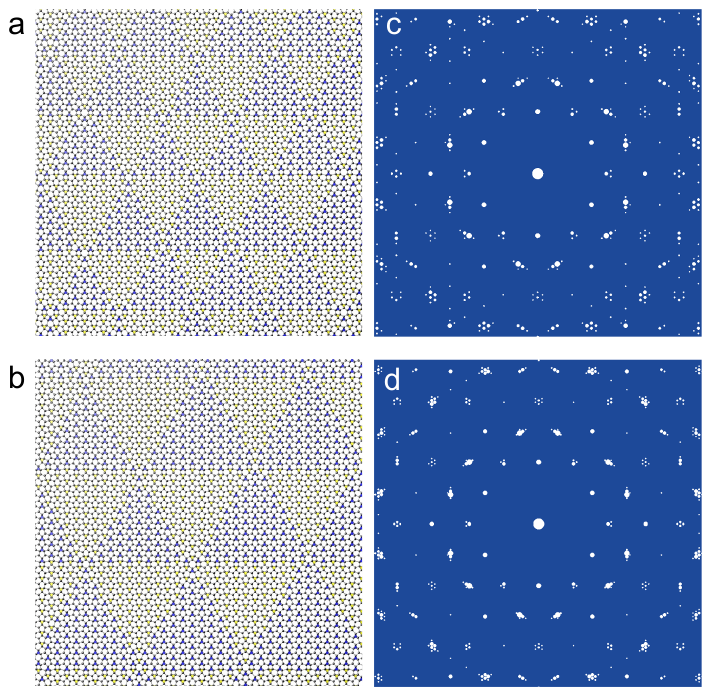}
    \caption{
    The decorated tilings ({\bf a}, $k=3$ and {\bf b}, $k=5$) and  the lattice structure factors ({\bf c}, $k=3$ and {\bf d}, $k=5$). The area is proportional to the intensity.
    }
    \label{fig: fourier_atoms}
  \end{center}
\end{figure*}

Supplementary Figure~\ref{fig: fourier_atoms} shows the decorated tilings and 
the corresponding lattice structure factors for $k=3$ and $k=5$.
We confirm that the peak structures of the decorated tilings are in a good agreement with
the MC results shown in Supplementary Fig.~\ref{fig: simulationimage}.

\bibliography{./suppl-refs}